\documentclass[a4paper,11pt]{article}
\usepackage{jinstpub} 

\title{\boldmath Response of the first POLAR-2 Prototype to Polarized Beams}

\author[a,b]{Merlin~Kole} 
\author[a,c]{Nicolas De Angelis}  
\author[d,e]{Ana Bacelj}
\author[a]{Franck Cadoux}  %
\author[f]{Agnieszka Elwertowska}
\author[a]{Johannes Hulsman}  
\author[g]{Hancheng Li} 
\author[f]{Grzegorz Łubian} 
\author[f]{Tomasz Kowalski} 
\author[g]{Gilles Koziol}  %
\author[f]{Agnieszka Pollo}  
\author[g]{Nicolas Produit}  
\author[f]{Dominik Rybka}  
\author[a]{Adrien Stil}  %
\author[h]{Jianchao Sun}  
\author[a]{Xin Wu}
\author[f]{Kacper Zezuliński} 
\author[h]{Shuang-Nan Zhang}  
\affiliation[a]{DPNC, University of Geneva, 24 Quai Ernest-Ansermet, CH-1205 Geneva, Switzerland}
\affiliation[b]{University of New Hampshire, Space Science Center, University of New Hampshire, Durham, NH 03824, USA}
\affiliation[c]{INAF-IAPS, via del Fosso del Cavaliere 100, I-00133 Rome, Italy}
\affiliation[d]{Max-Planck Institute for Extraterrestrial Physics, Giessenbachstr. 1, 85748 Garching, Germany}
\affiliation[e]{TUM School of Natural Sciences (Physics Department), Boltzmannstr. 10 85748 Garching}
\affiliation[f]{National Centre for Nuclear Research, ul. A. Soltana 7, 05-400 Otwock, Swierk, Poland}
\affiliation[g]{Geneva Observatory, ISDC, University of Geneva, 16, Chemin d’Ecogia, CH-1290 Versoix, Switzerland}
\affiliation[h]{Key Laboratory of Particle Astrophysics, Institute of High Energy Physics, Chinese Academy of Sciences, Beijing 100049, China}

\emailAdd{merlin.kole@unh.edu}

\abstract{POLAR-2 is a dedicated gamma-ray polarimeter currently foreseen to be launched towards the China Space Station around 2027. The design of the detector is based on the legacy of its predecessor mission POLAR which was launched in 2016. POLAR-2 aims to measure the polarization of the Gamma-ray Burst prompt emission within the $30-800\,\mathrm{keV}$ energy range. Thanks to its high sensitivity to gamma-ray polarization, as well as its large effective area, POLAR-2 will provide the most precise measurements of this type to date. Such measurements are key to improve our understanding of the astrophysical processes responsible for Gamma-Ray Bursts. The detector consists of a segmented array of plastic scintillator bars, each one of which is read out by a Silicon PhotoMultiplier channel. The flight model of POLAR-2 will contain a total of 6400 scintillators. These are divided into 100 groups of 64 bars each, in so-called polarimeter modules. In recent years, the collaboration has designed and produced the first prototypes of these polarimeter modules and subjected these to space qualification tests. In addition, in April 2023, the first of these modules were calibrated using fully polarized gamma-ray beams at the European Synchrotron Radiation Facility (ESRF) in France. In this work, we will present the results of this calibration campaign and compare these to the simulated performance of the POLAR-2 modules. Potential improvements to the design are also discussed. Finally, the measurements are used, in combination with the verified simulation framework, to estimate the scientific performance of the full POLAR-2 detector and compare it to its predecessor.\\}

\keywords{Gamma detectors, Polarimeters, Data analysis, X-ray detectors and telescopes}

\begin{document}
\maketitle
\flushbottom

\section{Introduction}\label{sec:intro}

Gamma--ray bursts (GRBs) were first discovered in 1967 and remain one of the most researched topics in astrophysics. A GRB consists of a gamma--ray component which lasts from seconds to minutes, called the prompt emission, followed by a longer--lasting afterglow which can be observed over a wide range of the electromagnetic spectrum. Thanks to over 50 years of research and the detection of over 10'000 GRBs, much has been learned about these phenomena. Insights into their origin were gained through precise measurements of their location in the sky which indicate that they have an extra--galactic origin \cite{Meegan:1992xg, Costa:1997obd, vanParadijs:1997wr, 1997Natur.387..878M}. Measurements of a second parameter of the gamma-rays, their time of arrival, has led to measurements of the duration of GRBs. These in turn allowed GRBs to be divided into two classes \cite{Kouveliotou:1993yx}: long and short GRBs. The class of long GRBs is defined as those with a prompt emission lasting over 2 seconds. Strong evidence exists that such GRBs are the result of the  death of massive stars \cite{Galama,Patat}. During such events the prompt emission is theorized to be emitted from within two highly relativistic jets powered by a central engine. Subsequent interactions of the jets with the interstellar medium later result in the afterglow emission. Short GRBs are defined as those GRBs which have a prompt emission time below 2 seconds. Such GRBs, which typically also have a harder energy spectrum compared to their longer counterparts, were theorized to be the result of the merger of two compact objects such as binary neutron stars \cite{Eichler:1989ve}. Strong evidence supporting this theory was found in 2017, through the joint detection of GRB 170817A with the gravitational wave (GW) event GW170817 \cite{LIGOScientific:2017vwq, LIGOScientific:2017ync}. Although all these measurements have provided important insights into the progenitor system of GRBs, much still remains unknown about these extreme phenomena. Especially questions regarding the nature of the relativistic jets, such as their structure, the presence of magnetic fields and how the gamma--rays are produced within them, remain poorly understood. While significant efforts have been made to gain a deeper understanding using measurements of the energy spectra, arrival time and direction of the photons, little progress has been made in recent decades. To break this stalemate, measurements of the gamma--ray polarization have been proposed in the last decades. Such measurements are theorized to be able to answer many open questions regarding GRBs \cite{Gill}. 

In this paper we will first provide a brief overview of the field of gamma--ray polarimetry. This is followed in section \ref{sec:polar-2_detector} by an overview of the POLAR-2 detector design with a particular focus on the polarimeter modules and how these are used to measure the polarization of the incoming photons. The POLAR-2 simulation framework will be introduced in section \ref{sec:simulations}. Section \ref{sec:esrf_campaign} will cover the calibration campaign performed at ESRF including an overview of the measurement and simulation results. Based on this outcome, suggestions for improvements to the detector will be presented in section \ref{sec:improvements} and finally, the foreseen performance of the full detector will be presented in section \ref{sec:extrapolated_performances}.

\section{Gamma-Ray Polarimetry}\label{sec:polarimetry}

\begin{figure}[!h]
  \centering
  \includegraphics[width=1.0\textwidth]{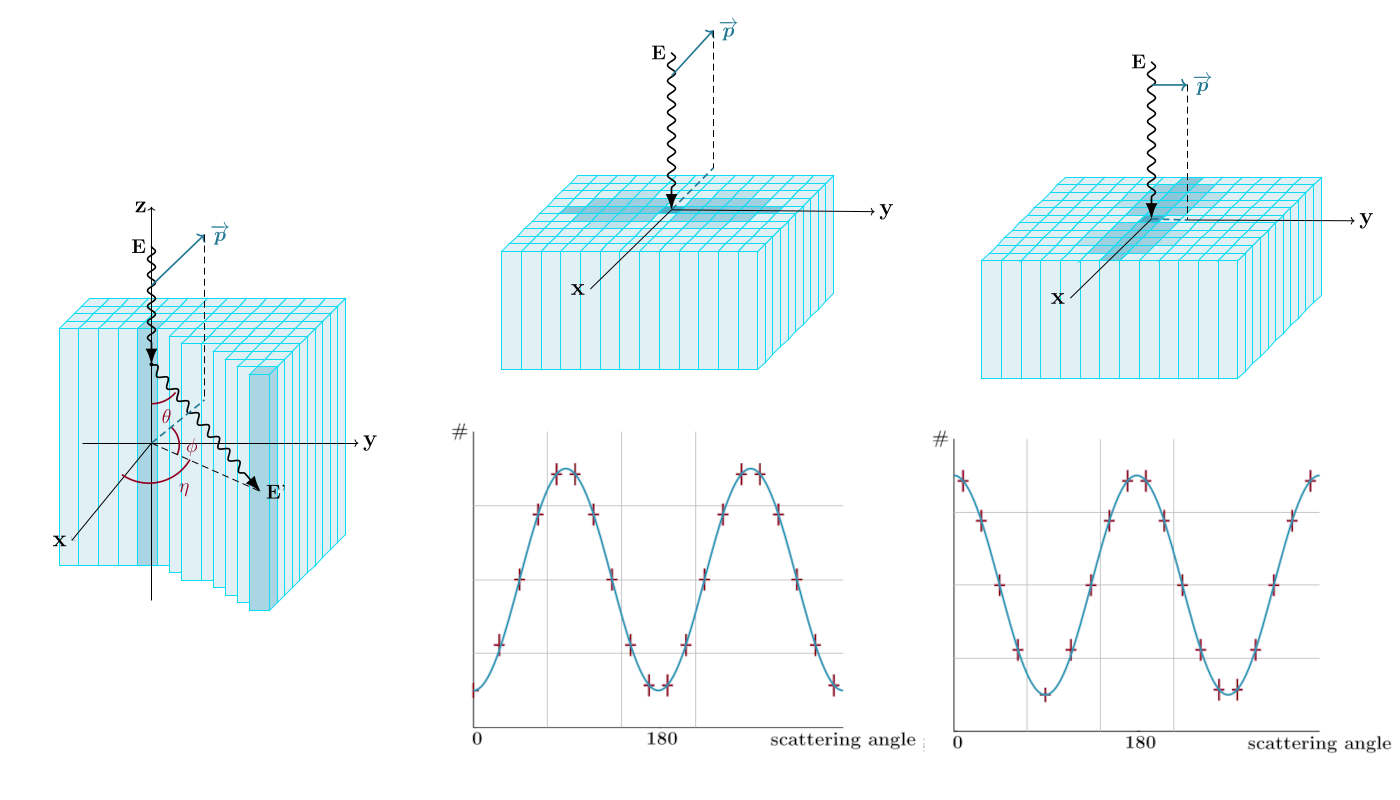}
  \caption{An illustration of how the polarization of an incoming photon can be measured using a segmented detector array. \textbf{\textit{Left}:} Schematic of a photon with incoming energy E, scatters with a polar angle $\theta$ and azimuthal angle $\phi$ before continuing with energy E'. The angle $\eta$, defined as the angle between the two detector segments with an energy deposition can be measured. \textbf{\textit{Middle:}} A photon with energy E interacting in the detector segment located at the origin. The colors of the surrounding segments indicate the probability for a second interaction to take place there based on the polarization vector $\vec{p}$ (darker color indicates higher probability). Below this the typical distribution of $\eta$ is shown for this scenario. \textbf{\textit{Right:}} The same as the middle but with $\vec{p}$ along the y-axis. }
  \label{fig:polarization_meas}
\end{figure}

The prompt emission of GRBs peaks in the $10-1000\,\mathrm{keV}$ energy range. Within this energy range, the polarization of the photons can be deduced by letting the photons Compton scatter in the detector and measuring their azimuthal Compton scattering angle. This is possible as a dependence of the azimuthal Compton scattering angle on the polarization vector of the incoming photon exists. This can be seen in the Klein-Nishina equation:

\begin{equation} \label{eq:1}
    \frac{d\sigma}{d\Omega} = \frac{r_o^2}{2}\frac{E'^2}{E^2}\left(\frac{E'}{E}+\frac{E}{E'}-2\sin^2\theta \cos^2\phi\right).
\end{equation}

\noindent where $r_0 = e^2/m_ec^2$ is the classical electron radius with $e$ the elementary charge, $E$ is the initial photon energy and $E'$ the photon energy of the outgoing photon. $\theta$ is the polar scattering angle and finally $\phi$ is the azimuthal angle between the polarization vector of the incoming photon and the velocity vector of the outgoing photon in the  plane normal to the momentum vector of the photon. This is illustrated in figure \ref{fig:polarization_meas} which also indicates how the dependency between the polarization and the azimuthal scattering angle can be used in a segmented detector array. In such a detector, an incoming photon can scatter in one detector element and then undergo a second interaction, which can be photo-absorption or Compton scattering, in a second detector segment. The relative position between the two elements can be used to measure the scattering angle. It should be noted that the $\cos^2$ term in the Compton scattering cross-section results in a $180^\circ$ symmetry. This means that it is not important to identify which one of the interactions was the first one. However, in case of more than 2 interactions in the detector, the order of the interactions is not known\footnote{unless one has extremely high timing precision but this is not realistic for space missions}. As a result it is not clear which interaction locations should be used to calculate the scattering angle. In addition, any photon which is directly absorbed can also not be used for polarimetry, nor can any photon which leaves the detector without a second interaction. As a result, the efficiency of gamma--ray polarimetry is typically low, thereby giving rise to the first difficulty in this field.

For photons where the angle is measured, the inferred scattering angle can be entered in a histogram to produce a scattering angle distribution, often referred to as a modulation curve. As can be seen from the Klein-Nishina equation, when using a perfect detector, the modulation curve will show a $180^\circ$ degree period for a $100\%$ polarized photon flux and will be flat when the incoming flux is unpolarized. The relative amplitude of this modulation, $\mu$, is therefore linearly related to the polarization degree (PD). The polarization angle (PA) can be extracted from the phase of the modulation as the location of the minimum in the curve.

Extracting the PD from the scattering angle distribution also requires the amplitude of the $180^\circ$ modulation for a $100\%$ polarized beam, here called $\mu_{100}$. Using this, the PD can be extracted using:

\begin{equation} \label{eq:2}
PD = \frac{\mu}{\mu_{100}}
\end{equation}

It should be noted that $\mu_{100}$ typically has a strong dependence on the incoming photon energy as well as on their incoming angle. Therefore the value of $\mu_{100}$ needs to be known for each energy and incoming angle. As it is not realistic to measure the value of $\mu_{100}$ for the full parameter space, this is typically derived through Monte Carlo simulations. Thereby making polarimetry highly reliant on simulations and making detailed validation of the used simulation framework vital for a reliable analysis.

It is important to note that the above description is only valid when one uses a perfect detector. In a real detector, the size, shape and uniformity of the sensitivity of the various detector elements, will induce systematic effects in the scattering angle distribution. As a result, a real scattering angle distribution will show a $180^\circ$ modulation induced by the polarization, as well as instrument-induced non-harmonic functions and harmonic functions which can have periods of $180^\circ$ as well. The latter can be mistaken for polarization, while the other effects will also affect the accuracy of deducing the polarization from the scattering angle distribution using a simple fit. 

As is discussed in detail in \cite{Kole_Sun} there are two analysis methods to get around this. In the first method, the instrumental effects are removed by dividing the measured scattering angle distribution by the one produced using an unpolarized flux. In the second, the instrumental effects are included in the instrument response and the measured distribution is fitted using forward folding. One can, in theory, produce such an instrument response, or the scattering angle distribution for an unpolarized flux, from calibration measurements. However, realistically, one cannot produce this for each possible incoming photon energy and incoming angle. As a result, the production of the response of the polarimeter needs to be produced using Monte Carlo (MC) simulations. As the analysis of gamma--ray polarimetry depends strongly on these MC simulations, any issues in these simulations will result in systematic errors in the measured PD. Systematic errors resulting from imperfections in the MC simulations are a second source of difficulty in the field of gamma--ray polarimetry \cite{Gill,Kole_Sun}.

Although highly complex, polarization measurements of the GRB prompt emission have been performed. A detailed overview of all the measurements published up to 2022 can be found in \cite{Gill}. The most precise measurements to date were produced by the POLAR mission. POLAR was a Chinese-European mission dedicated to measuring the polarization of the GRB prompt emission in the $50-500\,\mathrm{keV}$ energy range \cite{Produit2018}. This instrument was launched as part of the second Chinese spacelab, the Tiangong-2 in September 2016. It operated successfully from October 2016 until April 2017 at which point a problem in its high-voltage power supply did not allow for further scientific data taking. During the 6 months it operated POLAR detected a total of at least 55 GRBs, of which 14 were bright enough to allow for polarization measurements. The results of GRB polarimetry of POLAR, published in \cite{Zhang2019,Kole2020}, indicate that the polarization is typically below $40\%$ when integrating over the full prompt emission period. In addition, time-resolved analysis was performed during which hints were found that the PA varies rapidly during the emission, thereby smearing out the true PD value \cite{Burgess2019}. Finally, energy-resolved polarization studies were performed, however, no significant dependency on the energy was found \cite{NDA_2023_ICRC}. Apart from studying GRBs, the data from POLAR was also used to perform spectral polarimetric measurements of the Crab pulsar \cite{Li_crab}.

The measurement results from POLAR agree with the majority of the GRB emission models as these predict low polarization levels. Only models containing synchrotron emission from a toroidal magentic fields predict polarization degrees up to $50\%$ and therefore appear to disagree with the POLAR data \cite{Gill}. The low levels of PD currently predicted by the majority of the models, as well as the potentially varying PA, imply that significantly more precise measurements are required for GRBs. In addition, although the polarization measurements of the Crab pulsar did agree with previously published results, the large statistical errors did not allow to fully probe the emission mechanism \cite{Li_crab}. For these purposes the POLAR-2 project was initiated in 2018.

\section{The POLAR-2 Detector}\label{sec:polar-2_detector}

\begin{figure}[!h]
  \centering
  \includegraphics[width=.8\textwidth]{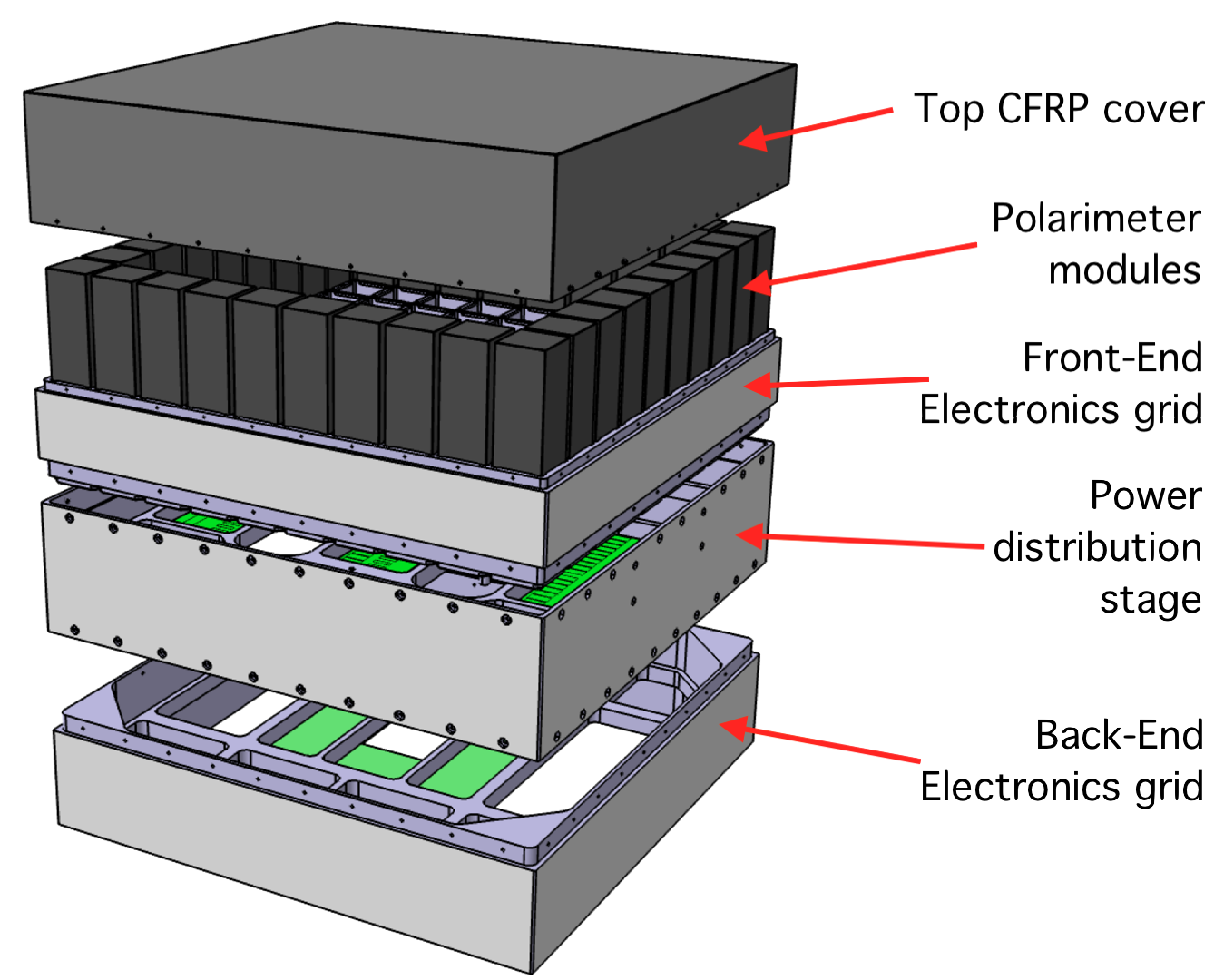}
  \caption{An exploded view of the POLAR-2 payload \cite{NDA_thesis}. The sensitive part of the payload consists of 100 polarimeter modules which can be seen below the carbon fiber cover. Below the 100 polarimeter modules, which are placed in an aluminum grid, two additional aluminum grids that house the low-voltage power supply and the back-end electronics can be seen.}
  \label{fig:POLAR-2}
\end{figure}

\subsection{Polarimeter Module: Mechanical Design}

\begin{figure}[!h]
  \centering
  \includegraphics[width=.8\textwidth]{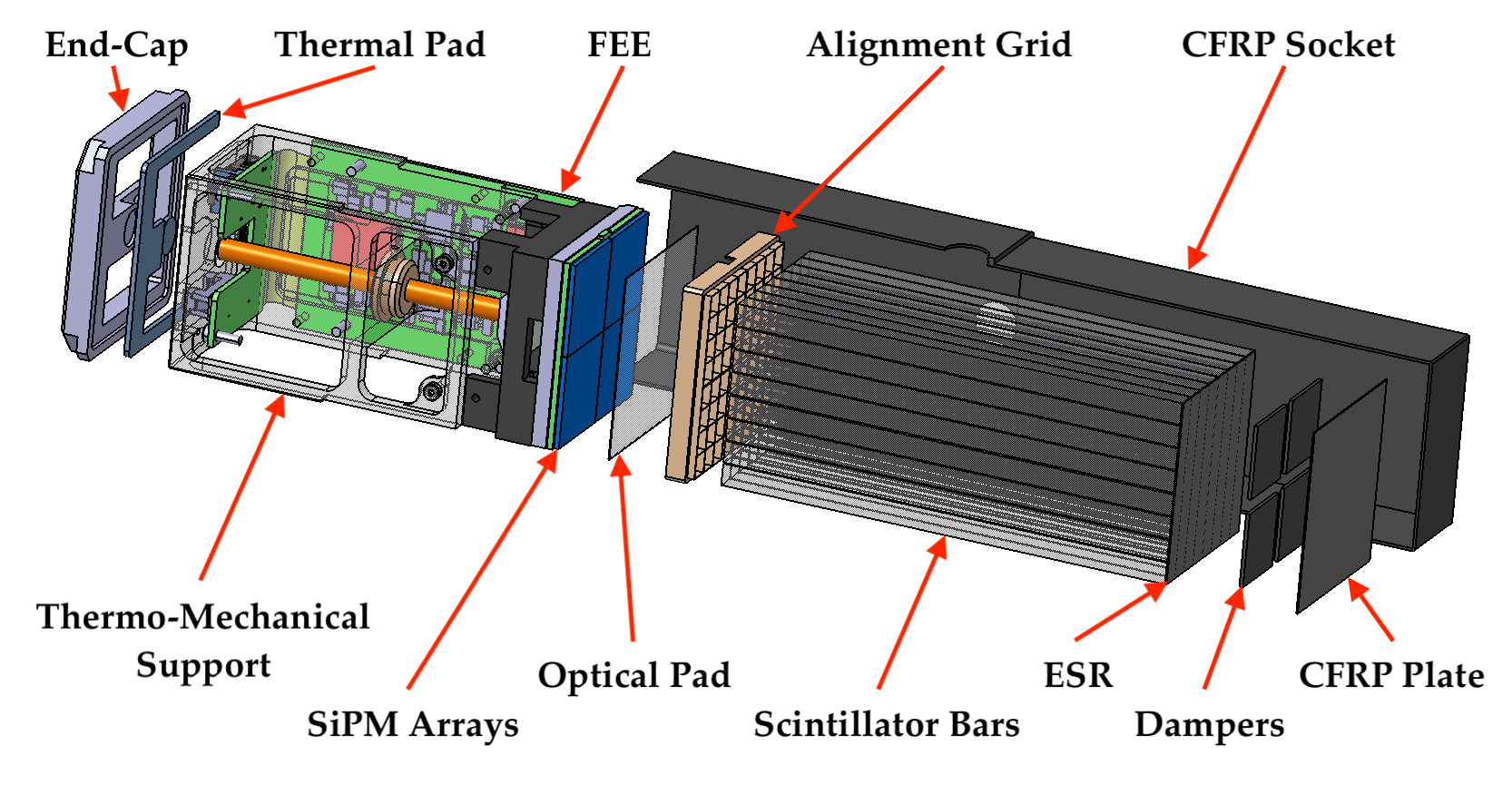}
  \caption{An exploded view of a POLAR-2 polarimeter module \cite{NDA_thesis}.}
  \label{fig:mod_des}
\end{figure}

The design of the POLAR-2 instrument is shown in figure \ref{fig:POLAR-2} while that of a polarimeter module is shown in figure \ref{fig:mod_des}. The photo-detection area of the module consists of 64 plastic scintillator bars with dimensions of $5.9\times5.9\times125\,\mathrm{mm^3}$. The material of the final design will be EJ-248M by Eljen \cite{EJ248}, however, for the polarimeter module tested at ESRF the plastic was EJ-200 \cite{EJ200}. The EJ-200 material used here was selected based on its high light yield compared to EJ-248M with 10 optical photons/keV for EJ-200 compared to  9.2 optical photons/keV for EJ-248M. This higher efficiency remains when convolving the emission spectrum of these scintillators, which both peak at 425 nm, with the photo-detection efficiency vs wavelength of the Silicon PhotoMultiplier (SiPM) which reads it out \cite{NDA_thesis}. However, recent studies, which are discussed in detail in \cite{NDA_thesis, NDA_opt}, indicate that, due to the smoother surface quality achievable for EJ-248M, the number of photons which reach the bottom of the scintillator is higher for EJ-248M, despite the difference in light yield. This effect was additionally verified during the tests described here by also irradiating a polarimeter module containing EJ-248M. Detailed comparisons of the performance, apart from the light yield comparison, were not possible as the electronics connected to the second polarimeter module were damaged prior to the test, resulting in several malfunctioning channels.

The scintillators are held in place using a plastic, 3D-printed grid, which ensures each scintillator is placed directly on top of a SiPM channel with a precision of 50~$\mu m$. Each scintillator is individually wrapped in reflective foils. To accomplish this, the first 4 sheets of Vikuiti by 3M with a thickness of 65~$\mu$m, also called Enhanced Specular Reflector film (ESR), \cite{3M} are used to cover the 4 sides of the scintillator, the combination is then wrapped in a layer of Claryl of 12~$\mu m$ \cite{Claryl} which provides a second optical insulation layer. An additional layer of Vikuiti, is placed at the top of the scintillators to increase the light collection further. Above this layer there is a 1 mm thick layer of sorbothane, a material which combines the properties of rubber, silicone and elastic polymers, purchased from a commercial vendor (IsolateIT), and a carbon fiber plate of 0.75 mm thick which together serve to provide pressure to the top of the scintillators to improve their coupling to the SiPMs. The SiPM readout consists of 4 arrays of S13361-6075NE-04 produced by Hamamatsu \cite{Hamamatsu}, each of which contains 16 channels with a size of $6\times6\,\mathrm{mm^2}$. The SiPMs were selected based on their high photo-detection efficiency (PDE), as high as $50\%$, and the possibility to get these in large area arrays. It should be noted that the newer, S14 family of Hamamatsu is also available. However, the S13 was selected based on the shorter decay time of the pulses of $\approx100\,\mathrm{ns}$ compared to $\approx600\,\mathrm{ns}$ for the S14. More details on the SiPM selection are presented in \cite{NDA_thesis}. Between the SiPM and the scintillators there is a 250~$\mu m$ thick layer of MAPSIL \cite{MAPSIL_paper, MAPSIL_website} which serves both as an optical light guide and, together with the sorbothane layer at the top, to protect the SiPM from damage during the vibration and shock sustained during the launch. The resistance of this design against vibrations and shocks was successfully tested during a dedicated test in 2022 \cite{NDA_thesis}. The SiPM array is mounted on an aluminium and plastic frame which holds the electronics, as shown in figure \ref{fig:mod_des}. This mechanics is designed to thermally insulate the SiPM arrays from the electronics, thereby keeping their temperature lower. During the tests discussed here, the SiPMs were operating at room temperature (measured in the middle of the array) while the electronics is around 35$^\circ$C as measured using a temperature sensor embedded in the electronics. 

The combination of the electronics and the scintillator array and its surrounding components are all placed inside a carbon fiber socket with a thickness of 0.75 mm. This socket provides the mechanical support for all the components inside and ensures the system is light tight. The electronics frame is finally supported from the bottom by an aluminium end-cap which provides pressure to the electronics frame and couples to the surrounding mechanics. For these tests, this consisted of an aluminium frame which is mounted on a translation table as described in section \ref{sec:esrf_campaign}.

\subsection{Electronics Design}

The 64 detector channels of each polarimeter module are individually read out using a front-end electronics board placed below the SiPM array and scintillator array. The readout consists of 1 FPGA, 2 CITIROC-1A ASICs (each with 32 channels), 4 ADCs, and a DC/DC converter (LT3482 by Linear Technology) used for providing the bias voltage to the SiPMs. 

The LT3482 is controlled from the FPGA, an IGLOOv2 by Microsemi, and supplies a common high voltage to the 64 SiPM channels. It is however known that the breakdown voltages of each channel vary. Therefore, the breakdown of each channel was measured prior to the detector assembly at different temperatures. It was found that the breakdown voltages vary within 0.1~V within one detector module. The CITIROC ASICs used for the SiPM readout can be used to correct for these variations using a DAC value which can be applied to the input from each channel on the ASIC. Using this DAC value the voltage over each channel can be modified by up to 2.5~V, thereby ensuring each channel is provided with an equal over-voltage. During the tests discussed here this over-voltage was set to be 3.0~V.

\begin{figure}[!h]
  \centering
  \includegraphics[width=.5\textwidth]{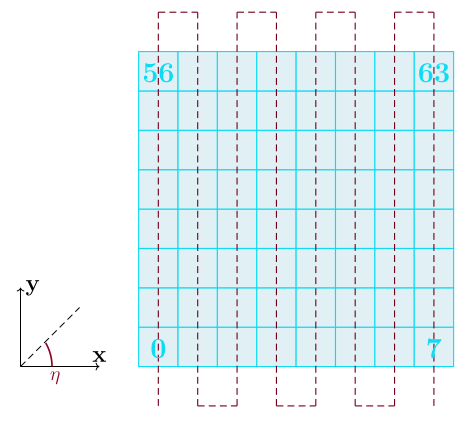}
  \caption{The channel numbering convention of the POLAR-2 polarimeter module as seen from the top. The scattering angle definition is also indicated on the left. The red dotted line indicates the path scanned by the synchrotron beam during an irradiation scan with the beam starting below channel 0 and ending below channel 7.}
  \label{fig:conv}
\end{figure}

The signal from each SiPM channel are directly fed into one of the two ASICs which handle the amplification and discrimination. In order to avoid a clear $180^\circ$ asymmetry in the detector, it was decided to alternate the ASIC used for each subsequent channel, as a result one ASIC reads out all odd numbered channels and the other all even numbered ones. The channel numbering convention is shown in figure \ref{fig:conv}. The ASICs are capable of analyzing the incoming signal either using a peak sensing algorithm as well as through a sampling and hold. Based on preliminary tests it was found that for the typically short signals from the combination of plastic scintillators and SiPMs, the peak sensing algorithm works well and was therefore selected. Within the ASIC, the peak sensing is applied to the signal produced by two amplifiers, one with a high gain and one with a low gain. The integration time is $\approx 100\,\mathrm{ns}$ for each channel. In the case of POLAR-2 the high gain will be used to measure the deposited energy in the $0-100\,\mathrm{keV}$ range, while that of the low gain is used to measure from approximately $50-800\,\mathrm{keV}$. In the case of POLAR-2 the signal from the high gain output is used in the ASIC trigger. Two separate discriminators are available for each of the 32 channels to give additional flexibility. Following the convention of the supplier\footnote{\url{https://www.weeroc.com/my-weeroc/download-center/citiroc-1a/89-citiroc1a-datasheet-v2-53/file}} the first discriminator is referred to as the charge discriminator, the second the time discriminator. While the time trigger can be used to provide detailed timing information for each of the individual 32 channels. The output from the 32 charge discriminators are coupled in an OR output only. For POLAR-2 the two available discriminators are used to provide two separate trigger signals. In the final POLAR-2 logic the 'charge threshold' is set higher than that of the 'time trigger' to give flexibility in the trigger logic.

The trigger patterns from the ASICs are provided to the FPGA which uses this to make a trigger decision. In the case of the ESRF test, only the charge trigger was used and all channels were read out if at least one channel is above the charge trigger threshold. In the logic of the final polarimeter the time trigger will be set at a lower voltage, and a readout of the module is performed when at least 2 channels are above this threshold level within a 100 ns time period. This short period ensures that contamination of the Compton scattering signal by chance coincidence events within the detector stays negligible.  Although this functionality was present at the ESRF test, the two thresholds were set at equal values in order to simplify the readout during this first irradiation campaign.

In case the FPGA starts a readout, the output from both the high gain and low gain amplifier of each of the 64 channels is provided to the ADCs. Each ASIC has 2 ADCs, one for the high gain output and one for the low gain. Therefore, a total of 128 ADC values are provided to the FPGA for each trigger. In addition, $2\times64$ booleans are provided containing the charge and the time trigger information of each channel. 

It should be noted that, although both the high gain and low gain output from each channel are read out for each triggered event, the ADC values for all channels which did not issue a charge trigger is typically 0. This is due to a setting on the ASIC which, by default, only starts the peak sensing on the input signal if a charge trigger is issued by this channel. If not, a baseline value is simply provided. It was found during the analysis phase of this campaign that this causes significant issues for the detector calibration and therefore this logic has since been updated in the POLAR-2 firmware, as will be discussed in section \ref{sec:improvements}.

During this irradiation campaign no prototype of the POLAR-2 back-end electronics was used. Instead, a General Purpose Input Output (GPIO) board produced by the electronics group of the DPNC department of the University of Geneva was used to communicate with the front-end electronics. The GPIO allows to configure the FPGA on the front-end and handles the data transmission from LVDS lines. The GPIO in turn provides the data to a desktop which, during the ESRF campaign, was placed outside of the hutch (the irradiation room with the beam at ESRF).

\subsection{Analysis Pipeline}\label{sec:data}

The data provided by the polarimeter module has to undergo several steps in order to convert the ADC output provided by the ASICs into keV. In this section the various steps of the analysis pipeline which is applied to the measured, as well as the simulation data, are presented.

\subsubsection{Pedestal Subtraction}

\begin{figure}[h]
\centering
  \hspace*{-0.5cm}\includegraphics[height=0.35\textwidth]{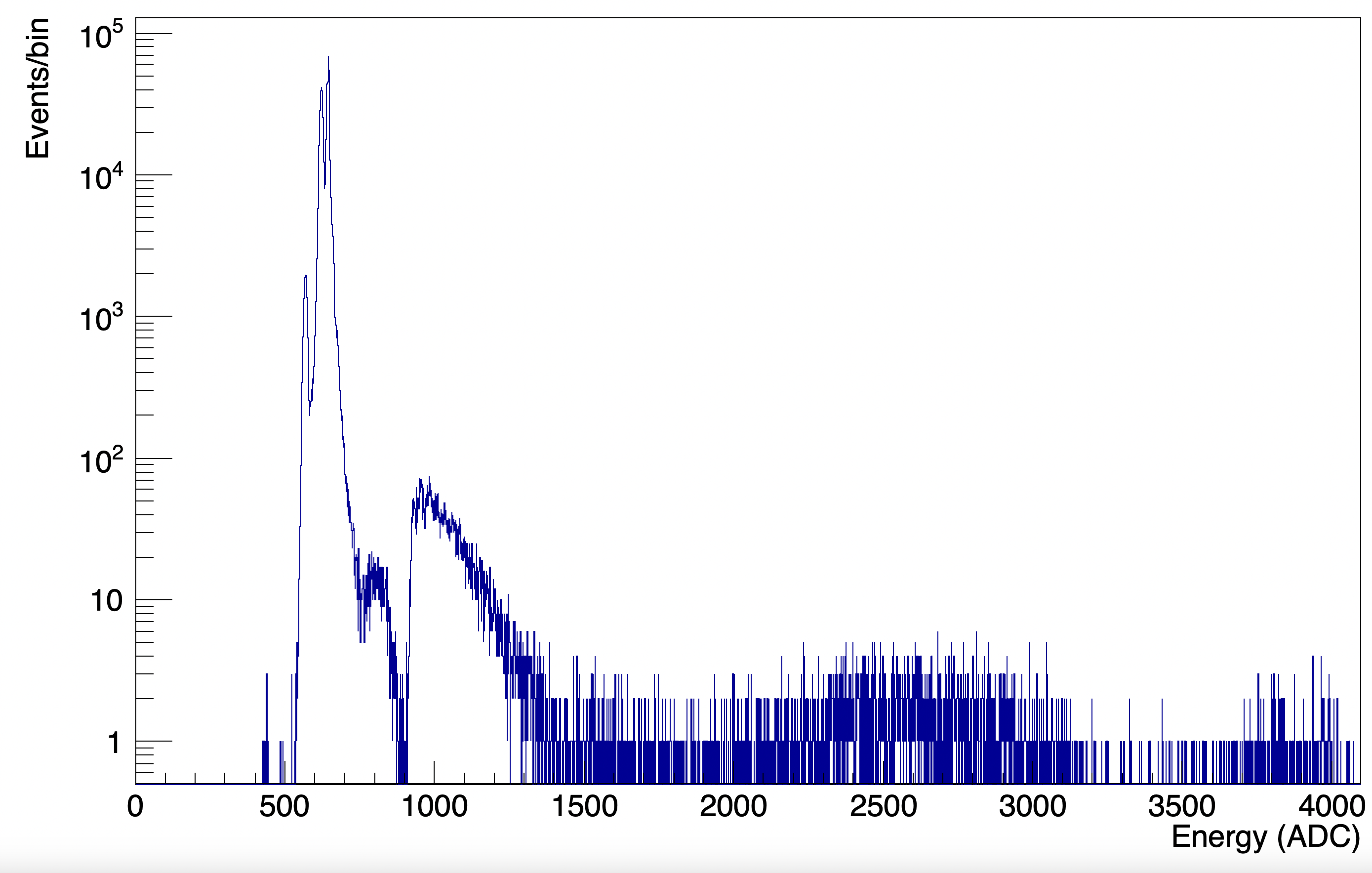}\hspace*{0.1cm}\includegraphics[height=0.35\textwidth]{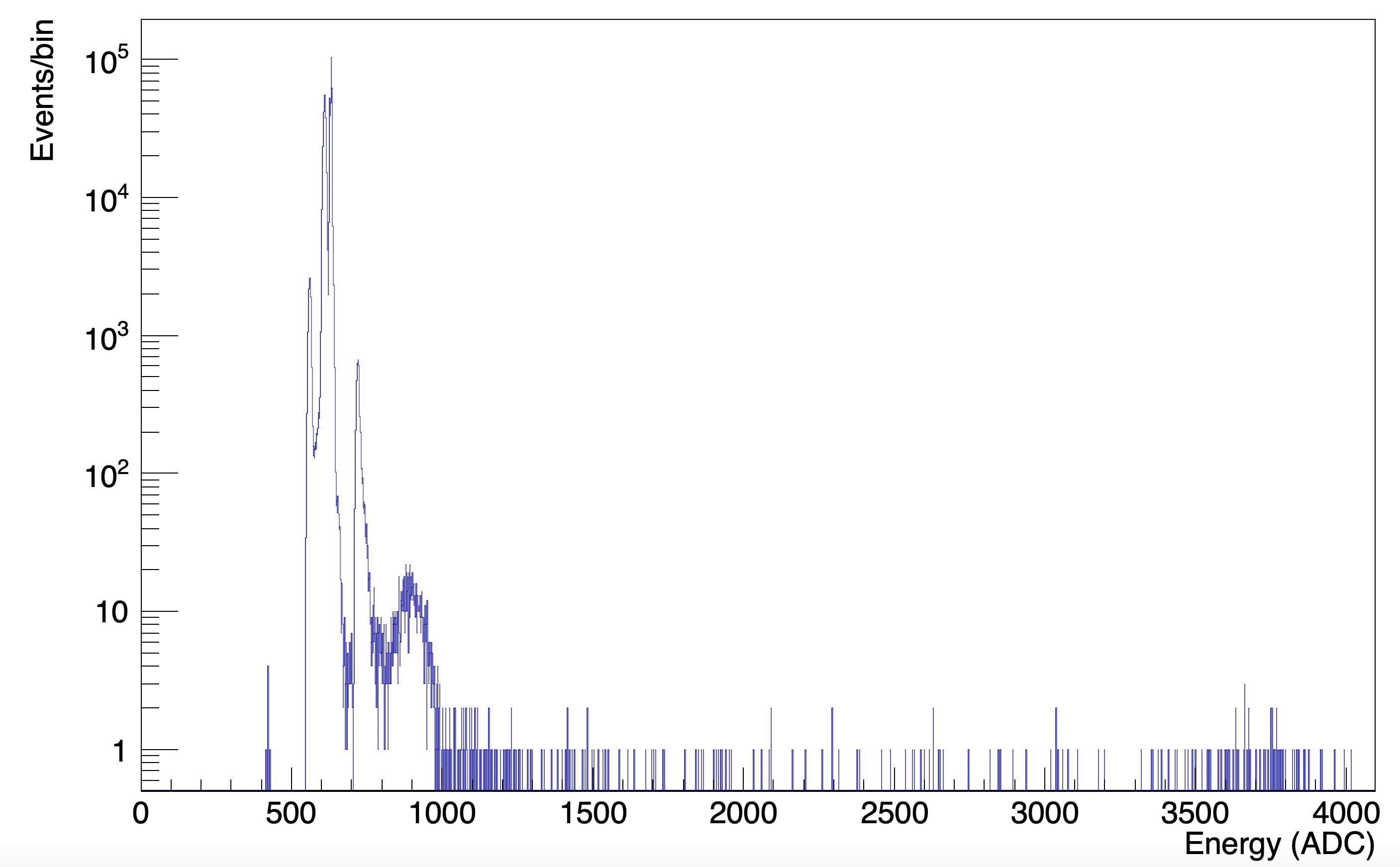}
  \caption{\textbf{Left:} The raw ADC spectrum from the high gain amplifier as measured for a 60~keV photon beam. The photo-peak can be seen around 2500 ADC while the trigger threshold (charge threshold) can be seen around 800 ADC. \textbf{Right:}. The raw ADC spectrum from the low gain amplifier as measured for a 60~keV photon beam. The same features as in \ref{fig:raw_chan8_high} can be observed however, due to the lower gain the ADC values are compressed.}
  \label{fig:raw_chan8_high}
\end{figure}

The left side of figure \ref{fig:raw_chan8_high} shows the typical 12-bit ADC spectrum from the high gain output taken with a $60\,\mathrm{keV}$ beam, while the right side shows that produced by the low gain. A clear shift in the baseline can be seen for both; a result of both the pedestal of the ADC as well as that from the ASIC output. The first step in the analysis procedure consists of subtracting this pedestal value from each of the 64 channels both for the high gain and the low gain. 

\begin{figure}[!h]
  \centering
  \includegraphics[width=.7\textwidth]{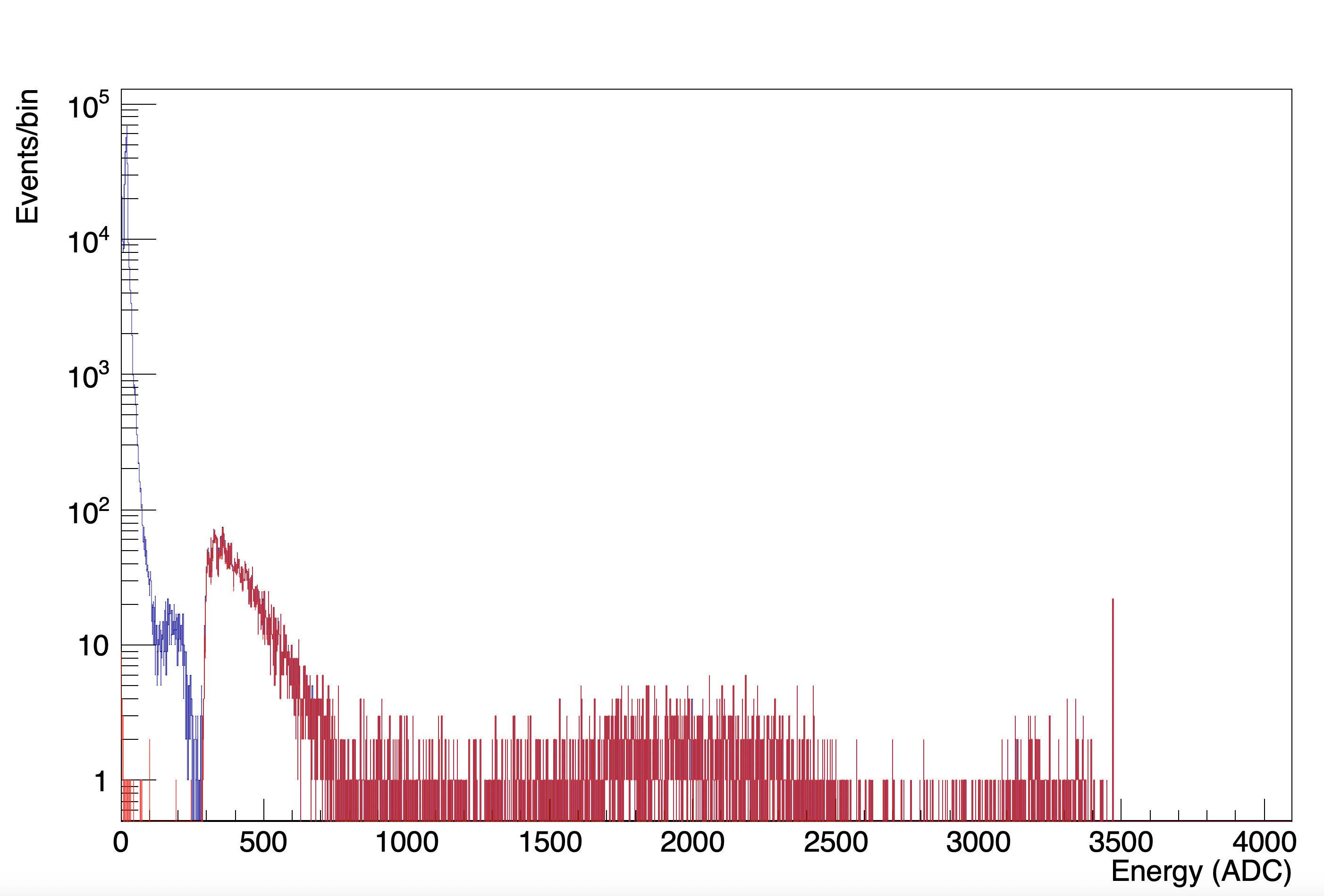}
  \caption{The raw ADC spectrum from the high gain amplifier as measured for a 60~keV photon beam after pedestal subtraction. The triggered events are shown in red, while the non triggering events are shown in blue.}
  \label{fig:raw_chan8_high_ped}
\end{figure}

In order to subtract this value the exact pedestal values need to be measured. For the data analyzed here these measurements were provided by forcing a readout of each channel while no high voltage was provided to the SiPM. This results in 2 distributions of ADC values for each channel (one for the high gain and one for the low gain output) which are well described by a normal distribution. The mean of this represents the pedestal value and the width, typically around 10 ADC, the electronic noise on this channel. The fit results of the mean are used for pedestal subtraction, while both the mean and the width are used in the simulations which will be described in the next section. The spectrum of figure \ref{fig:raw_chan8_high} after pedestal subtraction is shown in figure \ref{fig:raw_chan8_high_ped}. 

\subsubsection{Non-linearity Correction}\label{sec:nl}

\begin{figure}[!h]
  \centering
  \includegraphics[width=.7\textwidth]{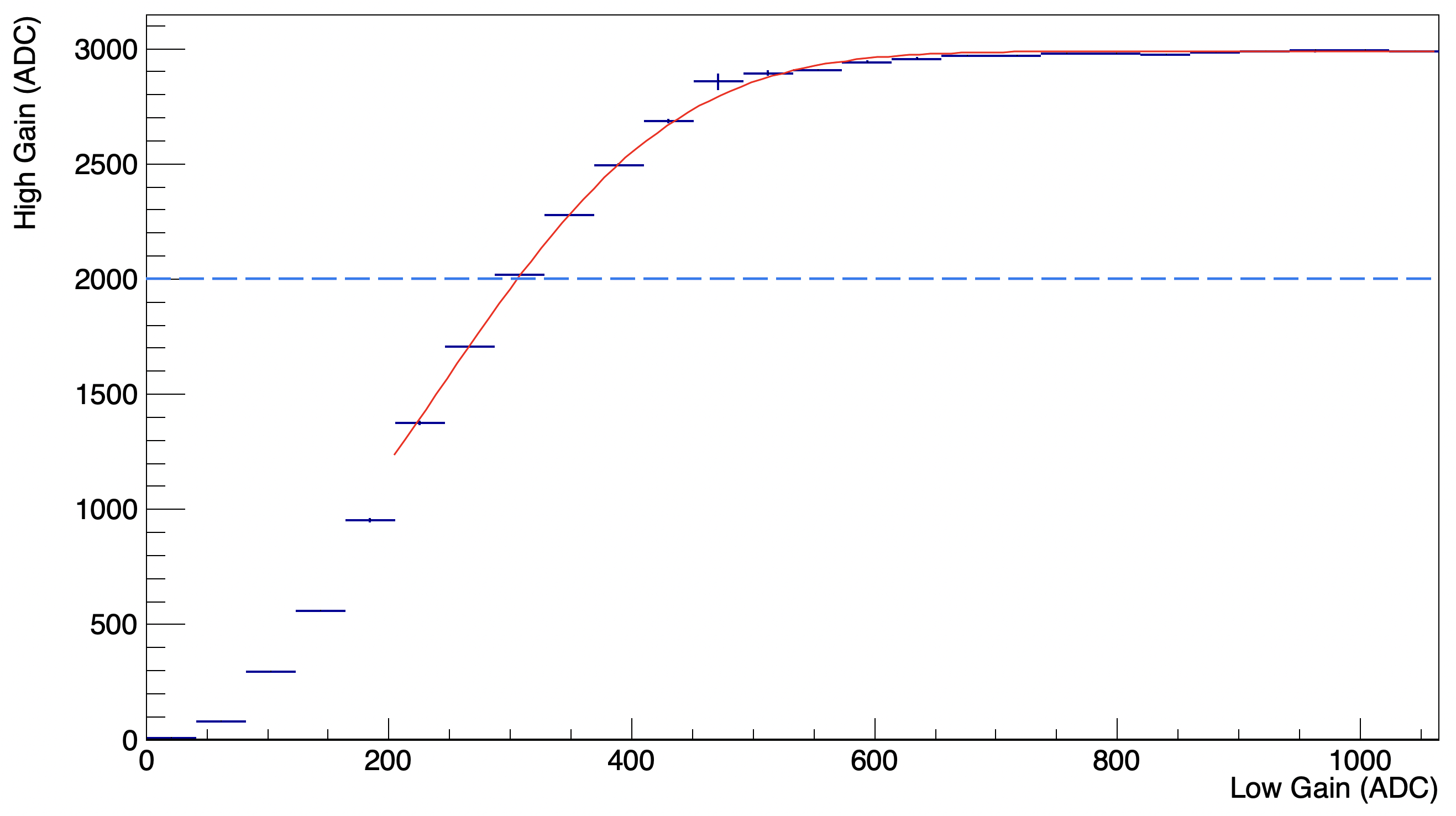}
  \caption{The pedestal subtracted output of the high gain vs the pedestal subtracted output of the low gain. A clear non-linear behaviour can be observed when the high gain ADC value starts to exceed 2500 ADC. The fit of a non-linear function is shown in red, while a blue dotted line indicates above which high gain ADC value the linear correction is applied. }
  \label{fig:high_v_low}
\end{figure}

In order to understand the correlation between the ADC values from the high gain and the low gain the two can be plotted against one another for each channel. The result can be seen in figure \ref{fig:high_v_low}. Although the correlation between the two is linear at low ADC values, when the ADC value in the high gain output exceeds $\approx\,2500$ ADC a non-linear behaviour is observed, this is likely a result of saturation of the high gain amplifier. This behaviour has been reported by other users of the CITIROC ASIC \cite{BabyMind}, and was also seen with the POLAR-2 front-end when injecting external charges directly into the ASIC.

\begin{figure}[!h]
  \centering
  \includegraphics[width=.7\textwidth]{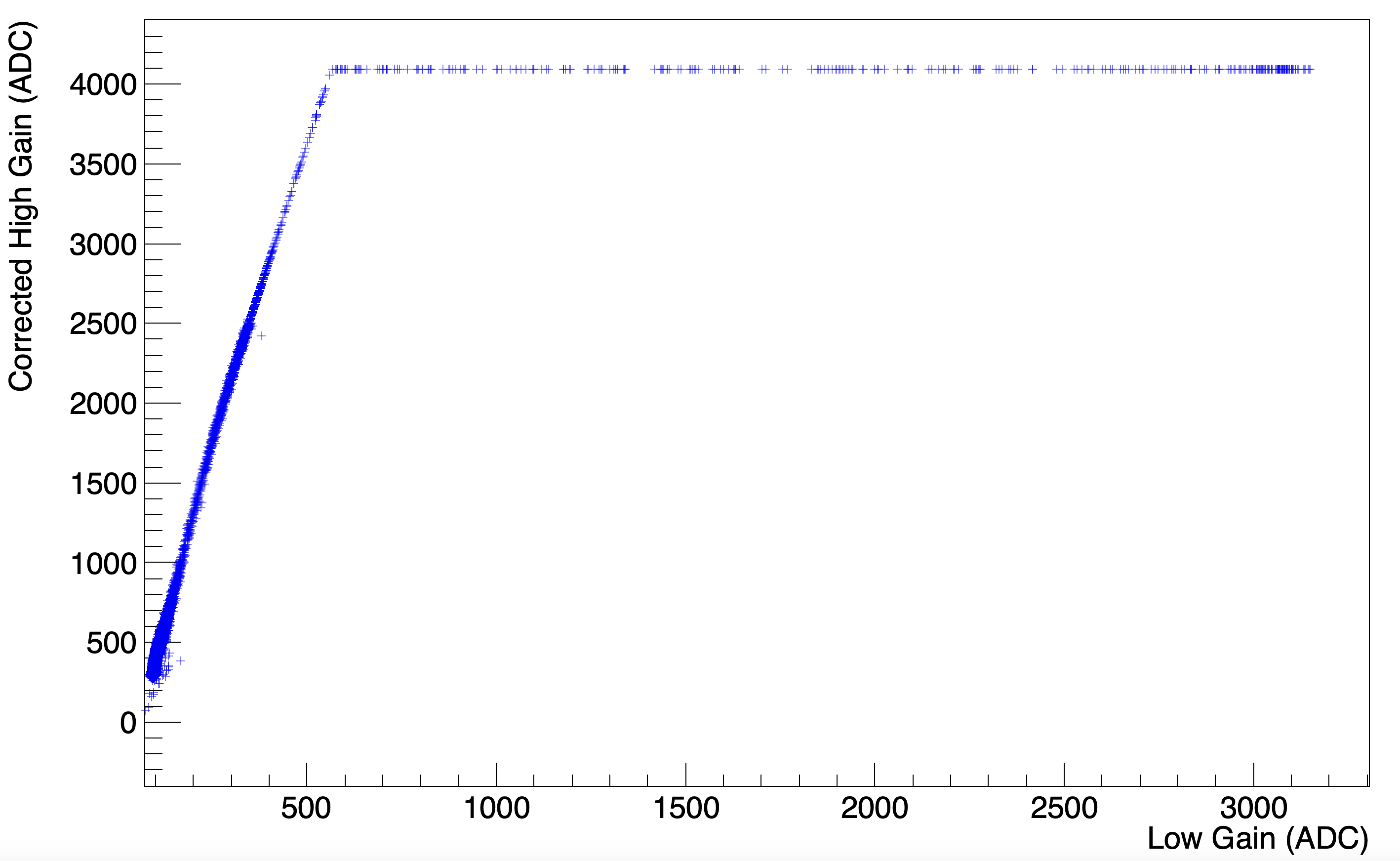}
  \caption{Equal to figure \ref{fig:high_v_low} but after applying a correction to the high gain ADC.}
  \label{fig:high_v_low_good}
\end{figure}

To ensure a better linearity of the high gain ADC value, its value is corrected in the analysis using the low gain ADC value. For this purpose, the linear relation was measured at low ADC values and parameterized using a first-order polynomial. For high gain ADC values which exceed 2000 ADC the corrected value is calculated based on the linear parameters from the fit and the low gain ADC value. In addition, in order to include this effect in the simulations, the correlation at values exceeding $\approx\,2000$ ADC was fitted using an error function. The linear parameters are used in the data analysis part to correct the high gain values when these exceed 2500 ADC. In case the correction exceeds 4095 ADC, it is fixed at this level. An example of the correlation between the two ADC values after correction is shown in figure \ref{fig:high_v_low_good}.

\subsubsection{Optical Crosstalk}\label{sec:Xtalk}

One of the major improvements in POLAR-2, compared to POLAR, regards to the reduction of optical crosstalk. In the design of POLAR optical crosstalk between different detector channels within a detector module was primarily a result of the use of the Multi-Anode Photo-Multiplier Tube (MAPMT) for the readout. In the MAPMT the 64 channels share one borosilicate window. This implies that optical photons from one scintillator bar can enter the anodes of a neighbouring channel, thereby producing a significant amount of crosstalk. A second issue of the use of a MAPMT is that the window is fragile, which for space instrumentation prompts the use of a damper material between the scintillators and the window. This damper material, which in the case of POLAR was a 3 mm thick layer of MAPSIL \cite{Produit2018}, induces additional optical crosstalk. Finally, the optical insulation between scintillators was achieved using a single layer of Vikuiti by 3M, which although highly efficient, will result in a small amount of additional crosstalk. This, however, in the case of POLAR is negligible compared to the optical crosstalk induced by the window and the damper.  As a result, the typical optical crosstalk from one channel to its direct neighbor (for example, channel 0 to 1) was between $10$ and $20\%$ \cite{POLAR:2018hqh}. Crosstalk to the diagonal neighbors (for example from channel 0 to 9) was around $5\%$ while that to the second neighbors (for example from channel 0 to 2) was of the order of $1\%$.

For POLAR-2 the use of a SiPM significantly reduces the amount of expected optical crosstalk. The issue of a thick shared entrance window is removed, while the less fragile nature of the SiPM significantly reduces the need for a damper material. Although no thick damper is required, some kind of material which serves both as a damper and an optical coupling between the scintillators and the SiPM is required. For this purpose a $250\,\mu m$ thick MAPSIL layer is molded directly on top of the SiPM \cite{NDA_thesis}. It should be noted that the MAPSIL layer is not optically insulated on the outside, as this is mechanically complex to do for an elastic material. This means some light, approximately $1\%$ will be lost on the outside of the module. Finally, as the two main sources of optical crosstalk are significantly reduced the third, related to photons penetrating the optical insulator materials, was further addressed in this design. Compared to POLAR, where one layer of Vikuiti was placed between scintillators, for POLAR-2 each scintillator bar is wrapped covered on all sides with a sheet of Vikuiti, this is then further wrapped in a layer of Claryl. As a result there are now 4 reflective layers between each scintillator compared to 1 in POLAR.

The expected level of optical crosstalk with this setup was simulated and is addressed in detail in a separate paper \cite{NDA_opt}. The simulations predict only significant crosstalk to the direct neighbors which is of the level of $2-3\%$. 

Due to the aforementioned issue where peak sensing is not enabled for channels which do not issue a charge trigger, only energy depositions above this threshold, at $\approx8\,\mathrm{keV}$, are recorded correctly. As a result of this, along with the low levels of crosstalk, the ESRF data could not be used to perform accurate measurements of the crosstalk. Instead, the crosstalk was measured after a firmware update, which will be described in section \ref{sec:improvements}, using the same polarimeter module and the same settings as employed during the ESRF beam test. 

To measure the optical crosstalk the polarimeter module was irradiated with a $^{137}Cs$ source which emits 662~keV photons. The data was then processed using the above steps with pedestal and non-linearity correction. Using the data in this format, the ADC value in channel x was plotted against that of channel y, with the only cut that channel y must issue a charge trigger. The resulting 2d histograms show, in case of neighboring bars, a linear distribution, which can be fitted with a first order polynomial which is forced to have an offset of 0.

It should be noted at this point that the correlation found using the above method is not purely the optical crosstalk as it is measured in ADC and the two channels do not have the same gain. As is described in detail in \cite{Hualin}, the conversion of energy depositions to electrical signals in the polarimeter module can be described using several matrices. We start with a vector $\Vec{E}_{\rm dep}$ which contains the energy depositions in the 64 scintillators. The number of optical photons which reach the bottom of a scintillator for a channel can be described as $N_{\rm bar} =s\, c\, E_{\rm dep}$ with $s$ the scintillation efficiency and $c$ the photon collection efficiency. As the scintillators here are all produced by the same manufacturer we can assume that $s$ is equal for all channels. For the full polarimeter module we can therefore write:

\begin{equation}\label{eq:3}
\Vec{N}_{\rm bar} = B\Vec{E}_{\rm dep} 
\end{equation}

with $B = s Diag(c1,c2,···,c64)$. We now have to additionally take into account that due to optical crosstalk, an energy deposition in channel x can also result in optical photons reaching the bottom of bar y. For this we introduce matrix $X$ which contains the crosstalk between all the channels. Here $x_{ij}$ is the crosstalk from channel i to channel j, with $x_{ij}$ being a value between 0 and 1, and $x_{ij}=x_{ji}$. The number of optical photons reaching the SiPM channel j from an energy deposition in channel i is then $N_{\rm SiPM}=x_{ij}N_{\rm bar}$, so:

\begin{equation}\label{eq:4}
\Vec{N}_{\rm SiPM} = X\Vec{N}_{\rm bar} 
\end{equation}

Finally, the number of optical photons which reach a channel x is transformed into an ADC value in this channel through the gain of the SiPM and the ASIC channel. The total gain can be described as $g$ and we can calculate the measured energy $\Vec{E}_{\rm meas}$ as:

\begin{equation}\label{eq:5}
\Vec{E}_{\rm meas} = G\Vec{N}_{\rm SiPM} 
\end{equation}

Where $G$ is a diagonal matrix consisting of elements $g_{ii}$. The complete path to convert the deposited energy to the measured energy can now be written as:

\begin{equation}\label{eq:6}
\Vec{E}_{\rm meas} = (GXB)\Vec{E}_{\rm dep} = R\Vec{E}_{\rm dep}
\end{equation}

The crosstalk measured using the method above is the ratio between the energies measured in ADC, so $\frac{E_{j}^{\rm meas}}{E_{i}^{\rm meas}}$. It contains contributions from the matrices $B$, $G$ and $X$. We can assume that the variations in the elements in $B$ are minimal, as each scintillator is produced in the same way. We then state that what is measured using our linear fit is:

\begin{equation}\label{eq:4}
f_{ij} = \frac{E_{j}^{\rm meas}}{E_{i}^{\rm meas}} = \frac{r_{ji} E_i^{\rm dep}}{r_{ii} E_i^{\rm dep}}=\frac{g_j x_{ji}}{g_i x_{ii}}
\end{equation}

So, for example, we can see that whereas we would expect the true optical crosstalk to be symmetric, so $x_{ij}=x_{ji}$, this is not the case for the ratio in the measured energies contained in the elements $f_{ij}$ which make up the matrix $F$.

Finally, what is required for the analysis here is the matrix $R$, such that one can perform the conversion:

\begin{equation}
    \Vec{E}_{\rm dep}=R^{-1}\Vec{E}_{\rm meas}
\end{equation}

As is again described in detail in \cite{Hualin}, we can retrieve matrix $R$ by measuring the matrix $F$, through our linear fits which produce elements $f_{ij}$ as well as a diagonal matrix $M$ which contains, the elements $r_{ii}$ of matrix R. The diagonal elements can be understood as the ADC to keV conversion for a single channel. These elements can be measured using an $^{241}{\rm Am}$ source which emits 59.5 keV photons resulting in a photo-peak in each channel. An example of this is shown later in this paper in figure \ref{fig:new_spec} which shows a clear photo-peak around 2000 ADC. By fitting this peak the value of $r_{ii}$ (as measured in ADC / keV) is retrieved for each channel. Finally, using the equations above, we can now retrieve matrix R using:

\begin{equation}
    R=F^{T}M
\end{equation}

\begin{figure}[!h]
  \centering
  \includegraphics[width=1.1\textwidth]{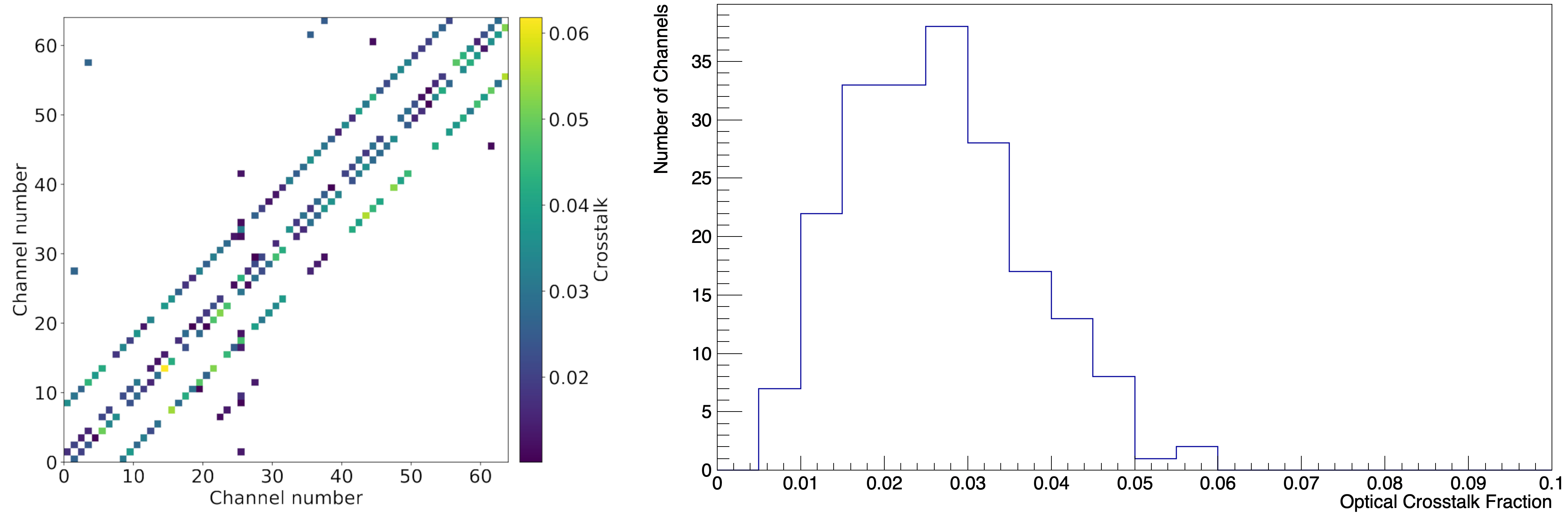}
  \caption{\textbf{Left:} The optical crosstalk matrix (in which we have set elements $x_{ii}$ to 0 for visualization purposes) produced using the above described method. It can be seen that, apart from a few small outliers, the cross talk is prominent from channels x to channels $x\pm1$ and $x\pm8$ which can be understood when looking at the channel numbering convention. \textbf{Right:} A histogram showing the amount of corrected optical crosstalk as measured to the direct neighbors of each channel (so typically 4 entries per channel).}
  \label{fig:xtalk_neigh}
\end{figure}

We can then also calculate the matrix $X$ containing the pure optical cross talk. An example of this is presented on the left in figure \ref{fig:xtalk_neigh}. A histogram showing the optical crosstalk to the direct neighbors for each channel is shown in figure \ref{fig:xtalk_neigh}. The histogram can be seen to peak between $2$ and $3\%$, the results therefore match the values predicted in \cite{NDA_opt}. 

As we now have the correct crosstalk as well as the gain for each channel, we can produce the matrix $R$. Then finally, we can apply the correction for the gain and the crosstalk to convert the pedestal subtracted ADC value to the energy in keV using:

\begin{equation}
    \Vec{E}_{\rm dep}=R^{-1}\Vec{E}_{\rm meas}
\end{equation}

\begin{figure}[!h]
  \centering
  \includegraphics[width=.5\textwidth]{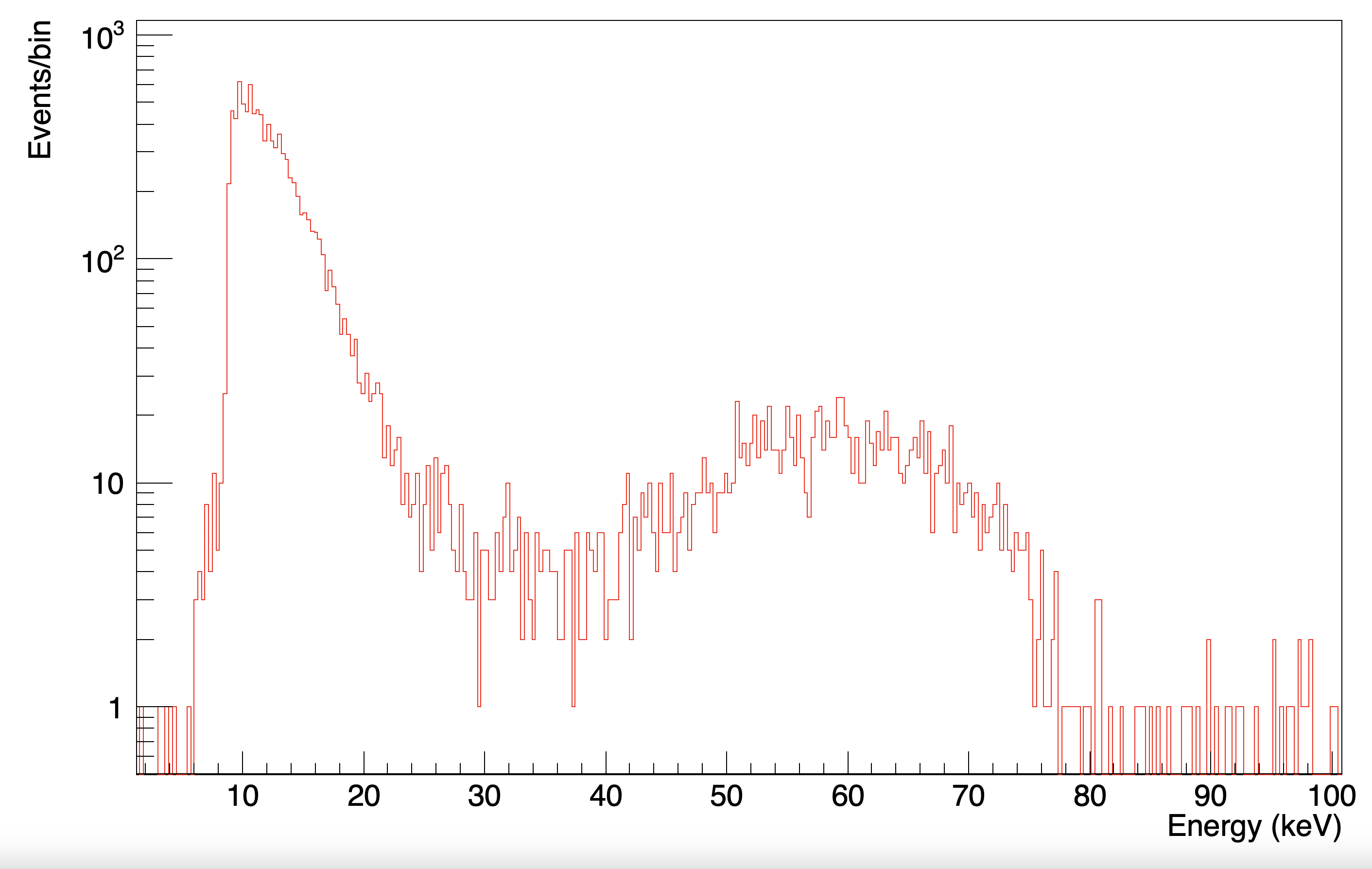}
  \caption{The energy spectrum as measured when irradiated by a 60~keV beam for a typical detector channel after pedestal subtraction, non-linearity correction, as well as the crosstalk and gain corrections.}
  \label{fig:corr_chan20_high}
\end{figure}

An example of the spectrum from irradiation by a 60~keV beam after this correction is shown in figure \ref{fig:corr_chan20_high}. The photo--peak can be seen to be correctly placed at 60~keV. 

\section{Simulations}\label{sec:simulations}

The simulation software developed for POLAR-2 consists of 2 steps. The physical interactions of the gamma--rays in the detector are simulated using Geant4 \cite{G4}. The output of Geant4 is then passed on to a digitization Monte Carlo (MC) which handles the conversion of the deposited energies to an ADC signal. The final result is a data file which has the same format as the real data prior to pedestal subtraction. 

The decision to divide the simulation into two steps was taken to simplify the production of detector responses in the future. This is because the detector response changes due to factors such as bias voltage settings and detector temperature. As such effects are only taken into account in the digitization, the Geant4 part will only have to be performed once, while this output can then be used to produce the detector responses for various instrument configurations. Both steps are presented in detail below.

\subsection{Geant4}

\begin{figure}[!h]
  \centering
  \includegraphics[width=.3\textwidth]{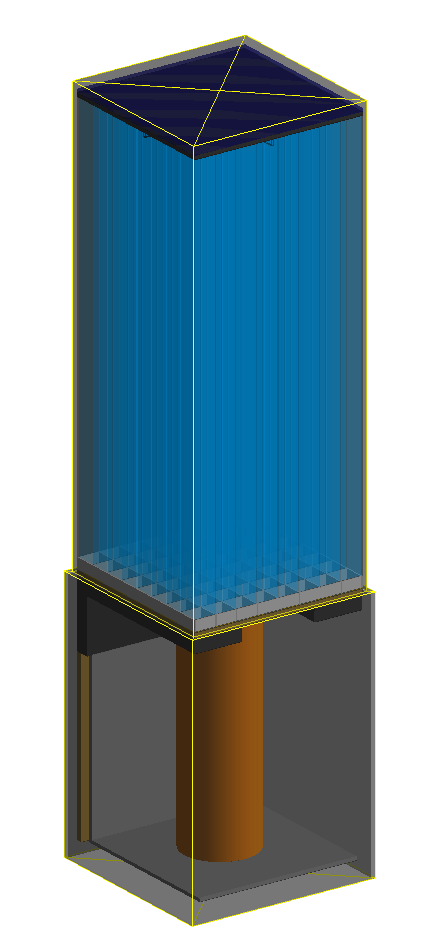}
  \caption{The mass model of a POLAR-2 polarimeter module as implemented in Geant4. The scintillators are shown in blue held by a plastic grid at the bottom. The carbon fiber housing is made transparent here to show the contents. Below the scintillators the thin SiPM array is shown in brown with below this the mechanical support structure of the electronics. Finally the brown cylinder is a copper bar which serves to extract heat from the SiPM array.}
  \label{fig:module}
\end{figure}

The Geant4 part of the simulation includes a detailed model of the POLAR-2 detector. For the purpose of the ESRF analysis only a single polarimeter module was simulated placed inside of an aluminium grid which can hold 9 polarimeter modules. The module as simulated is shown in figure \ref{fig:module}. This mass model was implemented in Geant4 using standard Geant4 objects rather than using a GDML input. Although this approach is more time consuming to develop, it makes for a human readable code which can be debugged. The simulated module includes all the components present in figure \ref{fig:mod_des} including the plastic scintillators, the damping materials and carbon fiber plate above it, the carbon fiber sockets, the plastic grids which hold the scintillators in place, the optical pad, and the SiPMs. A simplified version of the front-end electronics and their support structure is also present, although, for the purpose of these simulations these will not have a significant effect on the results. It was decided not to include the reflective foils (Vikuiti and Claryl) in the simulations after initial tests showed these have no significant effect on the results. As they do slow down the simulations significantly, it was decided to remove these from the simulation geometry.

The incoming particles are simulated using the General Particle Source (GPS) function which allows for a large flexibility. The physical interactions of the photons inside of the detector is handled using the G4EmLivermorePolarizedPhysics list. This physics list was previously found to handle the Compton scattering, including the polarization, correctly \cite{POLAR:2017iip}. As the conversion of the energy deposition to the electrical signal is handled in the digitization, only the amount of energy deposited, the detector channel where this occurred and the position of the interaction within the scintillator are stored. The latter is stored in case a dependency of the light yield on this position is found in the analysis. It should be noted that one correction to the deposited energy is already applied within Geant4. This regards the Birks' effect \cite{Birks} where a Birks' constant of $0.143\,\mathrm{mm/MeV}$ is used, as taken from \cite{Polar_Birks}.

For each primary photon the  Birks' corrected energy depositions, the channel number and the position within the scintillator are stored within a ROOT file which is passed to the digitizer.

\subsection{Digitization}

The ROOT file produced by Geant4 is read-in by the digitizer along with several configuration files which contain information on the various properties of each channel. For each primary photon the following steps are performed:

\begin{itemize}
    \item For each energy deposition ($E$) the deposited energy is converted to a number of optical photons. This conversion is done using the light yield (LY) provided in the data sheet for the EJ-200 material which is 10.0 optical photons / keV. It should be noted that in the operational temperature range applicable for these tests (room temperature) and its operation in space ($-0^\circ C$) the light yield does not change for this material \cite{EJ200}. To take into account the statistical fluctuations the value of E/LY is used as the $\lambda$ parameter in a Poisson distribution.
    \item The light collection efficiency, defined as the fraction of the produced optical photons which reach the SiPM entrance window, is taken into account by multiplying the total number of produced photons by a factor of $0.30$ which is based on optical simulations performed of the system \cite{NDA_thesis}. In the case of POLAR the number of photons which reached the PMT was dependent on where these photons are produced in the bar. However, for POLAR-2, optical simulations did not show such a dependence, which was also verified by measurements which will be presented in section \ref{sec:pos_LY}. This difference can be attributed to the uniform shape of the scintillator bars in POLAR-2 compared to those used in POLAR which were tapered. The tapering causes a dependence on the light yield for interactions in these tapered areas. The uniform shape of the bars used in POLAR-2, along with the uniform wrapping and their small size compared to the optical absorption length of the scintillator material (380 cm) \cite{EJ200}, allows for their position independent light yield. A detailed discussion can be found in \cite{NDA_thesis}.
    \item Before converting the optical photons into photo-electrons in the SiPM the optical crosstalk has to be applied. For this purpose the true optical crosstalk, called $x_{ij}$ before, is required. In order to calculate this, the matrix $F$, as presented in section \ref{sec:Xtalk}, is used along with the gain matrix. The matrix $F$ is provided along with the gain for each channel as an input configuration file. The matrix $X$ is calculated once, at the start of the simulations, and subsequently applied to distribute the number of produced optical photons to the 64 scintillators. The number of optical photons which reaches channel j from channel i is calculated by multiplying the total number of optical photons by $x_{ij}$ and using this as the $\lambda$ in a Poisson distribution.
    \item The optical photons which reach the SiPM are now converted into photo-electrons. For this purpose the number of photo-electrons is picked from a Poisson distribution centered around the number of optical photons multiplied by the photo-detection efficiency of the SiPM. For the Hamamatsu SiPMs used here, for the applied bias voltage, this value is 0.5.
    \item Subsequently, the intrinsic SiPM crosstalk is added to the number of photo-electrons for each channel. The intrinsic crosstalk is defined as the probability for a photo-electron to induce a second photo-electron. For the type of SiPMs used here this is approximately $15\%$. The probability for $N$ photo-electrons to be added to the signal is $0.15^N$. To take this into account a random number is taken between 0 and 1 and compared to $0.15^N$ to pick the value of $N$. 
    \item The number of photo-electrons can now be converted to an ADC value. For this purpose a configuration file containing, for each channel, the distance between the photo-electron peaks ($\rm P.E._{dist}$) and the width of each peak, is read in to provide these values. This value allows to calculate the mean ADC value for the given number photo-electrons which is used as the mean of a normal distribution. The width of the photo-electron peak, $\sigma_{N_{\rm P.E.}}$, is described as: 

    \begin{equation}\label{eq:6}
        \sigma_{N_{\rm P.E.}} = \sqrt{\sigma_1^2 N_{\rm P.E.} + \sigma_e^2}
    \end{equation}

    where $N_{\rm P.E.}$ is the number of photo-electrons, $\sigma_e$ is the electronic contribution to the width and, $\sigma_1$ the gain contribution. More details on this parameterization can be found in \cite{NDA_master}. The ADC value is calculated by picking a random number from a normal distribution centred around the number of photo-electrons multiplied by the $\rm P.E._{dist}$ and with a width of $\sigma_{N_{\rm P.E.}}$. The values of $\rm P.E._{dist}$, $\sigma_1$ and $\sigma_e$ are produced by fitting, in the data from each channel, the first 4 p.e. peaks using the sum of 4 normal distributions with a fixed distance $\rm P.E._{dist}$ and a width of $\sqrt{\sigma_1^2 N_{\rm P.E.} + \sigma_e^2}$. The typical values of $\rm P.E._{dist}$, $\sigma_1$ and $\sigma_e$ are $\approx 40,\,2$ and $4\,\mathrm{ADC}$ respectively. It should be noted that the approach used here simulates a fully linear correlation between the number of p.e. and the ADC value. While the measurements presented here are within the linear response part of the SiPM, some non-linear behaviour is seen in the ASIC. This is taken in account by one of the subsequent steps in the digitization.

    \item The ADC value calculated in the previous step is that for the high gain output. Data from a configuration file which contains the gain non-linearity parameters mentioned in section \ref{sec:nl} are then used to calculate the low gain output. For this the linear parameters which correlate the two ADC values are used.
    \item The high gain ADC value is now corrected to account for the non-linear gain by using the non-linear parameters (those which describe an error function).
    \item Electronic noise is now added to each channel. The level of this noise is based on a measurement of the width of the baseline for each channel. The noise added is taken from a normal distribution centred around 0 with a width taken from this configuration file.
    \item The ADC values (in the high gain) are now used to produce the charge triggers (as well as the time triggers although these are not relevant in the ESRF analysis). The threshold values are again measured from the real data. For this purpose the measured threshold position along with the width of this threshold is read-in from another configuration file. These values are measured by taking the ratio of the spectrum with only triggered events over the spectrum with all events. Such a distribution can be fitted using an error function, from which the mean position and the width of the threshold can be extracted. This procedure was previously applied successfully in \cite{POLAR:2017iip}.

    \item The ADC values undergo one final modification by adding the pedestal, which, again is taken from a configuration file based on measurement data.
    \item Finally, the trigger logic is applied where the event is kept only if at least one channel exceeds the charge threshold or 2 channels for the time threshold.
    \item To reproduce the ESRF data accurately one additional step is applied, where the ADC values for all channels which do not exceed the charge trigger threshold are picked from a normal distribution centred around 0 and a width equal to the electronic noise. This is done to reproduce the issue found with the peak sensing mentioned earlier. For future simulations this step can be removed.
\end{itemize}

\begin{figure}[!h]
  \centering
  \includegraphics[width=.7\textwidth]{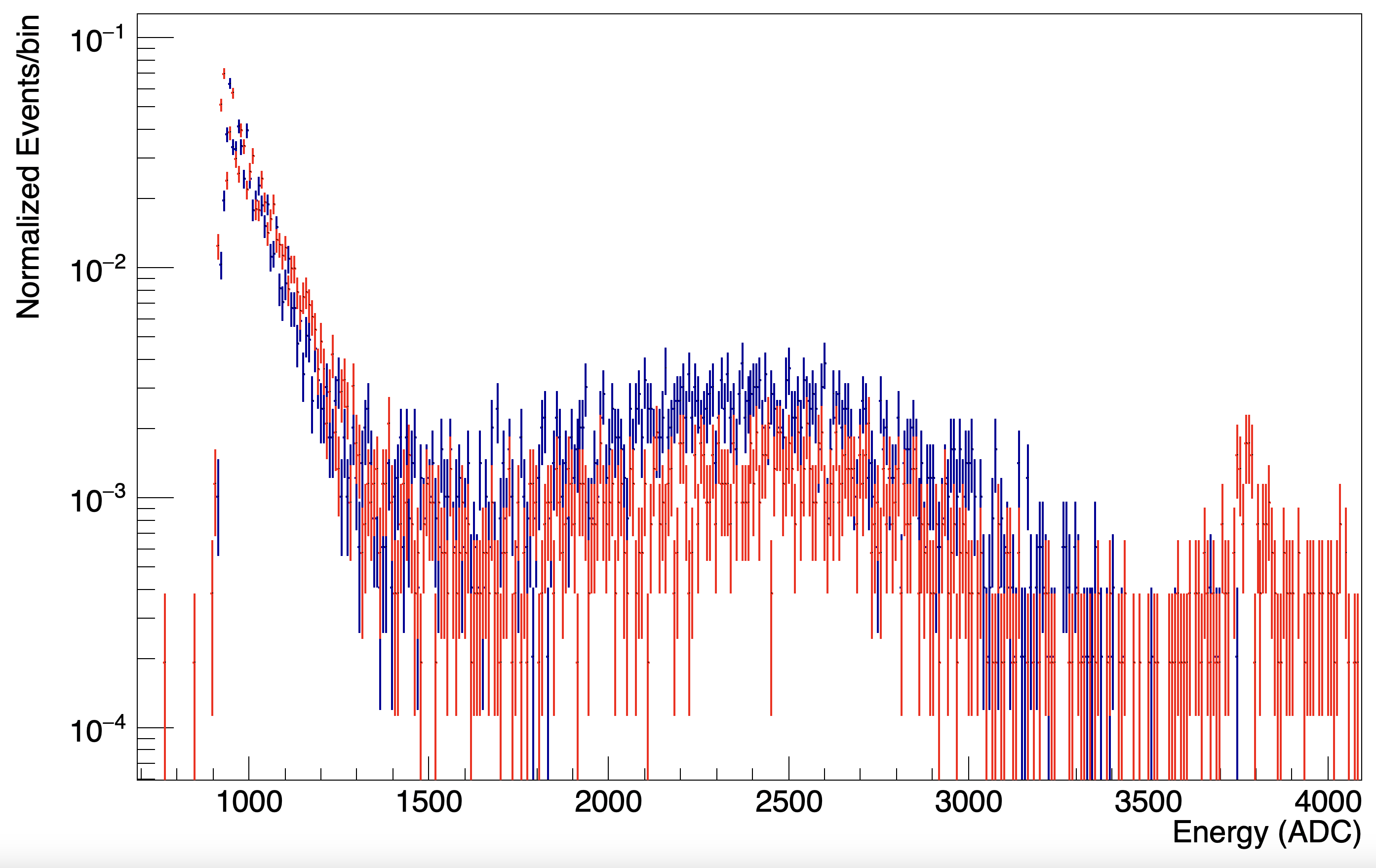}
  \caption{A simulated spectrum (blue) resulting from the digitization software compared to the measured raw data from the same detector channel (red) as resulting from a 60~keV beam.}
  \label{fig:sim_60_raw}
\end{figure}

\begin{figure}[!h]
  \centering
  \includegraphics[width=.7\textwidth]{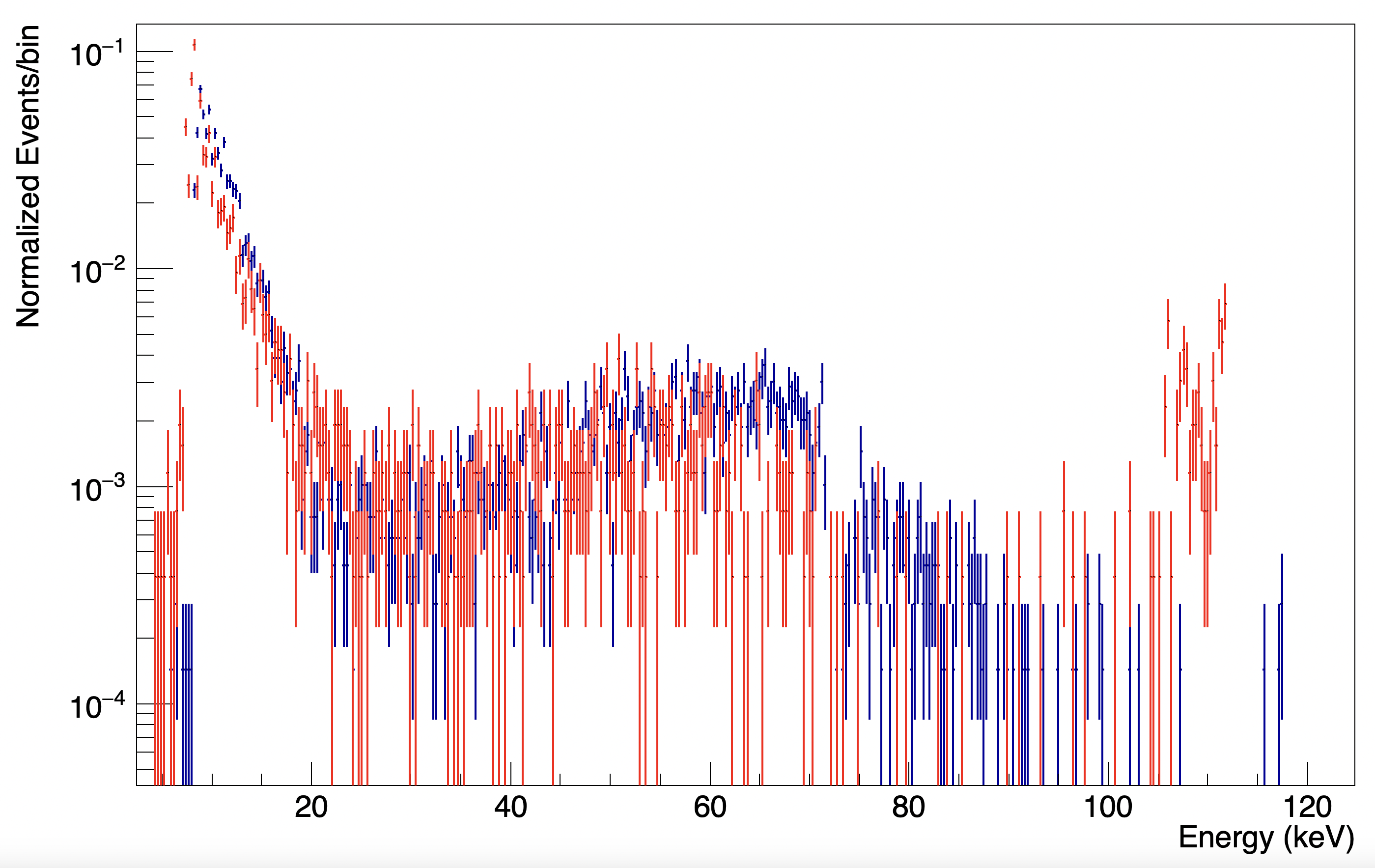}
  \caption{The same as figure \ref{fig:sim_60_raw} but after non-linearity, crosstalk and gain corrections.}
  \label{fig:sim_60_corr}
\end{figure}

\begin{figure}[!h]
  \centering
  \includegraphics[width=.7\textwidth]{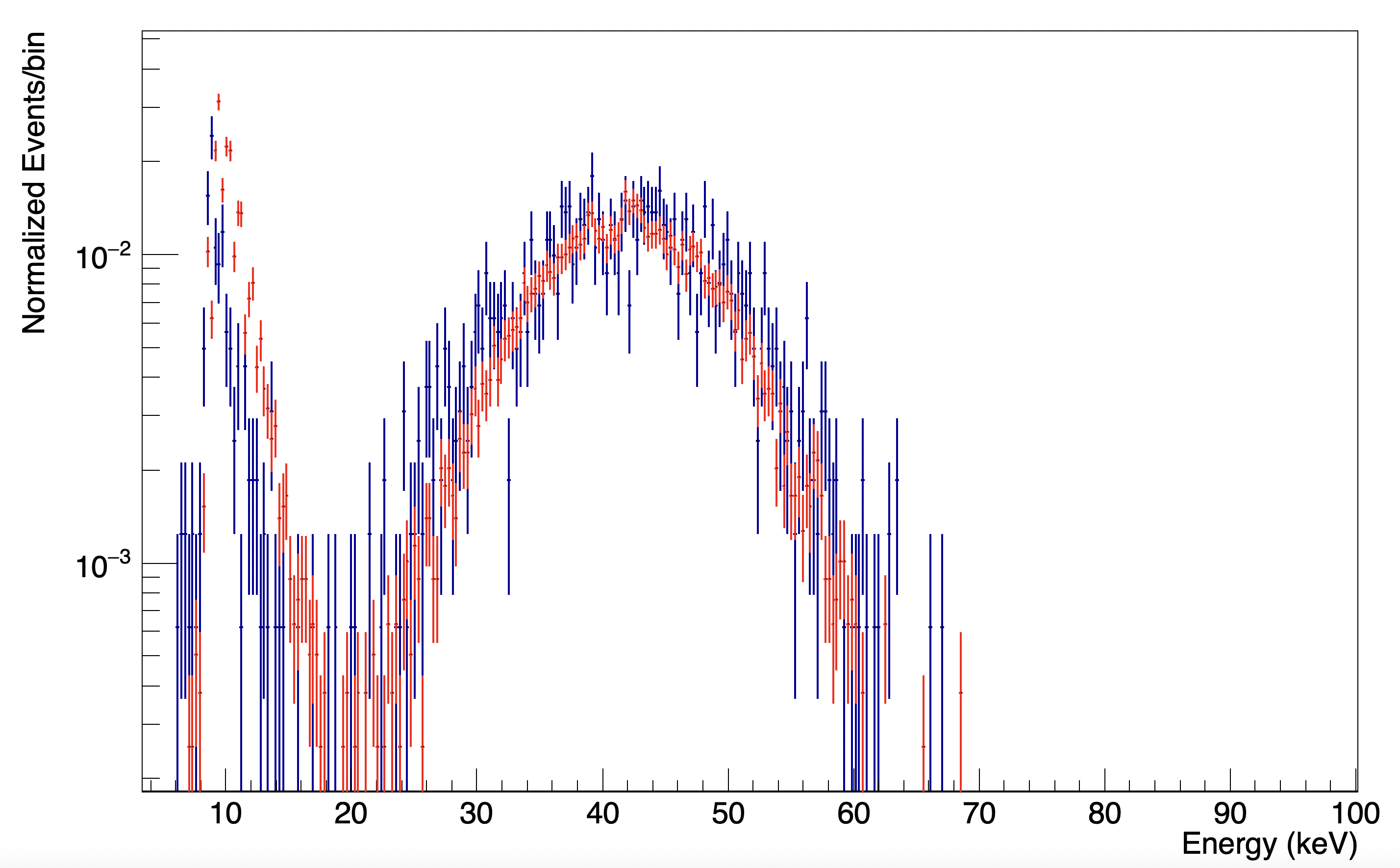}
  \caption{The same as figure \ref{fig:sim_60_corr} but for a 40~keV beam.}
  \label{fig:sim_40_corr}
\end{figure}

After the above steps ADC values (both high gain and low gain) as well as trigger information are available for each channel. These are stored in ROOT files which follow the data format of the measured data, thereby allowing this data to pass through the same analysis chain. An example of a normalized measured and simulated spectrum (from a 60~keV beam) from one channel is shown in figure \ref{fig:sim_60_raw}. This figure shows the simulated data (in blue) directly after the digitization, as a result the baseline is not yet subtracted and the output is in ADC. It should be noted that the measured spectrum has some energy depositions at high ADC values. These are a result of cosmic rays interacting in the detector during the data taking (which took several minutes). Figure \ref{fig:sim_60_corr} shows the same spectrum, again for measurements and simulations, but now after both the data sets have passed through the analysis chain described in the previous section. Finally, figure \ref{fig:sim_40_corr} shows the simulated and measured spectrum after passing through the analysis chain but for a 40~keV beam. It can be observed that the simulation results match the measurements well. For example the energy resolution at both 40 and 60~keV match the data while, unlike a commonly used approach, the energy resolution is not given as a parameter in the simulations. Rather this energy resolution results naturally by simulating the various steps, such as the production of optical photons and their conversion into photo-electrons, in the digitization process. Due to the lack of data below the threshold the photo-electron peaks are not clearly visible in these spectra. However, as will be discussed in section \ref{sec:improvements}, updates to the firmware implemented after the ESRF campaign have allowed to make these more clearly visible. As will also be discussed in that section through comparison with measurement data, the photo-electron peaks and their widths are accurately simulated.

\section{ESRF Campaign}\label{sec:esrf_campaign}

This section will start with an overview of the setup used during the ESRF campaign. The data taken there \cite{ESRF_data} will be described in detail in the subsequent subsections.

\subsection{Irradiation Setup}

\begin{figure}[!h]
  \centering
  \includegraphics[width=.5\textwidth]{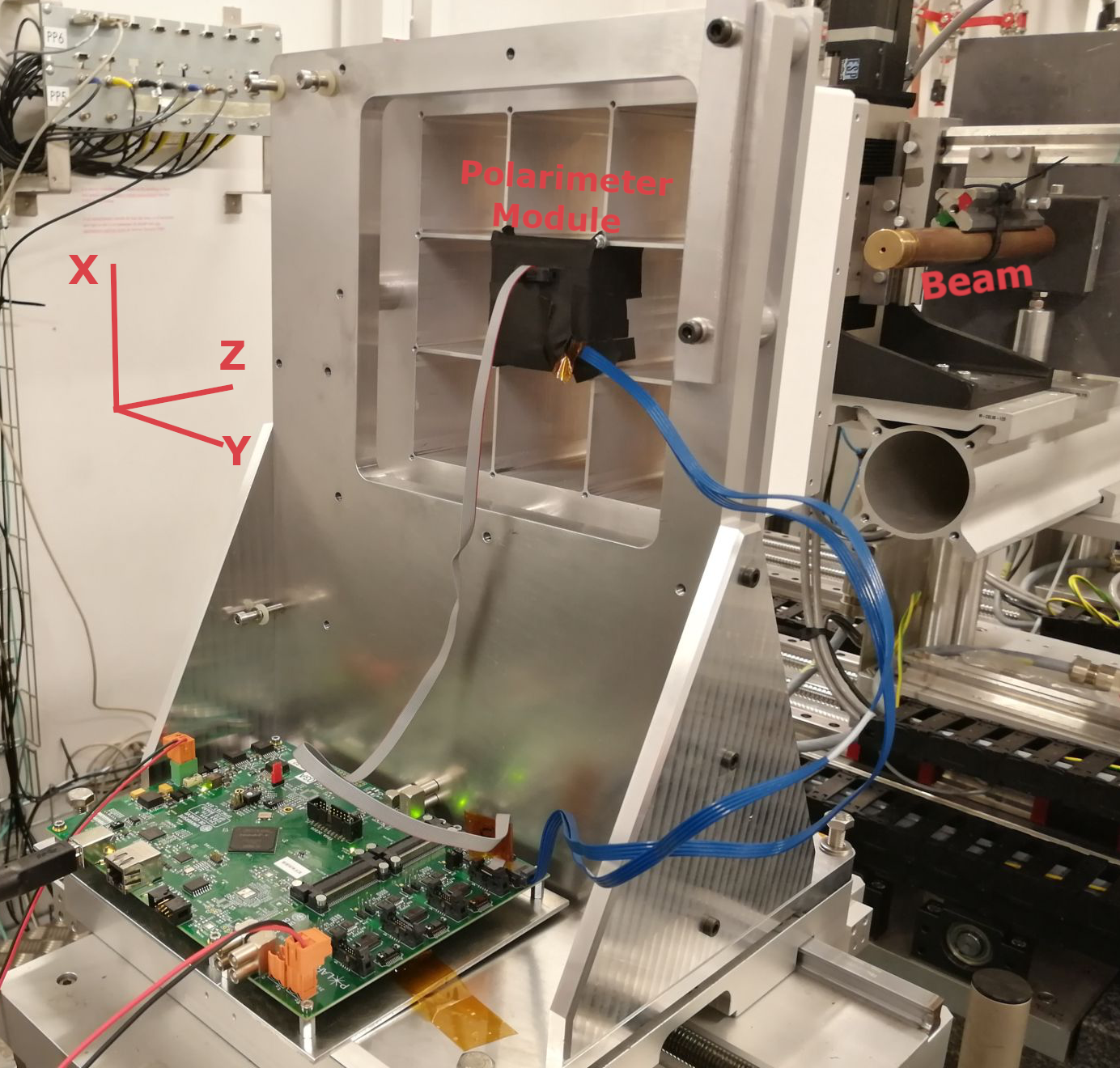}
  \caption{The setup as used at the ESRF beamline along with the definitions used for the x, y and z axis. The polarization of the beam is always along the y-axis. A single polarimeter module is placed within a scaled down version of the module grid (a $3\times3$ grid instead of $10\times10$ as will be used in the full detector). The readout electronics are placed behind the mechanical grid along with a cold plate used to keep the setup at around room temperature. All this is placed on an x-y translator capable of moving the detector with respect to the beam position.}
  \label{fig:setup}
\end{figure}

For the purpose of this calibration campaign the polarimeter module was placed in an aluminium frame which can support a total of 9 modules. This frame is a scaled down version of the final grid which will be used in the full detector, capable of holding 100 modules. This grid, which can be seen in figure \ref{fig:setup}, was connected to a custom made aluminium piece which connects it to a translation table. This translation table is capable of moving the detector within the plane perpendicular to the incoming beam direction. In addition, it can be used to rotate the detector around the x-axis (the definition of which is indicated in figure \ref{fig:setup}). The data from the polarimeter module is provided by a dedicated board developed by the University of Geneva which can be seen at the bottom of figure \ref{fig:setup} which also provides the power to the polarimeter module electronics. This board in turn provides the data to a PC which is placed outside of the experimental hutch through a single ethernet cable. The operation of the translation table, as well as the control of the beam is performed using a Python based interface provided by ESRF.

The beamline ID15A of ESRF was used for this test, details on the beam can be found in \cite{ID15}. The synchrotron beam is collimated to a size of 1~mm diameter, and can be adjusted in the energy range of 40 to 140~keV. During the tests performed here, irradiations were performed at beam energies of 40 and 60~keV as well as 100 and 120~keV. As will be discussed in section \ref{sec:improvements}, issues induced by the high rate experienced at 100 and 120~keV were found during the analysis phase which means only the data from 40 and 60~keV is analyzed here. It should be noted here that although the energy resolution of the beam is below $1\%$ \cite{ID15}, third level harmonics resulting from the beam manipulation are present for both 40 and 60~keV irradiation. As these harmonics are of the level of $1-2\%$ they were measured prior to the irradiation campaign using a Germanium detector provided by ESRF. In addition, the polarization of the beam is, based on simulations provided by the ID15A beamline team, above $99\%$ and is along the y-axis in figure \ref{fig:setup}.

As a uniform irradiation of the full detector surface cannot be achieved with the beam directly, the detector is moved with respect to the beam during irradiation. Each detector scan consists of moving the beam over the 8 rows of scintillator as indicated in figure \ref{fig:conv}. In the setup shown in figure \ref{fig:setup}, channel 0 is located at the bottom (x=0) and in the positive y direction, channel 56 is located in the same x location in the negative y direction.

To ensure the beam irradiates each scintillator for an equal amount of time, the movement in the y direction starts 10 mm outside of the module and ends 10 mm away from the opposite side. When reaching this point a movement in the x position is initiated to the centre of the scintillators in the row above, before moving backwards in the negative y direction. This is repeated until all 8 rows are scanned. The detector is moved at 1 mm per second, as a result an on-axis scan takes approximately 10 minutes. As the beam intensity, which is linearly proportional to the electron flux in the synchrotron ring varies by several $\%$ during this period, the electron flux is stored in the data. This allows to correct for fluctuations in the beam intensity during the scan. Finally, it should be noted that the synchrotron beam has to be refilled every few hours, thereby producing a sudden increase of several percent in beam intensity. Although one can correct for this, it was decided to avoid taking data during these sudden changes in beam intensity. To achieve this, the time before a refill is automatically monitored in our scanning software which operates the table movement. During each translation in the x-direction (so in between the scans of the rows), it is checked if a refill will occur during the scan of the next row. If so, the start of the y-translation is paused until after the refill has occurred.

Off-axis scans are performed by rotating the detector around the x-axis, where the rotation angle is referred to as $\theta$. To accommodate for the larger detector surface the translation in the y-direction is increased for such scans. Finally, in order to change the polarization angle, the polarimeter module is rotated by 90 degrees around the z-axis.

\subsection{Light Yield}\label{sec:LY}

Using the data analysis pipeline discussed in section \ref{sec:data} the spectra of all 64 channels inside the polarimeter module are processed. The spectrum, after pedestal and non-linearity correction, from a 60~keV irradiation can be fitted using a Gaussian function. This provides the gain as measured in ADC/keV. In addition, the photo-electron peaks, or fingers, can be fitted in order to extract the distance, as measured in ADC, between them. This second fit provides the ADC/photo-electron ratio. Using the combination of these two ratios one can finally calculate the photo-electrons/keV ratio, which gives an indication of the detection efficiency or light yield detection. It should finally be noted that, as explained for the optical crosstalk, the true gain cannot directly be measured as it is affected by the crosstalk. We therefore finally correct for the optical crosstalk to produce the true light yield for each channel. The results for this measurement are shown below in figure \ref{fig:EJ200}. While the uncertainty from the fit of the photo-peak is of the order of $1\%$, the fit on the distance between the fingers has a more significant error of $\approx 10\%$. Overall, the systematic error on each entry in the distribution is therefore around $10\%$, indicating that some intrinsic differences between the channels do exist. Such differences are, for example, the differences in the quality of the scintillators and their optical coupling to the SiPM. 

The light yield detection presented here is measured using a 60~keV beam. It is important to note that the Birks effect remains non-negligible at these energies. At higher energies, the light yield detection will therefore be approximately $10\%$ higher.

\begin{figure}[!h]
  \centering
  \includegraphics[width=.5\textwidth]{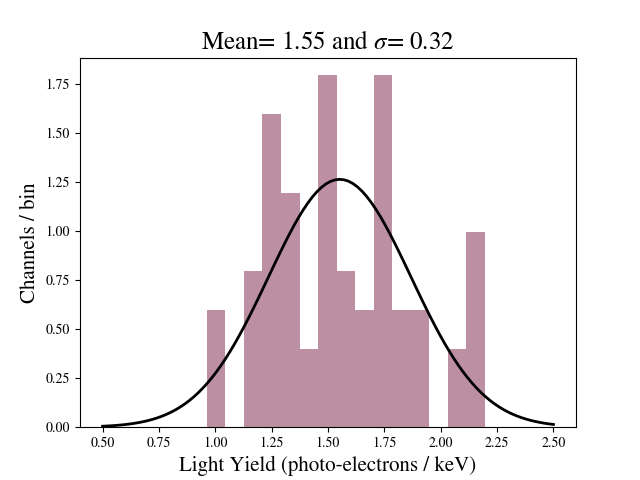}
  \caption{The distribution of the light yield as measured for the polarimeter module which uses EJ-248M as the scintillating material. The distribution is fitted with a Gaussian and the mean and width are presented.}
  \label{fig:EJ248M}
\end{figure}

Using optical simulations presented in \cite{NDA_thesis} the light yield detection, including the optical crosstalk, was predicted both for a polarimeter module with EJ-200 and for one with EJ-248M. The results in figure \ref{fig:EJ200} are for one with EJ-200, which based on the predictions from \cite{NDA_thesis} is expected to be\footnote{The values provided in \cite{NDA_thesis} are for higher energies and therefore need to be corrected for the Birks effect, resulting in decrease of $\approx10\%$. They also already include the optical crosstalk which is corrected for in this work and result in an opposite correction of $10\%$.} $1.31\pm0.16$ photo-electrons/keV. Measurements with a second polarimeter module (which contained several damaged SiPM channels and was therefore not used for polarization analysis) which uses EJ-248M as the scintillating material are shown in figure \ref{fig:EJ248M}. The light yield can be seen to be $\approx 1.5$ photo-electrons/keV which again matches the simulations presented in \cite{NDA_thesis} that expects $1.43\pm0.15$ photo-electrons/keV. The larger spread in these results compared to that in figure \ref{fig:EJ200} can be attributed to the lower quality of this polarimeter module. As mentioned several SiPM channels were damaged and could therefore not be used, while others contained minor scratches. As a result, both the intrinsic differences between the channels are larger, while in addition, the typical errors on the various fits required to produce the light yield are larger. Overall, we do however, see an improvement in the light yield consistent with simulations, although it is clear that more detailed tests are required in the future.

To put this number into comparison, for POLAR, this value was only $0.3$ photo-electrons/keV. The large difference between these two can mainly be attributed to the difference in the scintillator shape, where the POLAR scintillators were tapered towards the bottom to a size of $5\times5\,\mathrm{mm^2}$ in order to reduce optical crosstalk. Despite this, a second improvement comes still from the significant reduction in the optical crosstalk, which in case of a POLAR detector, induced a loss which could exceed $50\%$ of the light to neighboring channels. For POLAR-2 this has been reduced to $\approx10\%$. Finally the increase in the photo-detection efficiency of the detector of $50\%$ against $20\%$ for POLAR, further improves this. 

The importance of this value, in the case of a SiPM based detector, can be explained in the form of the dark noise. As SiPMs suffer from significant dark noise, the trigger threshold for a channel should exceed a certain number of photo-electrons, or fingers, in order to reduce the dark noise induced trigger rate. The number of photo-electrons produced for a dark noise event scales with the intrinsic crosstalk of the SiPM, which in the case of the SiPMs used here is of the order of $10\%$. As a result, $10\%$ of the dark noise induced events have more than 1 photo-electron, while a fraction of $0.10^{N+1}$ exceed N photo-electrons. With a typical dark noise of 1~MHz per channel, reducing this to 1~Hz correlates to a trigger threshold exceeding 7 photo-electrons. Using the typical light yield of a POLAR-2 channel, this translates to approximately\footnote{The value of 5~keV used here is calculated using a linear correlation between p.e. and keV. However, at low energies the Birks' effect should be taken into account. The real threshold for energy deposition is therefore $\approx6\,\mathrm{keV}$.} 5~keV. For the measurements performed during the tests here, the threshold was conservatively set to 8~keV, thereby making the dark noise contribution negligible during these measurements.

\begin{figure}[!h]
  \centering
  \includegraphics[width=.5\textwidth]{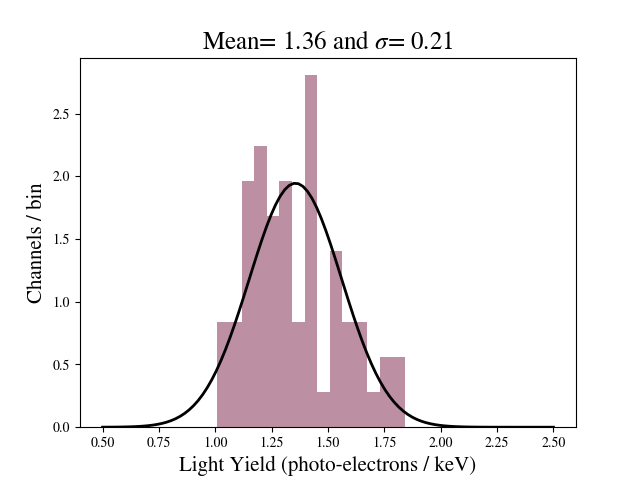}
  \caption{The distribution of the light yield as measured for the polarimeter module which uses EJ-200 as the scintillating material. The distribution is fitted with a Gaussian and the mean and width are presented.}
  \label{fig:EJ200}
\end{figure}

\subsection{Position Dependent Light Yield}\label{sec:pos_LY}

The results presented in section \ref{sec:LY} were acquired by using a 60~keV beam and irradiating the scintillator bars from the top.  In the case of an off-axis irradiation the distribution of the photo-absorption position in the bar would be different from that of the on-axis case. In case a dependence between the interaction position and the light yield exists, the results presented in section \ref{sec:LY} would only be valid for an on-axis beam. In order to understand the response of the POLAR-2 instrument, where GRBs are very unlikely to come directly from the zenith, it is therefore vital to know whether the light yield depends on the interaction position within a scintillator bar. 

\begin{figure}[!h]
  \centering
  \includegraphics[width=.5\textwidth]{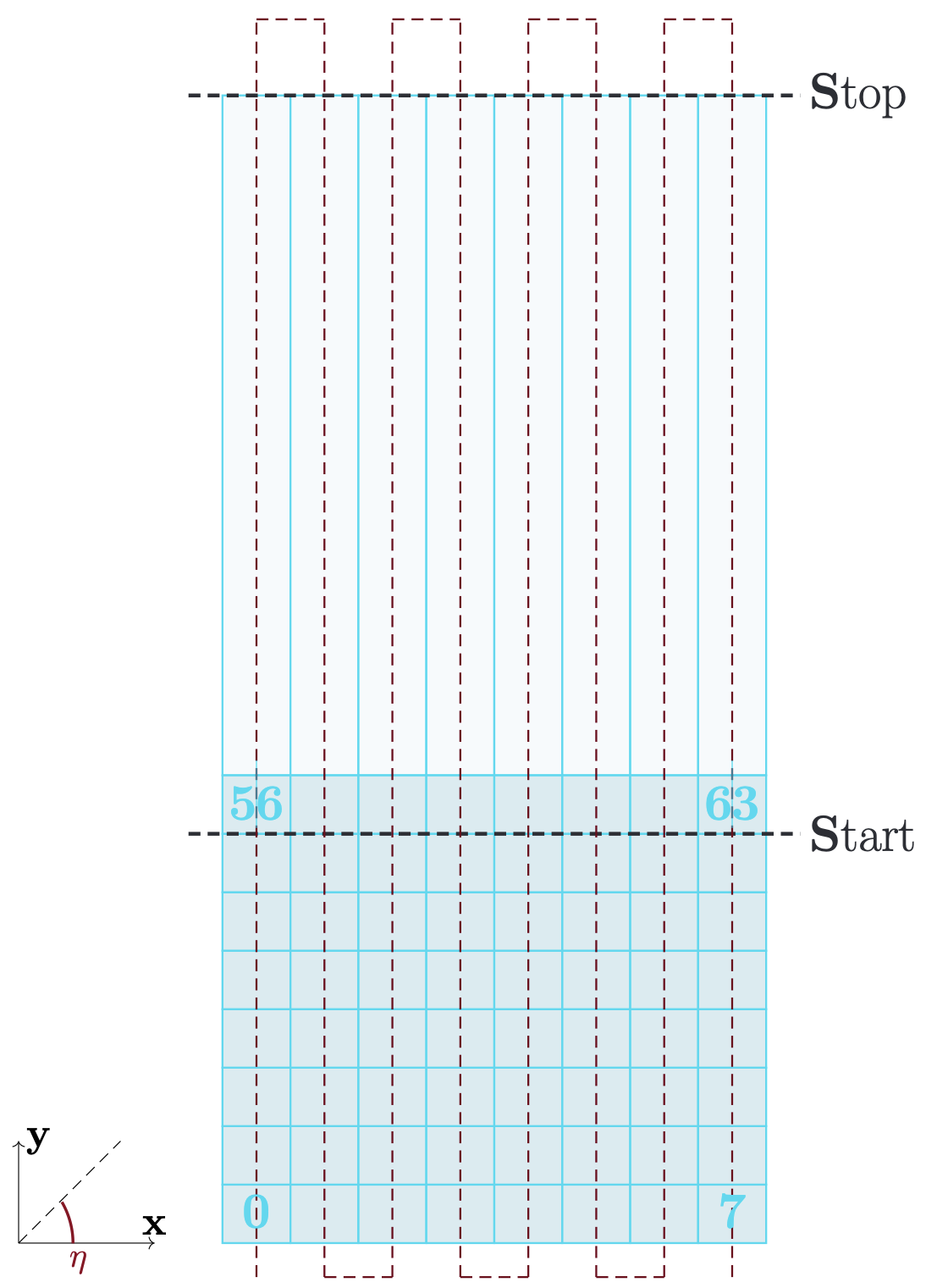}
  \caption{The same as figure \ref{fig:conv} but now tilted along the x-axis to indicate an off-axis scan. The black dotted start and stop lines correspond to when the beam moves in and out of the scintillators used for studying the position dependent light yield.}
  \label{fig:off_axis}
\end{figure}

While optical simulations presented in \cite{NDA_opt} indicate that the dependence is minimal, this needs to be verified using measurements. For this purpose we use the data from one of the off-axis irradiation runs where the polarimeter module was tilted with an angle of $30^\circ$ with respect to the 60~keV beam. During such a scan, the beam irradiates the majority of the bars for a short period, while the bars on one of the outer edges are all irradiated for a total of $\approx65$ seconds. This is illustrated in figure \ref{fig:off_axis}, which shows the scintillators, as in figure \ref{fig:conv} but tilted along the x-axis. During the 65 seconds the beam also scans the side of scintillator channels 56 to 63. Therefore producing a high probability for photo-absorption to occur at the top during one part of the scan, and a high probability for this to occur at the bottom during a different part. A dependence of the light yield on the interaction position would result in a shift in the photo-absorption peak.

\begin{figure}[!h]
  \centering
  \includegraphics[width=.5\textwidth]{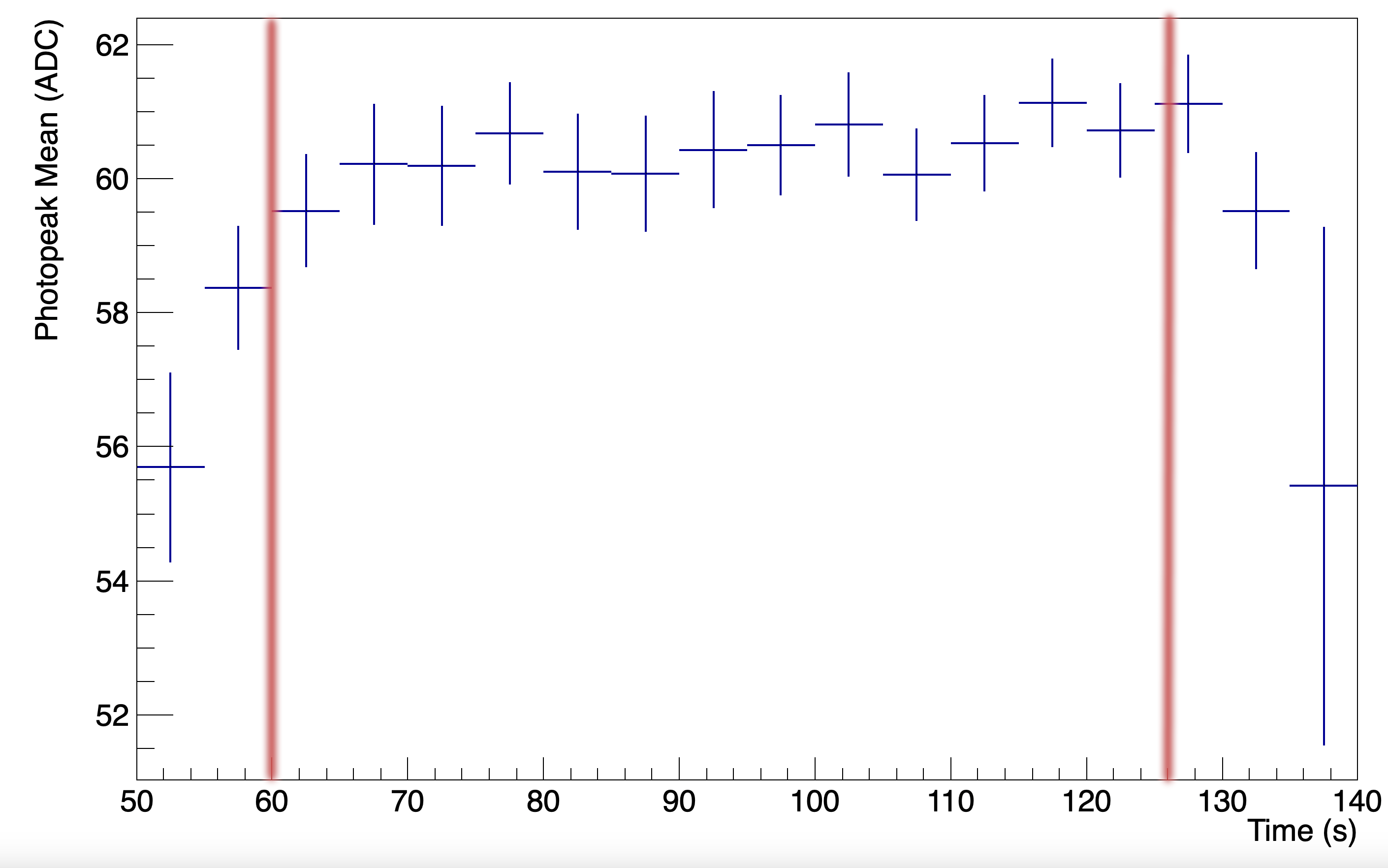}
  \caption{The mean position of the photo-absorption peak as a function of time during an off-axis scan of one channel with 60~keV photons. The time at which the beam enters the bar is indicated with the red line around 60 seconds, while the moment it leaves is indicated with the second red line.}
  \label{fig:mean_photo_56}
\end{figure}

To study this we use the data from channel 56 from a scan where the beam enters the top of the bar at around 60 seconds (indicated as the black start line in figure \ref{fig:off_axis}) and leaves the bottom of the bar at around 125 seconds (indicated as the Stop line in figure \ref{fig:off_axis}). The mean position of the photopeak is shown as a function of time in figure \ref{fig:mean_photo_56}. The two red lines indicate when the beam enters and leaves the bar. It can be observed that while the beam is not pointing at the bar the photo-peak is slightly below 60~keV. This can be explained by the fact that an incoming photon cannot deposit its full 60~keV photon in this bar as it first needs to scatter in a neighboring bar, thereby losing some of its energy. While the beam is irradiating the scintillator bar directly we see that all measurements are compatible with 60~keV and we do not see any trend in the position of the peak position. Similar results were found for all the 8 bars studied in this way for this scan. The results therefore support the simulations presented in \cite{NDA_opt} and we will continue to assume that the light yield in a bar does not depend on the interaction position.

\subsection{On-Axis Modulation}

Using both the scanned data, as well as the simulated one, scattering angle distributions can be produced for various energies. As not every triggered event contains energy depositions in two or more scintillator bars an event selection has to be applied before. In the study performed here this event selection is kept simple. The scattering angle, measured as the angle from the x-axis as defined in figure \ref{fig:conv}, is calculated if at least two energy depositions are found where both have issued a charge trigger. The requirement for a charge trigger is needed as, as discussed before, the peak sensing on the ASIC was not initiated for channels in which no charge trigger is issued. As a result, the ADC values in these channels are not correlated to actual energy depositions. The scattering angle is calculated from the bar with the second highest energy deposition to the bar with the highest energy deposition. It should be noted that the order used here is arbitrary as inverting it will only produce a phase shift of $180^\circ$. Furthermore, to calculate the angle, a random position is selected within the two scintillator bars. This is required to account for the lack of information on the interaction position within the bars and to avoid a discrete set of scattering angles to appear in the scattering angle distribution.

The selection applied here is the most simple one, while future studies can be used to optimize the selection to increase, for example, the $\mu_{100}$ or the number of selected events. Such an optimization study will need to be applied on the full detector, so with 100 instead of 1 module. We therefore restrict ourselves here to the simple analysis with the main purpose of validating that the instrument is indeed sensitive to polarization and, more importantly, to validate the MC simulation software.

It should be stressed that, before applying the event selection on both the simulated and measured data, both types of data go through the same analysis pipeline which includes pedestal subtraction, non-linearity correction, cross talk and gain correction. Two minor corrections are however applied to the measured data which are not applied to the simulated data. The first of these is the application of a cut on the data to only include events recorded while the beam was pointing at the detector module. Although, in normal circumstances only a fraction of the data is recorded when the beam is not irradiating the detector, during some scans the synchrotron electrons were refilled during a scan. The motor operating algorithm was written such that during such a refill of the synchrotron ring, which typically lasts several minutes, the beam was moved out of the detector while this happened. As a result, some scans include several minutes of additional background data which is removed from the analysis through this cut.

The second is a minor correction to account for variations in the beam intensity during irradiation. During a typical scan of the detector the beam intensity varies by a few $\%$ as the synchrotron electron beam slowly decreases in intensity. The intensity of the beam was recorded in the house-keeping data and each event entered in the scattering angle distribution is weighted by the relative beam intensity, thereby ensuring a fully uniform irradiation of the detector surface.

\subsubsection{Modulation for a 60~keV beam}\label{sec:60keV}

\begin{figure}[!h]
  \centering
  \includegraphics[width=.7\textwidth]{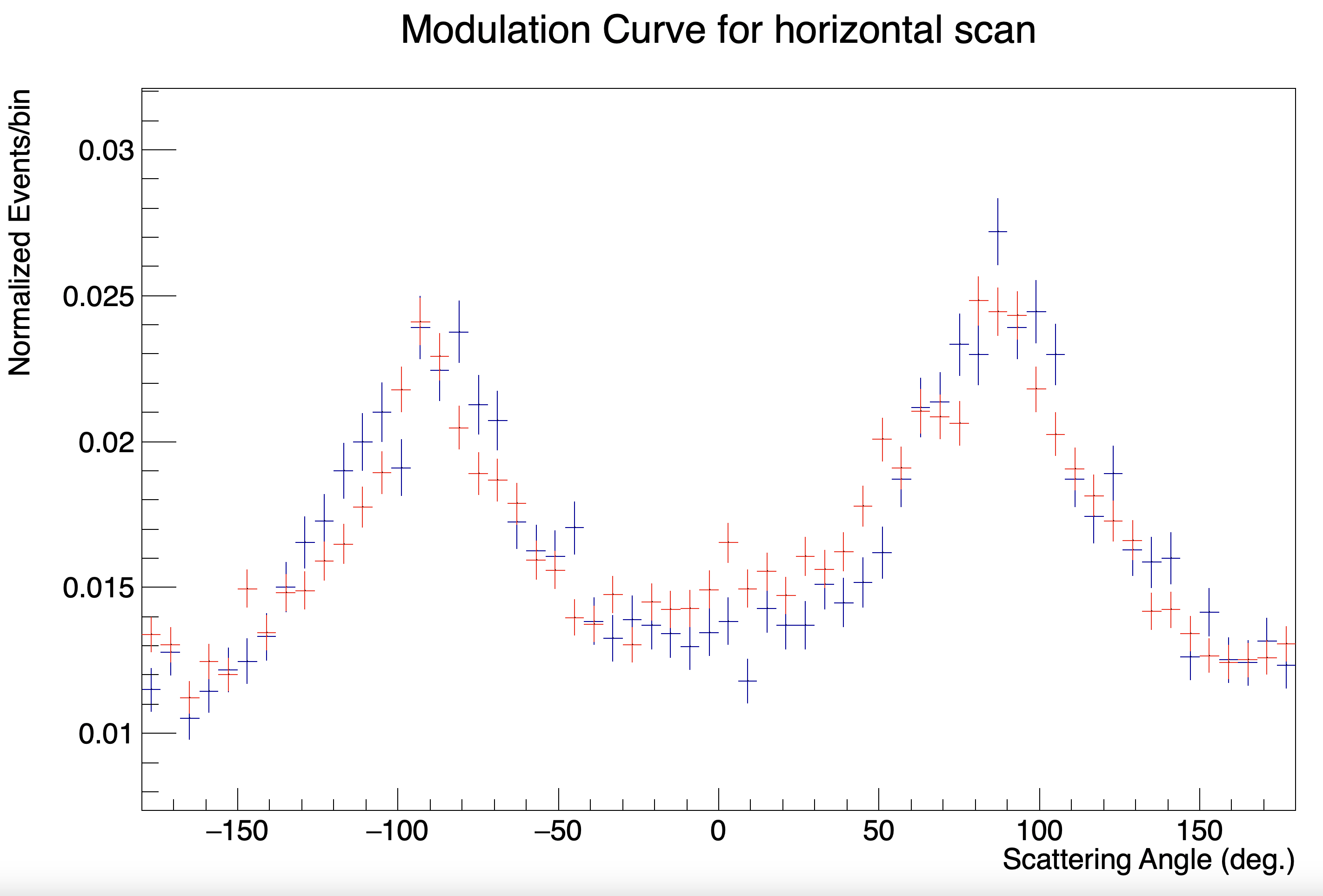}
  \caption{The normalized scattering angle distributions produced using simulations (blue) and measurements (red) for a 60~keV beam with PD = $100\%$ and PA=$0^\circ$ (polarization vector along the x-axis of the polarimeter module).}
  \label{fig:mod_60_raw_hor}
\end{figure}

\begin{figure}[!h]
  \centering
  \includegraphics[width=.7\textwidth]{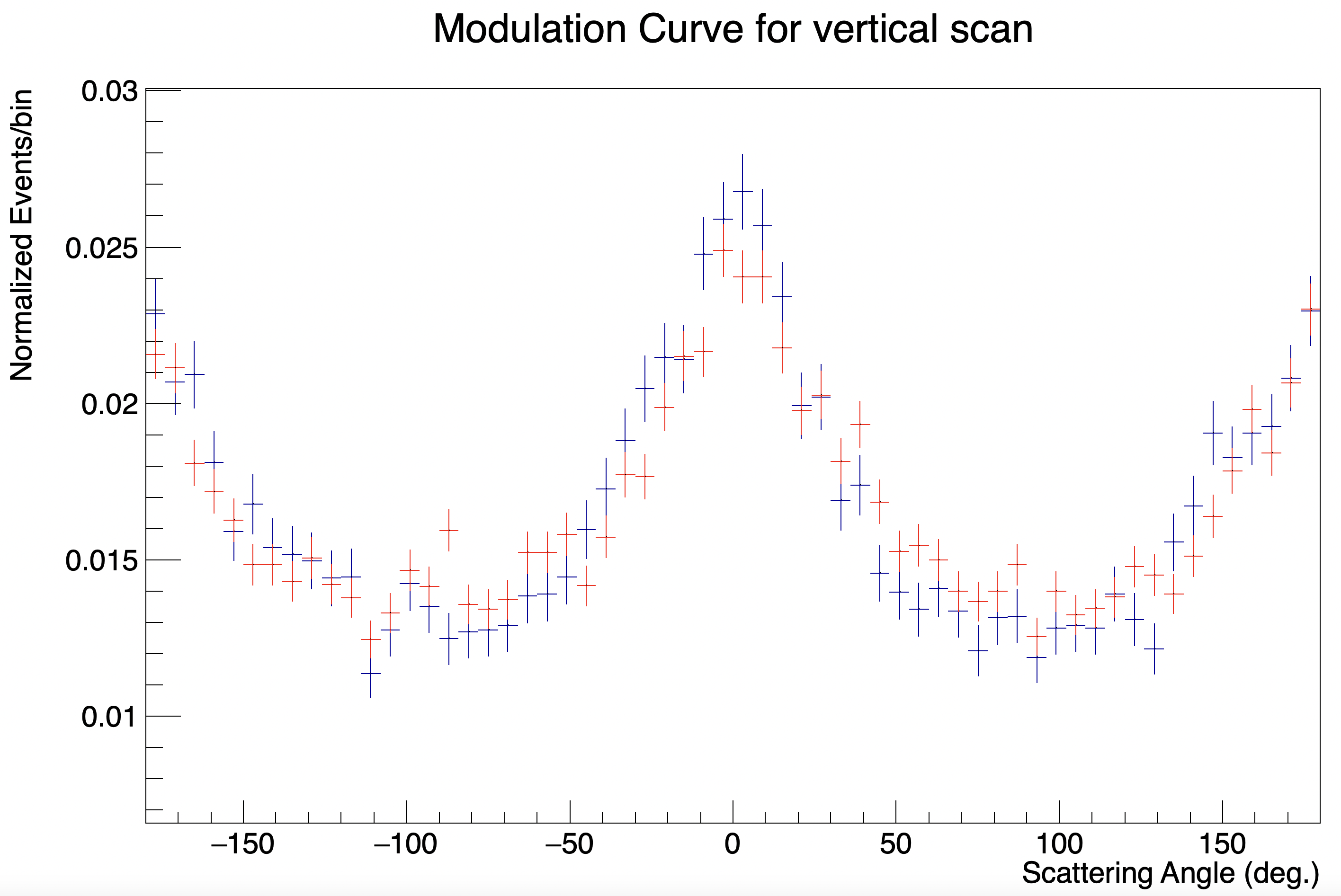}
  \caption{The normalized scattering angle distributions produced using simulations (blue) and measurements (red) for a 60~keV beam with PD = $100\%$ and PA=$90^\circ$ (the polarization vector aligned with the y-axis of the module).}
  \label{fig:mod_60_raw_vert}
\end{figure}

Figures \ref{fig:mod_60_raw_hor} and \ref{fig:mod_60_raw_vert} show the normalized measured and simulated scattering angle distributions resulting from a 60~keV beam with orthogonal polarization angles. The polarization of the beam is, in both cases along the Y-axis of figure \ref{fig:setup}. As the module is rotated in between scan (along the z-axis), for figure \ref{fig:mod_60_raw_hor} the x-axis, as defined in the polarimeter module, is along the y-axis of the setup, while in figure \ref{fig:mod_60_raw_vert} the polarimeter module is rotated such that the x-axes of the two systems are aligned. Both of the scattering angle distributions show a clear $180^\circ$ modulation where the phase is shifted by $90^\circ$ as expected. In addition to the polarization-induced effect, clear instrumental effects can be seen. These become easier to spot when we combine the two curves. By combining two polarized beams with an orthogonal polarization an unpolarized beam is created. This can be illustrated through Stokes parameters for a polarized beam of intensity 'I' added to one with equal intensity but orthogonal polarization:

\begin{equation}\label{eq:7}
    {\displaystyle {\begin{pmatrix}I\\I\\0\\0\end{pmatrix}} + {\begin{pmatrix}I\\-I\\0\\0\end{pmatrix}} = {\begin{pmatrix}I\\0\\0\\0\end{pmatrix}}}
\end{equation}  	

\begin{figure}[!h]
  \centering
  \includegraphics[width=.7\textwidth]{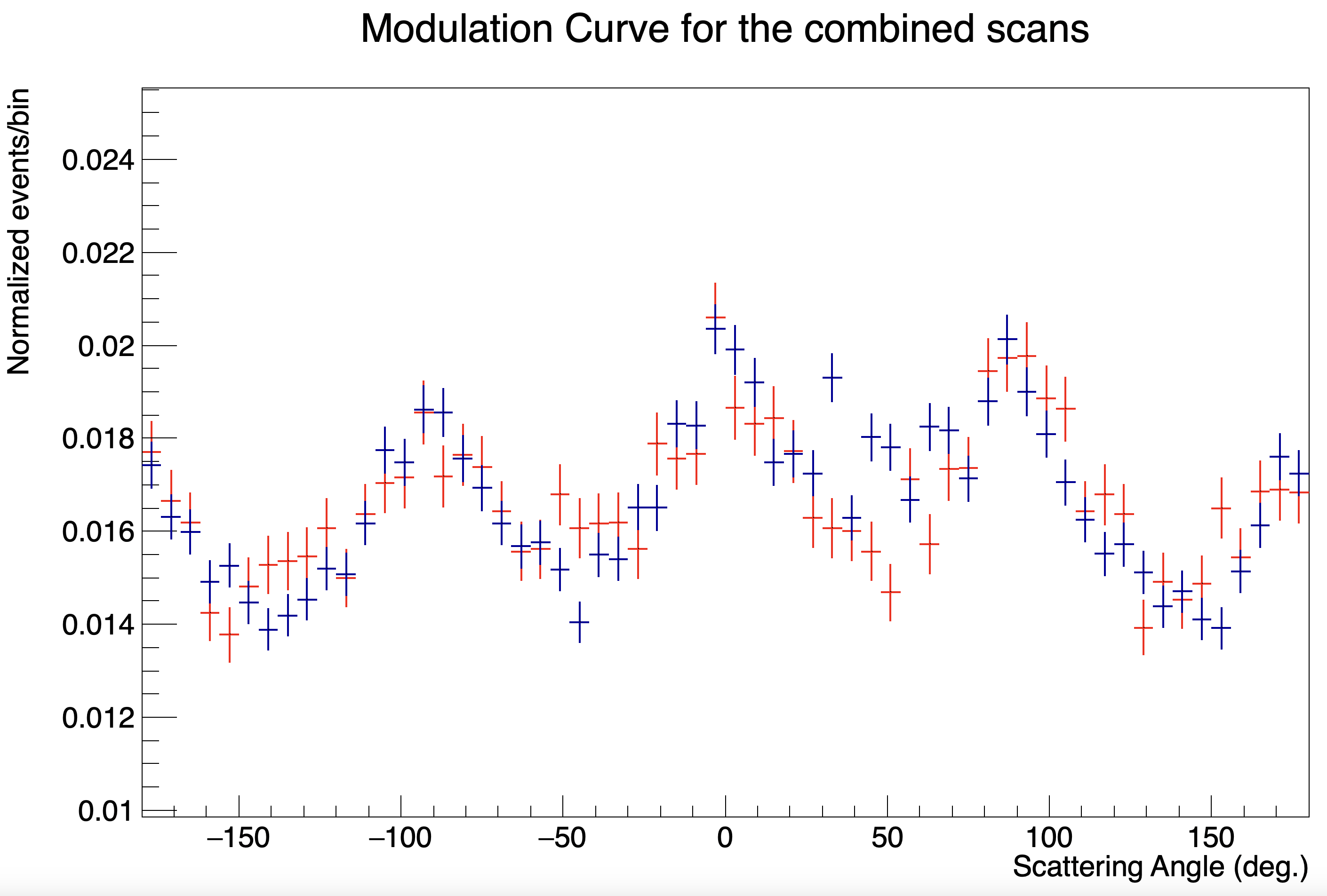}
  \caption{The normalized scattering angle distributions produced using simulations (blue) and measurements (red) for a 60~keV beam with PD = $0\%$.}
  \label{fig:mod_60_raw_upol}
\end{figure}

Therefore by summing and using a weighting correction for the beam intensities a fully unpolarized beam response can be produced. The result, both from simulations (blue) and measurements (red) can be seen in figure \ref{fig:mod_60_raw_upol}. The figure shows clear peaks at every $90^\circ$ resulting from the square geometry of the detector module and the scintillator bars. In addition, it can be observed that not all of the peaks have the same height, resulting from a non-uniformity in the sensitivity of the various detector channels. These come, for example, from the difference in the gain and threshold of each channel. Although these effects indicate that the detector does not have a fully uniform response, it is important to note that the non-uniformity is also seen in the simulated distribution, indicating that these are modeled correctly in the simulations.

\begin{figure}[!h]
  \centering
  \includegraphics[width=.5\textwidth]{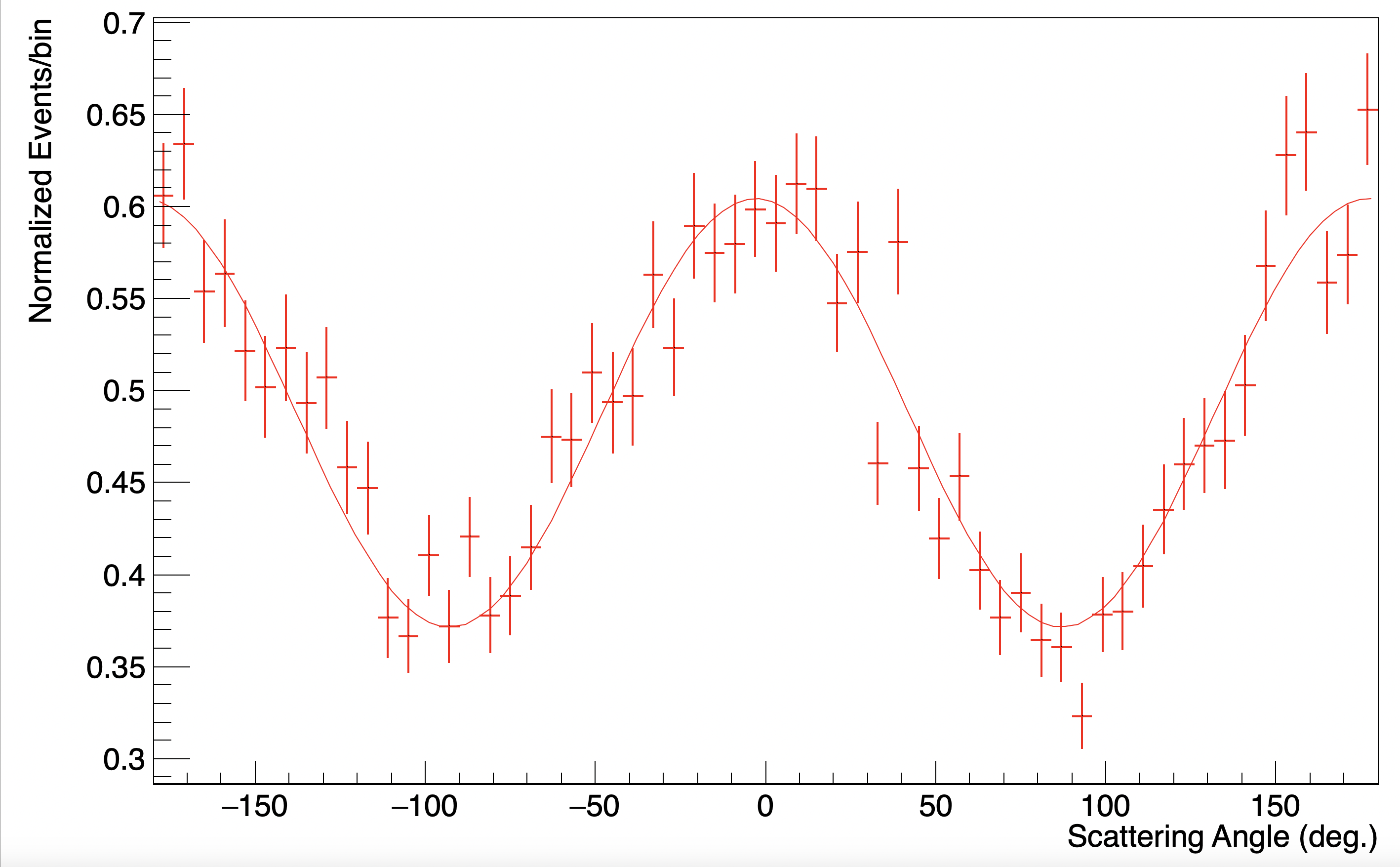}
  \caption{The normalized corrected scattering angle distributions from measurements for a 60~keV beam with PD = $100\%$ and PA = $90^\circ$. The $\mu_{100}$ extracted from here is $(23.9\pm0.9)\%$.}
  \label{fig:60keV_mod_hor}
\end{figure}

The unpolarized response can now also be used to remove the instrumental effects thereby leaving only the polarization-induced effects. An example from measurements is shown in figure \ref{fig:60keV_mod_hor}. The resulting distributions, fitted with a $180^\circ$ harmonic function, represent the response of the detector to a $100\%$ polarized beam and can therefore be used to extract the $\mu_{100}$. The values of $\mu_{100}$ from measurement were found to be $(23.9\pm0.9+0.2)\%$ for PA = $90^\circ$ and $(24.6\pm1.2+0.2)\%$ for  PA = $0^\circ$. The second, asymmetric error here stems from the uncertainty on the beam polarization which is known to be > $99\%$. From simulations, we find $(24.3\pm2.0)\%$ and  $(26.0\pm2.4)\%$ respectively. The simulations therefore match the measurements within the error. 

\begin{figure}[!h]
  \centering
  \includegraphics[width=.7\textwidth]{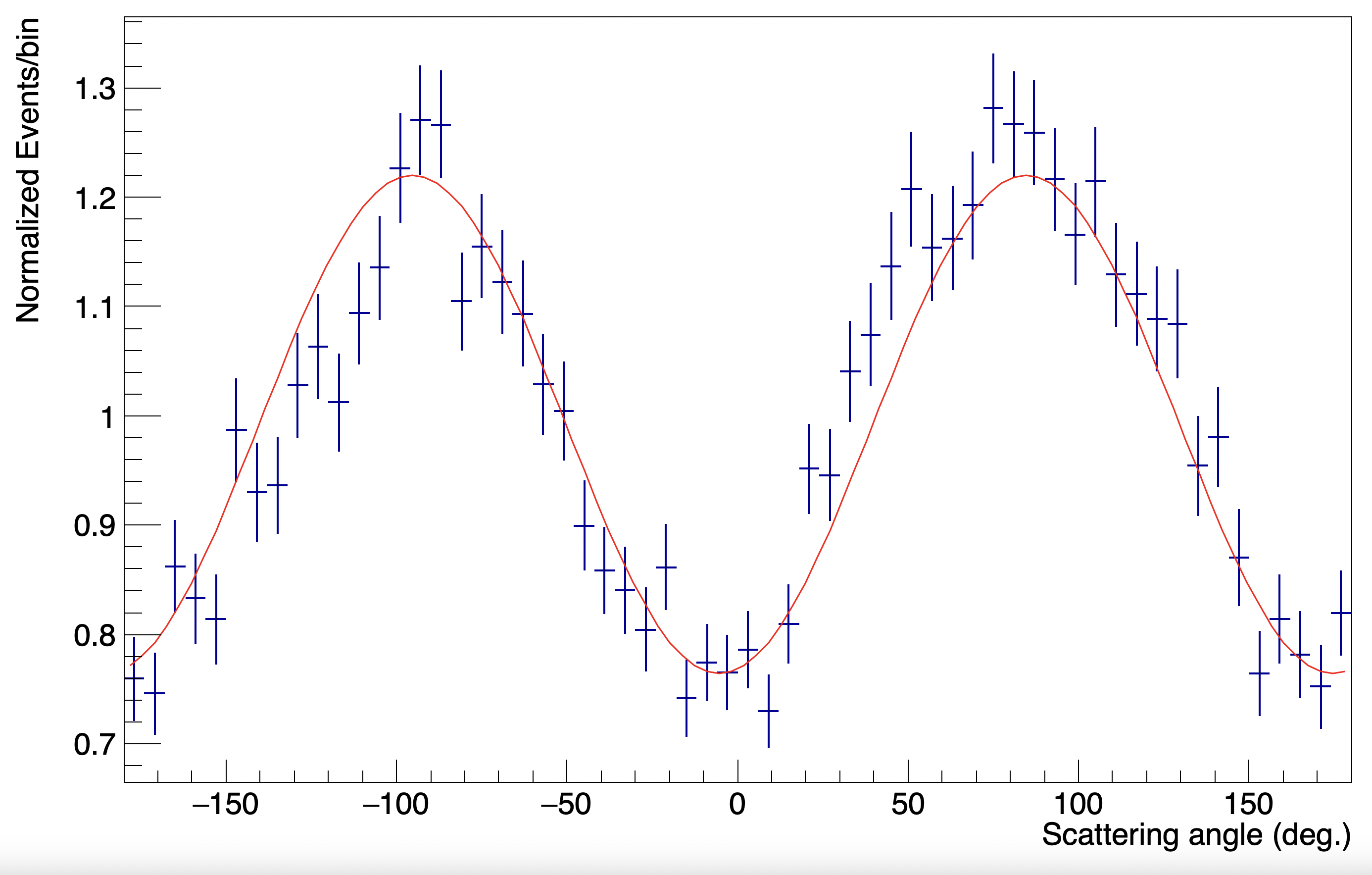}
  \caption{The normalized  measured scattering angle distributions corrected using an unpolarized scattering angle distribution produced using simulations for measurements for a 60~keV beam with PD = $100\%$ and PA = $0^\circ$. The $\mu_{100}$ extracted from here is $23.0\pm0.9\%$.}
  \label{fig:corr_sim}
\end{figure}

While the data taken for this campaign can be corrected directly using other measured data to extract the value of $\mu_{100}$, this will not be the case for science data. The scattering angle distributions measured from, for example a GRB, will be the result of a complex spectral shape which, for practical reasons, cannot be reproduced in the lab for each measured GRB. The correction using the unpolarized scattering angle distribution therefore relies on the production of such a curve using simulations. Inaccurate simulations will firstly not remove all the instrumental effects from the modulation curve, therefore not producing a $180^\circ$ modulation. Secondly the amplitude of this modulation can be wrong, therefore producing an inaccurate measurement of the polarization of the GRB. We therefore additionally tested the correction of the measured data using an unpolarized scattering angle distribution produced using the simulations. The results for $E=60\,\mathrm{keV}$, PD = $100\%$ and PA = $90^\circ$ are shown in figure \ref{fig:corr_sim}. It can be seen that the result accurately reproduces the $\mu_{100}$ with a value of $(23.0\pm{0.9})\%$.

It should finally be noted that although the beam is calibrated to produce a $60\,\mathrm{keV}$ beam with a FWHM of $\approx 0.5 \,\mathrm{keV}$, some contamination from third order harmonics is present in the beam. For the 60~keV beam the relative strength of this $180\,\mathrm{keV}$ line was about $3\%$ and was included in the simulations.

\subsubsection{Modulation for a 40~keV beam}

\begin{figure}[!h]
  \centering
  \includegraphics[width=.7\textwidth]{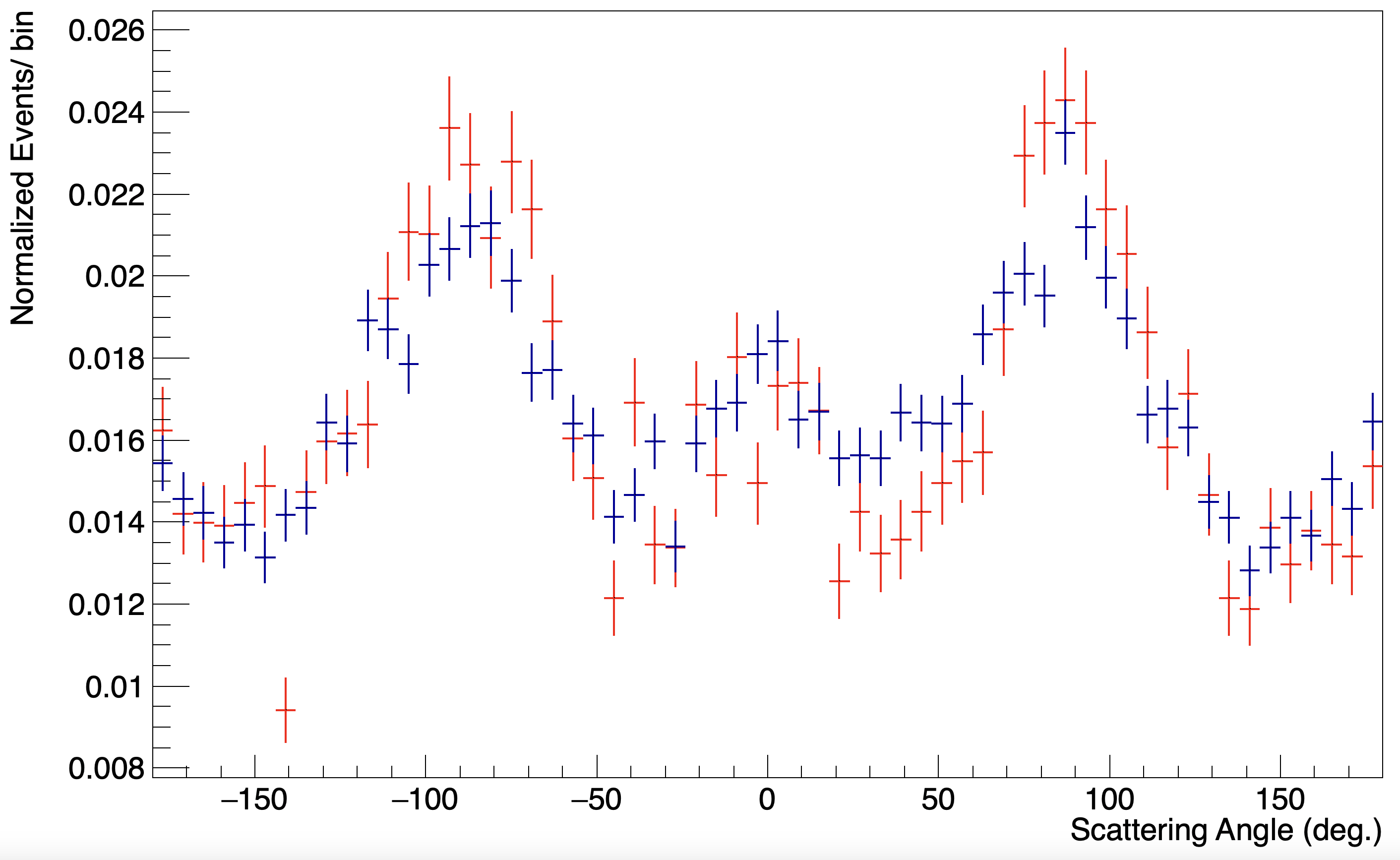}
  \caption{The normalized scattering angle distributions produced using simulations (blue) and measurements (red) for a 40~keV beam with PD = $100\%$ and PA = $0^\circ$ .}
  \label{fig:mod_40_raw_hor}
\end{figure}

\begin{figure}[!h]
  \centering
  \includegraphics[width=.7\textwidth]{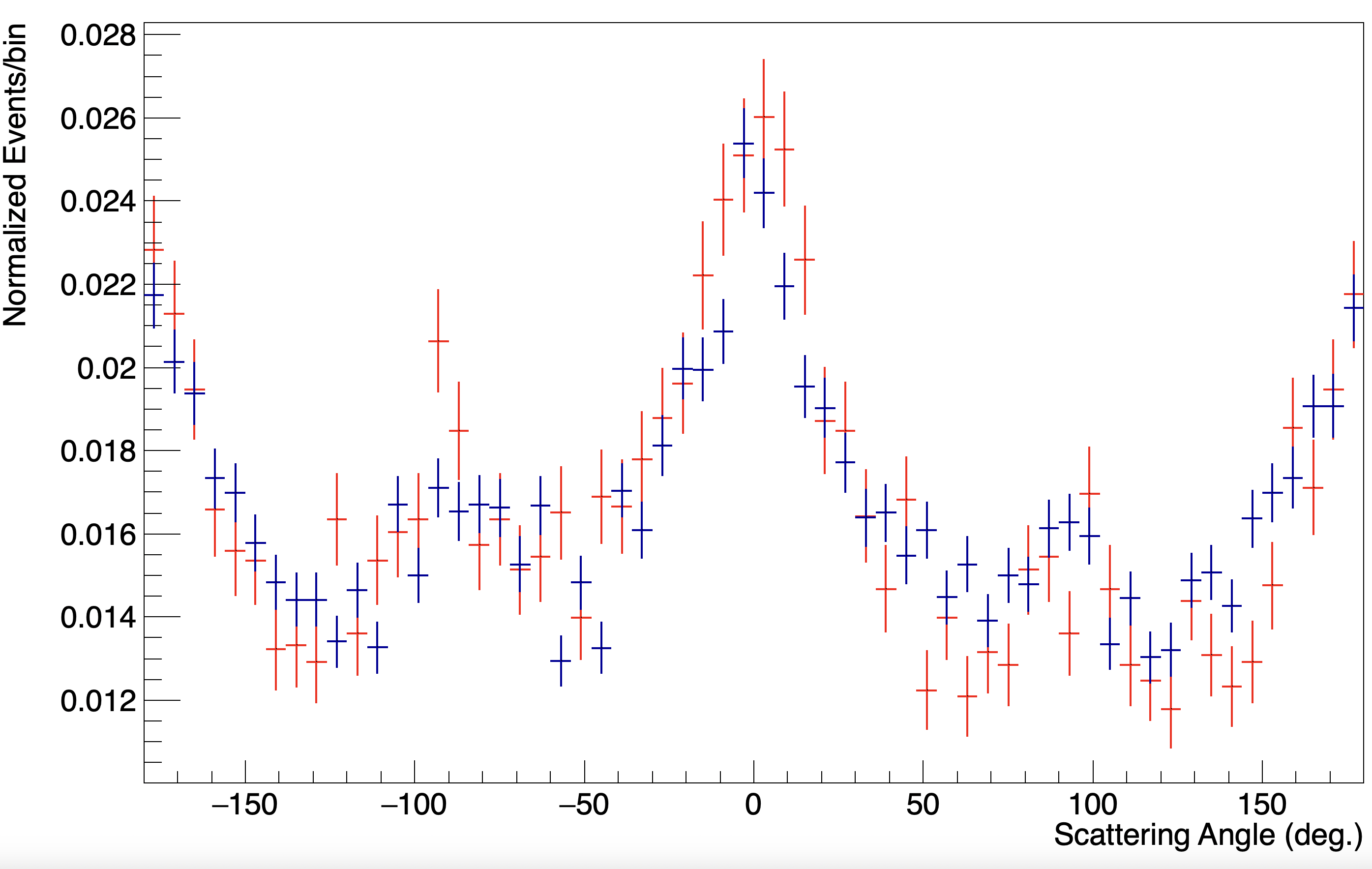}
  \caption{The normalized scattering angle distributions produced using simulations (blue) and measurements (red) for a 40~keV beam with PD = $100\%$ and PA = $90^\circ$ .}
  \label{fig:mod_40_raw_vert}
\end{figure}

\begin{figure}[!h]
  \centering
  \includegraphics[width=.7\textwidth]{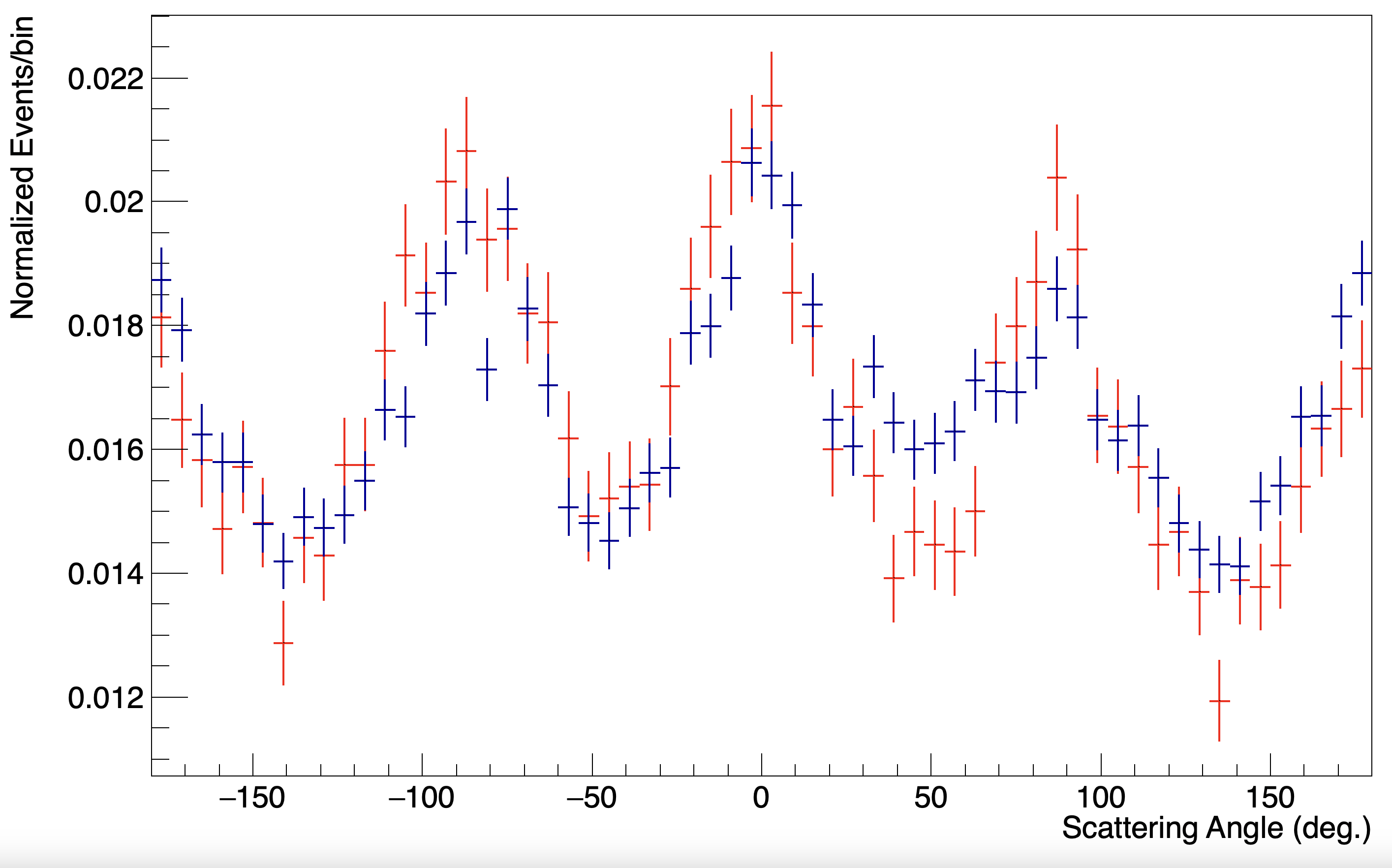}
  \caption{The normalized scattering angle distributions produced using simulations (blue) and measurements (red) for a 40~keV beam with PD = $0\%$.}
  \label{fig:mod_40_raw_upol}
\end{figure}

Figures \ref{fig:mod_40_raw_hor} and \ref{fig:mod_40_raw_vert} show the measured and simulated scattering angle distributions resulting from a 40~keV beam with orthogonal polarization angles. The unpolarized response, produced again by summing the orthogonal polarization results with a weighting factor to account for beam intensities, is shown in figure  \ref{fig:mod_40_raw_upol}. Again the simulated responses can be seen to match the measured ones well. The presence of a third order harmonic was taken into account in the simulations. From the figures we can clearly see that the relative effect of the polarization is significantly smaller than at $60\,\mathrm{keV}$. This is reflected in the $\mu_{100}$ which were calculated to be $(14.7\pm0.9+0.1)\%$ and $(13.9\pm1.1+0.1)\%$ for PA = $0^\circ$ and PA = $90^\circ$ from measurements. From the simulations we find $(15.9\pm1.3)\%$ and $(15.0\pm1.2)\%$ for these values of PA. The results are therefore again in agreement. The results are therefore again in agreement. An overview of all the results is presented in table \ref{tab:overview}.

\begin{table}[h!]
\centering
\begin{tabular}{||c | c c c c||} 
 \hline
 Energy (keV) & $\mu_{100}$ Meas. PA$=0^\circ$ & $\mu_{100}$ Sim. PA$=0^\circ$ & $\mu_{100}$ Meas. PA$=90^\circ$ & $\mu_{100}$ Sim. PA$=90^\circ$ \\ [0.5ex] 
 \hline\hline
 40 & $(14.7\pm0.9+0.1)\%$ & $(15.9\pm1.3)\%$ & $(13.9\pm1.1+0.1)\%$ & $(15.0\pm1.2)\%$ \\ 
 \hline
 60 & $(24.6\pm1.2+0.2)\%$ & $(24.3\pm2.0)\%$ & $(23.9\pm0.9+0.2)\%$ & $(26.0\pm2.4)\%$ \\
  \hline
\end{tabular}
\caption{Overview table of the on-axis results.}
\label{tab:overview}
\end{table}

\subsection{Off-Axis Modulation}

Due to the transient nature of GRBs, POLAR-2 needs to observe a large portion of the sky in order to have a larger chance of detecting GRBs. As the detector response changes for GRBs entering the detector off-axis it is important to also study the scattering angle distribution for such scenarios.

It should be noted here that it is not possible to produce an unpolarized beam for off-axis measurements. This, as changing the PA requires rotating the detector around the beam axis. As for an off-axis measurement the detector is additionally rotated along the x-axis, reproducing an unpolarized beam by summing two orthogonally polarized beams is not possible here. Therefore, to reproduce the $\mu_{100}$ here one relies on the simulations, which, was shown in section \ref{sec:60keV} to be accurate.

Both for 40 and 60~keV, off-axis scans were performed at angles of $15^\circ$ and $30^\circ$ where the detector was rotated along the x-axis. As a result, the scanning distance in the y direction was increased to compensate for the larger area seen by the beam. Channels 56 to 63 were therefore irradiated directly for a longer time, allowing to produce the results presented in section \ref{sec:pos_LY}.

\begin{figure}[!h]
  \centering
  \includegraphics[width=.7\textwidth]{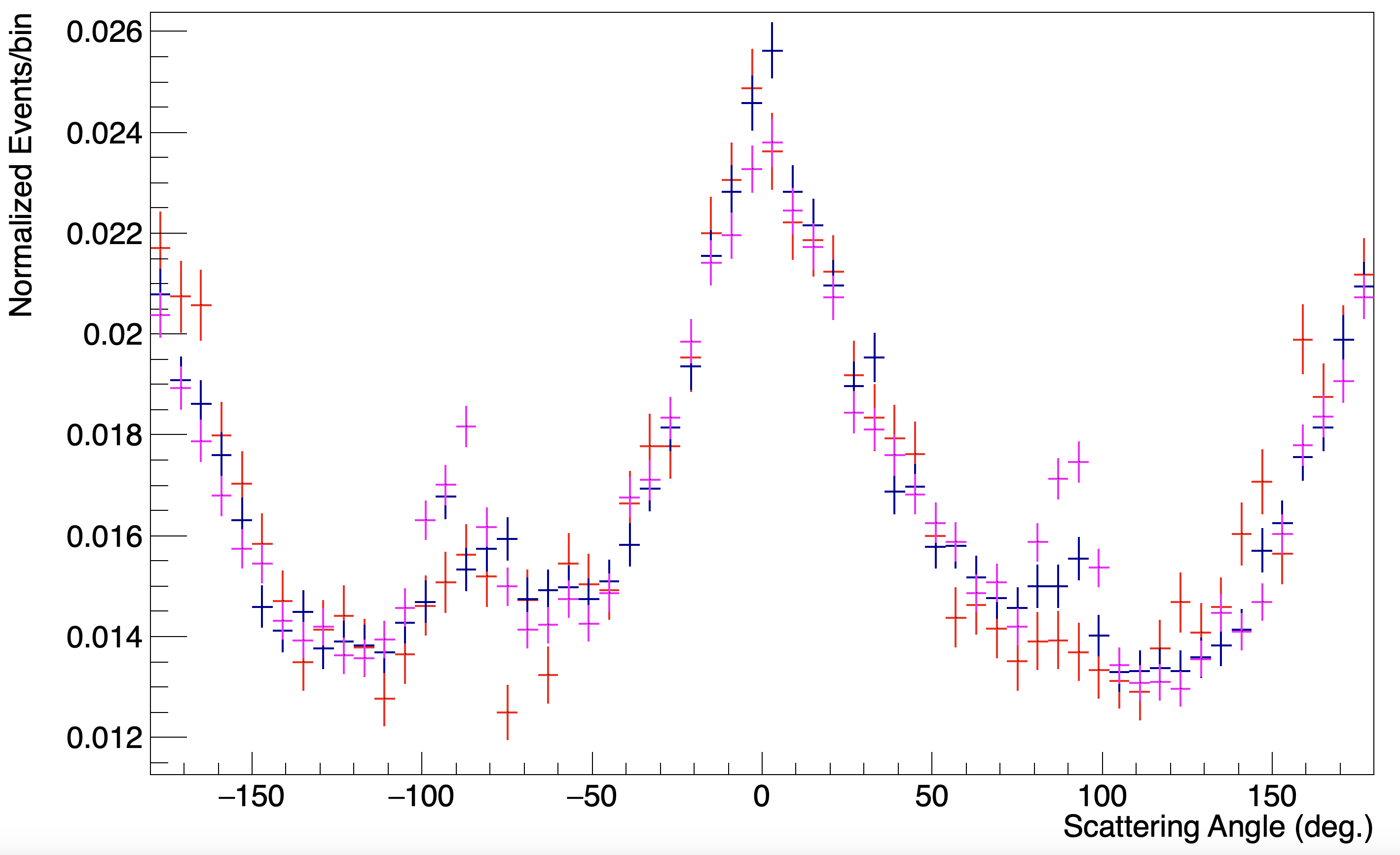}
  \caption{The measured scattering angle distributions produced for a 60~keV beam with PD = $100\%$ and PA = $90^\circ$ using an on-axis beam (blue) and $\theta=15^\circ$ off-axis beam (red) and $\theta=30^\circ$ off-axis beam (purple).}
  \label{fig:off-axis_meas}
\end{figure}

\begin{figure}[!h]
  \centering
  \includegraphics[width=.7\textwidth]{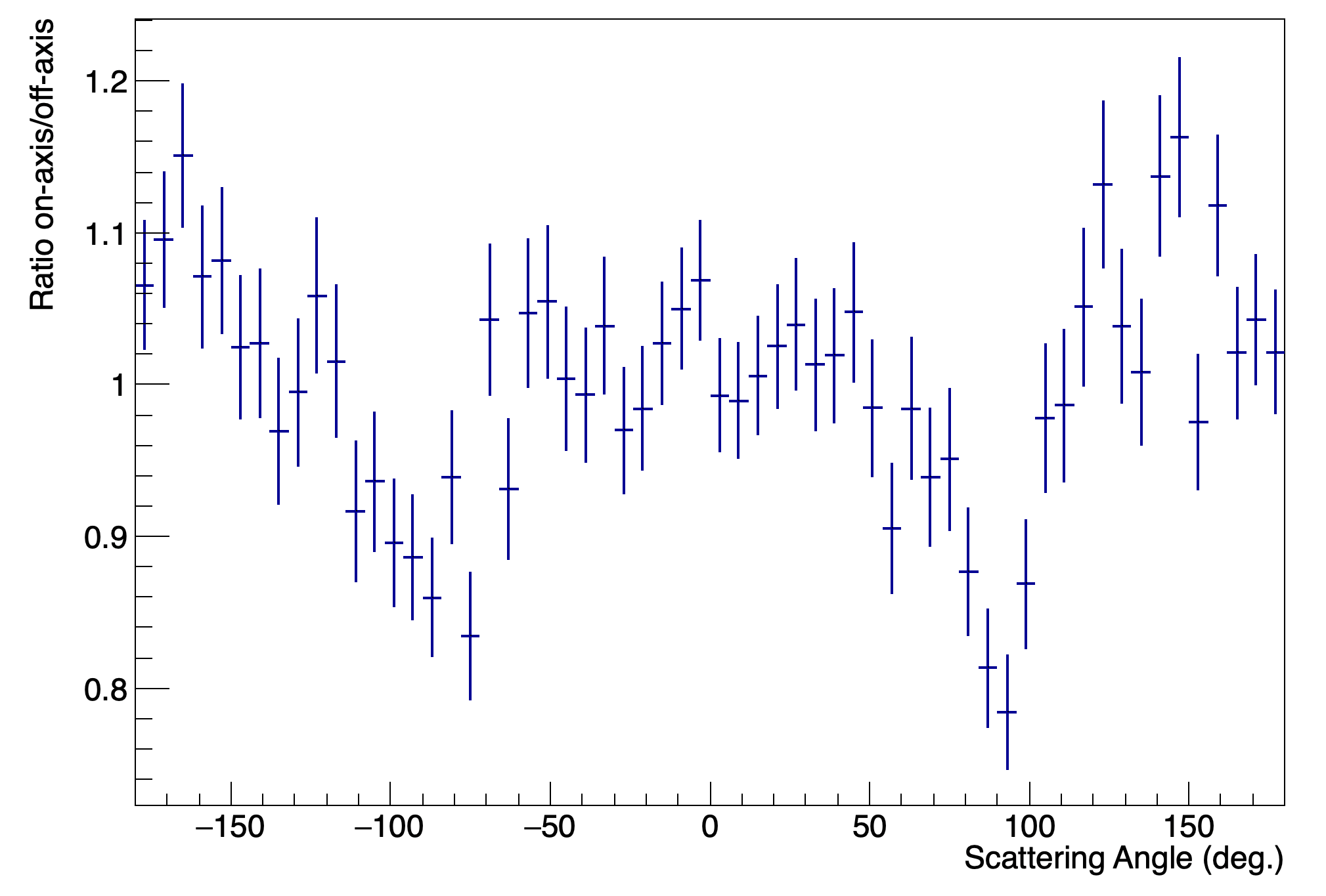}
  \caption{The ratio of the measured scattering angle distributions for on-axis and $30^\circ$ off-axis beam with an energy of 60~keV. A clear increase at both $90^\circ$ (for example from channel 0 to 56) and $-90^\circ$ (for example from channel 56 to 0) can be observed when moving off-axis.}
  \label{fig:ratio_60}
\end{figure}

\begin{figure}[!h]
  \centering
  \includegraphics[width=.7\textwidth]{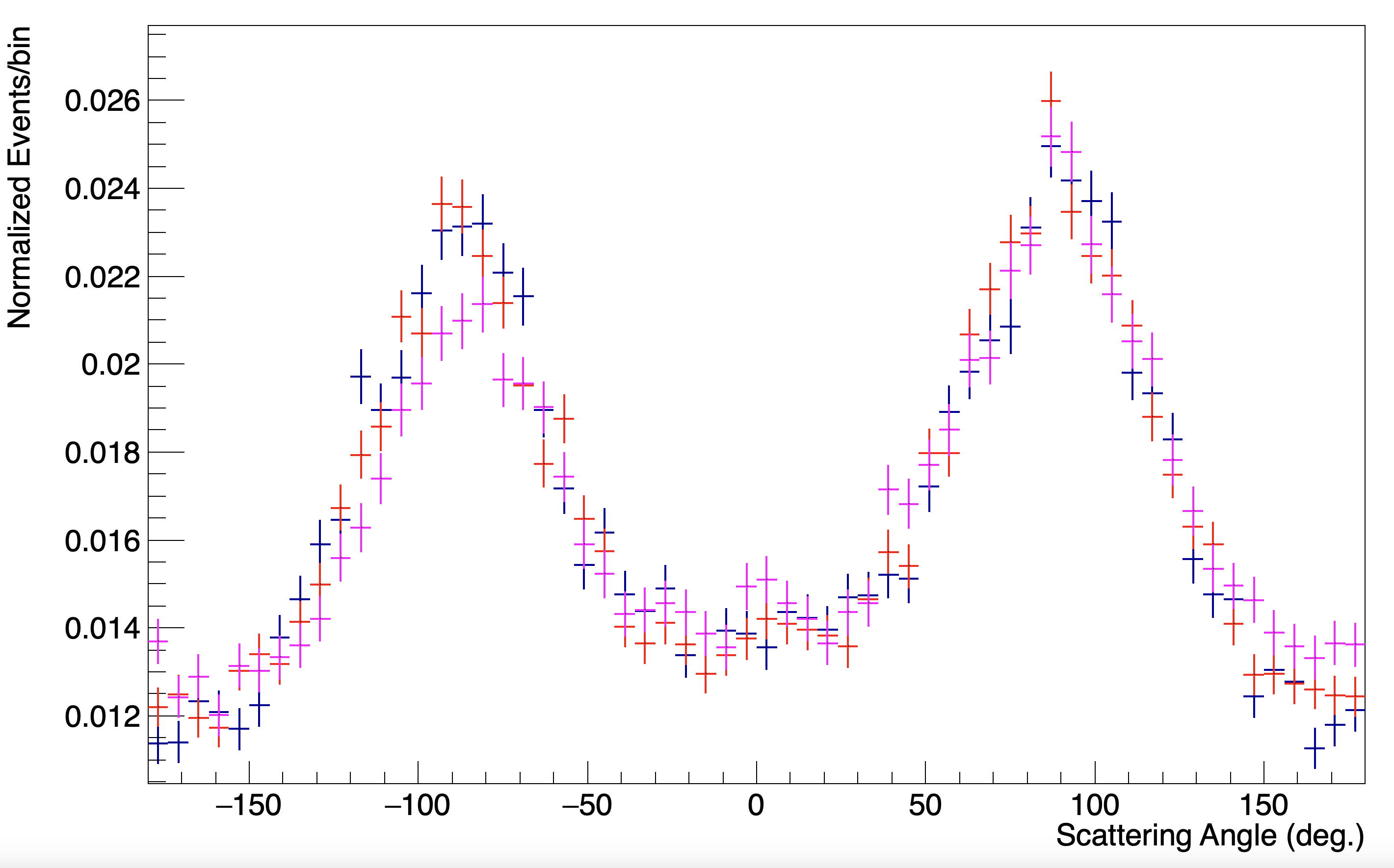}
  \caption{The simulated scattering angle distributions produced for a 60~keV beam with PD = $100\%$ and PA = $0^\circ$ using an on-axis beam (blue), $15^\circ$ off-axis beam (red) and a $30^\circ$ off-axis beam (purple).}
  \label{fig:off-axis_sim}
\end{figure}

For the POLAR detector, both in calibration data and from measurements in-orbit and simulations, see for example \cite{Kole_Sun}, it was found that the modulation curve changed significantly when changing the incoming angle of the photons. Here we only observe minor differences which do not result in a significant change in $\mu_{100}$. The scattering angle distributions as measured using a 60~keV beam for on-axis, $15^\circ$ and $30^\circ$ off-axis are shown together from measurements in figure \ref{fig:off-axis_meas}. When studying the ratio between the on-axis observation and the $30^\circ$ off-axis case the differences can be seen more clearly, see figure \ref{fig:ratio_60}. It can be observed that when moving off-axis, scattering at angles of $0^\circ$ and $180^\circ$ becomes less likely. In addition, scattering at $-90^\circ$, meaning for example from bar 56 in the direction of 0, becomes more likely as can be expected with this rotation. Using simulations we can simulate a system where the detector is instead rotated around the y-axis and with a perpendicular PA. This is shown in figure \ref{fig:off-axis_sim}, the same effects can be seen here.

\begin{figure}[!h]
  \centering
  \includegraphics[width=.7\textwidth]{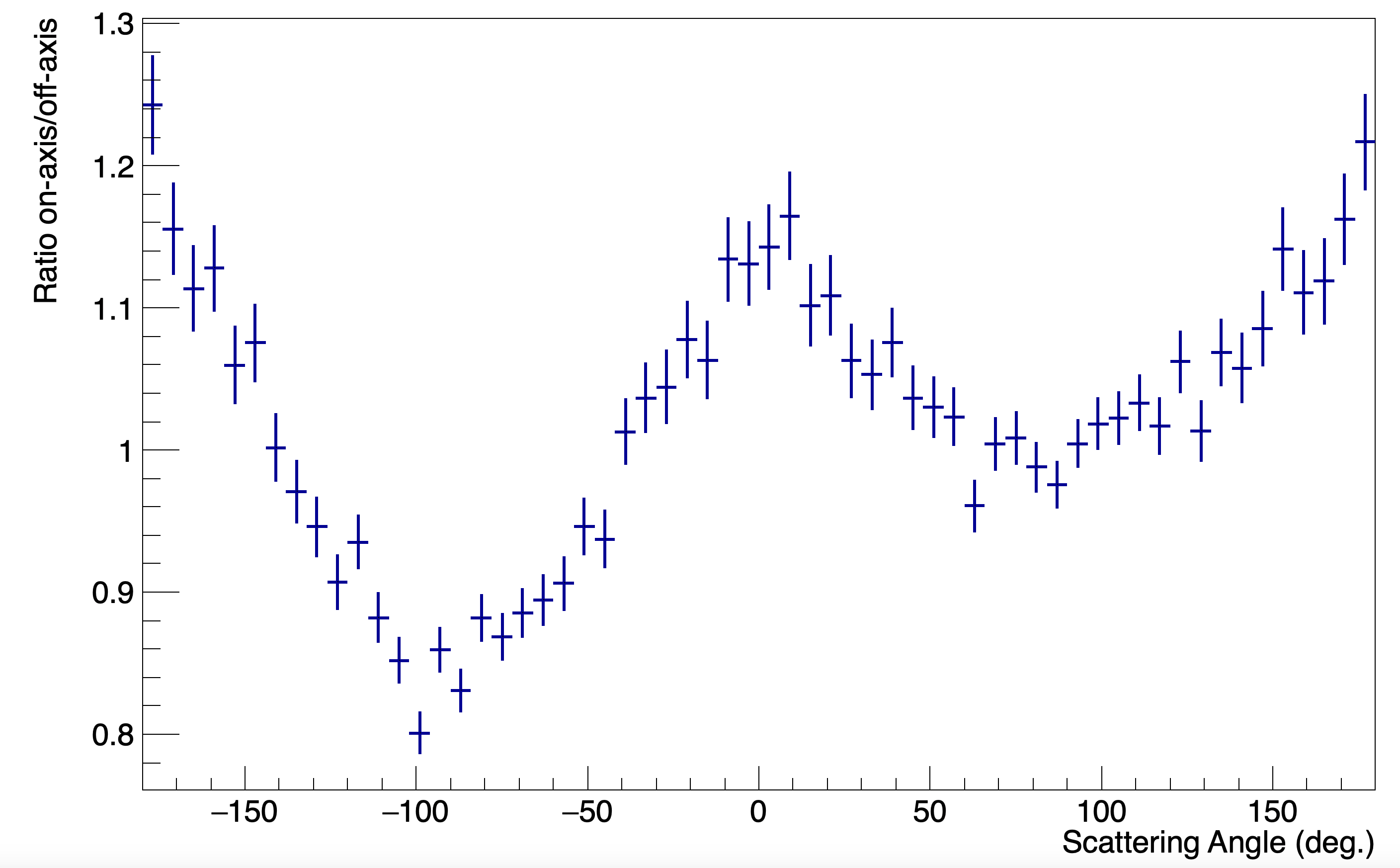}
  \caption{The ratio of the simulated scattering angle distributions measured for on-axis and off-axis ($30^\circ$) for a 120~keV beam.}
  \label{fig:120_ratio}
\end{figure}

For 40~keV irradiation the difference was found to be even less significant and as a result the values of $\mu_{100}$ did not change significantly. Only when increasing the energy of the incoming photons does the off-axis effect become significant, this was tested with simulations of a $120\,\mathrm{keV}$ beam for 0 and $30^\circ$ off-axis. The ratio of the modulation curves for both is shown in figure \ref{fig:120_ratio}, where the same trend as in figure \ref{fig:ratio_60} can be seen but significantly more pronounced. 

The small difference between the off-axis scattering angle distributions at low energies can be explained by the small energies deposited for photons scattering at small polar angles. For an on-axis beam photons scattering at small polar angles will have a large probability of interacting a second time in the same scintillator, thereby not allowing to measure their scattering angle and thus not contributing to the scattering angle distribution. When moving off-axis such photons will likely scatter to a different scintillator, and thereby contribute to the scattering angle distribution. It is such photons which would result in a difference in the scattering angle distribution. However, for low incoming photon energies, such at at 40 keV, the energy depositions for such interactions are, mostly, below the trigger threshold so they remain undetected. Therefore, the scattering angle distributions for on-axis and off-axis remain largely similar at 40 keV. When increasing the energy to 120 keV, small polar scattering angles are instead detected and therefore contribute to the off-axis scattering angle distribution while, for the on-axis case the large probability to interact in the same scintillator still does not allow for the calculation of a scattering angle. As a result, the scattering angle distributions at larger energies do start to differ more significantly.

\section{Required Improvements}\label{sec:improvements}

As discussed before, the DAQ used in the detector module during this irradiation campaign is not yet finalized and was found to not be optimal yet. Three main issues with the DAQ were identified:

\begin{enumerate}
\item Peak sensing was only enabled for detector channels which issues a charge trigger.
\item Too high data rates caused timing issues in the data readout.
\item Inaccurate measurement of the dead time.
\end{enumerate}

The last of these was due to issues in the firmware where the dead time of the electronics for each trigger was not accurately measured. This means that, with the high data rates experienced at ESRF, no correction for variations in the dead time of the detector could be applied to the analysis. Although the changes in the beam intensity were included in the analysis, variations in the dead time had to be ignored. Such variations could occur due to small differences in the threshold of the various channels. Due to such issues additional systematic effects could appear in the modulation curves. However, as no significant deviations from the simulated curves were seen this issue is likely negligible. For future measurements this issue should be resolved to not only confirm that no systematic errors are induced in the modulation curve, but also to accurately measure the fluences of GRBs. In future beam tests of the detector, the flux of the incoming beam should be measured accurately to verify an accurate calculation of the dead time. This will, as could not be done here, allow to calibrate the effective area for polarization of the detector as well.

\begin{figure}[!h]
  \centering
  \includegraphics[width=.7\textwidth]{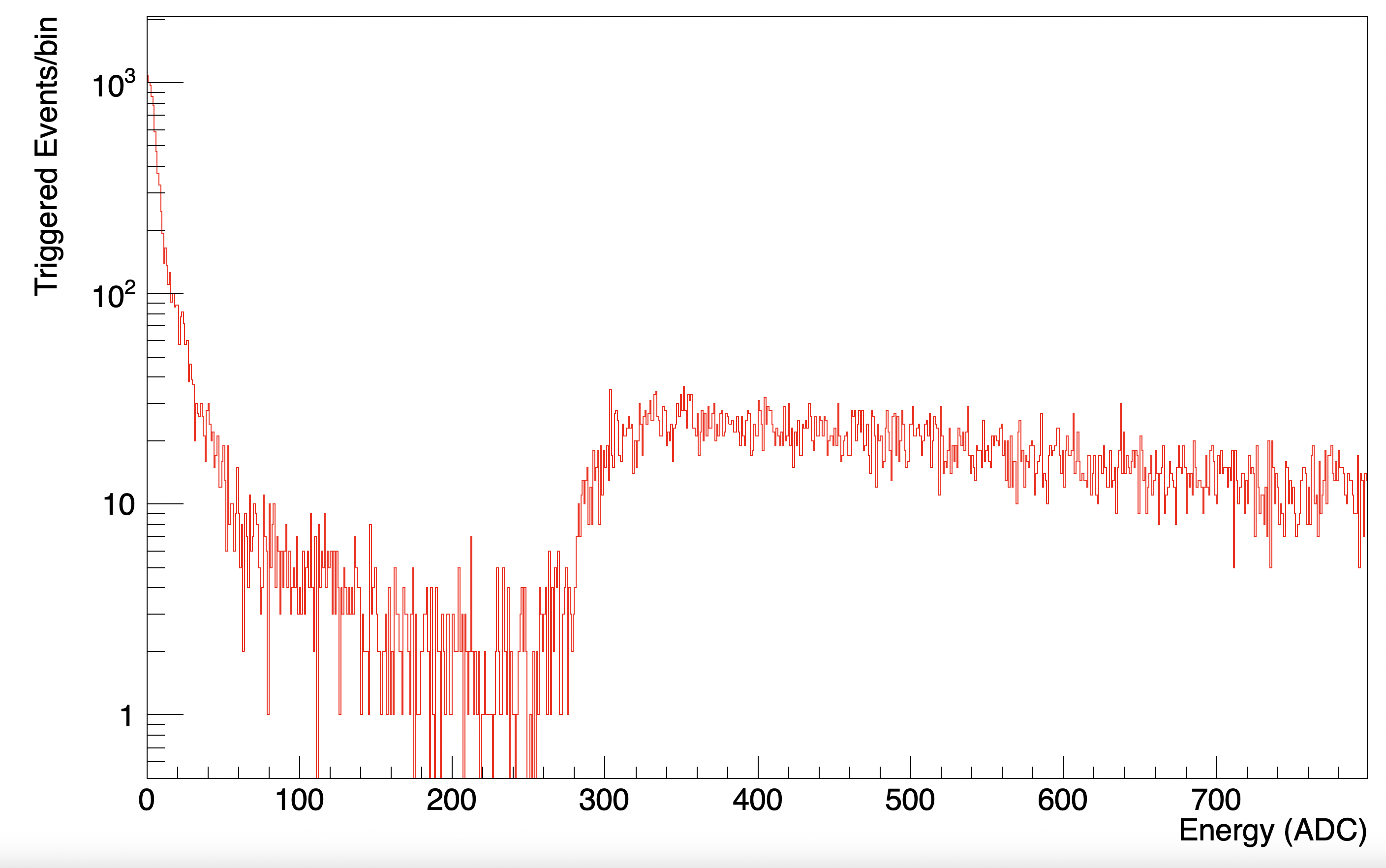}
  \caption{A spectrum as measured by one channel during a scan with a $120\,\mathrm{keV}$ beam. Only events which issued a charge trigger are shown, it can be observed that a significant amount of events can be seen far below the trigger threshold which is around 280 ADC.}
  \label{fig:bad_spec}
\end{figure}

The second of these issues was apparent during data taking with beams of 100~keV and 120~keV. The issue can be seen in figure \ref{fig:bad_spec}. Although this histogram only shows events where a charge trigger was issued, many events can be seen below the trigger threshold. This is, potentially, a result of data packages being corrupted during the high data taking rate ($\approx 1800$ triggers / second) experienced with the beam. During the 60~keV irradiation this did not cause significant issues, as the corrupted packets make up only a very minor fraction of the data as can be seen in figure \ref{fig:corr_chan20_high}. Although the spectra taken at the higher energies looked good upon initial inspection, a more detailed study of this data found that a significant amount of the data was corrupted leading to a significant amount of events which appear to trigger below the threshold. This data was therefore not analyzed in greater detail. 

\begin{figure}[!h]
  \centering
  \includegraphics[width=.7\textwidth]{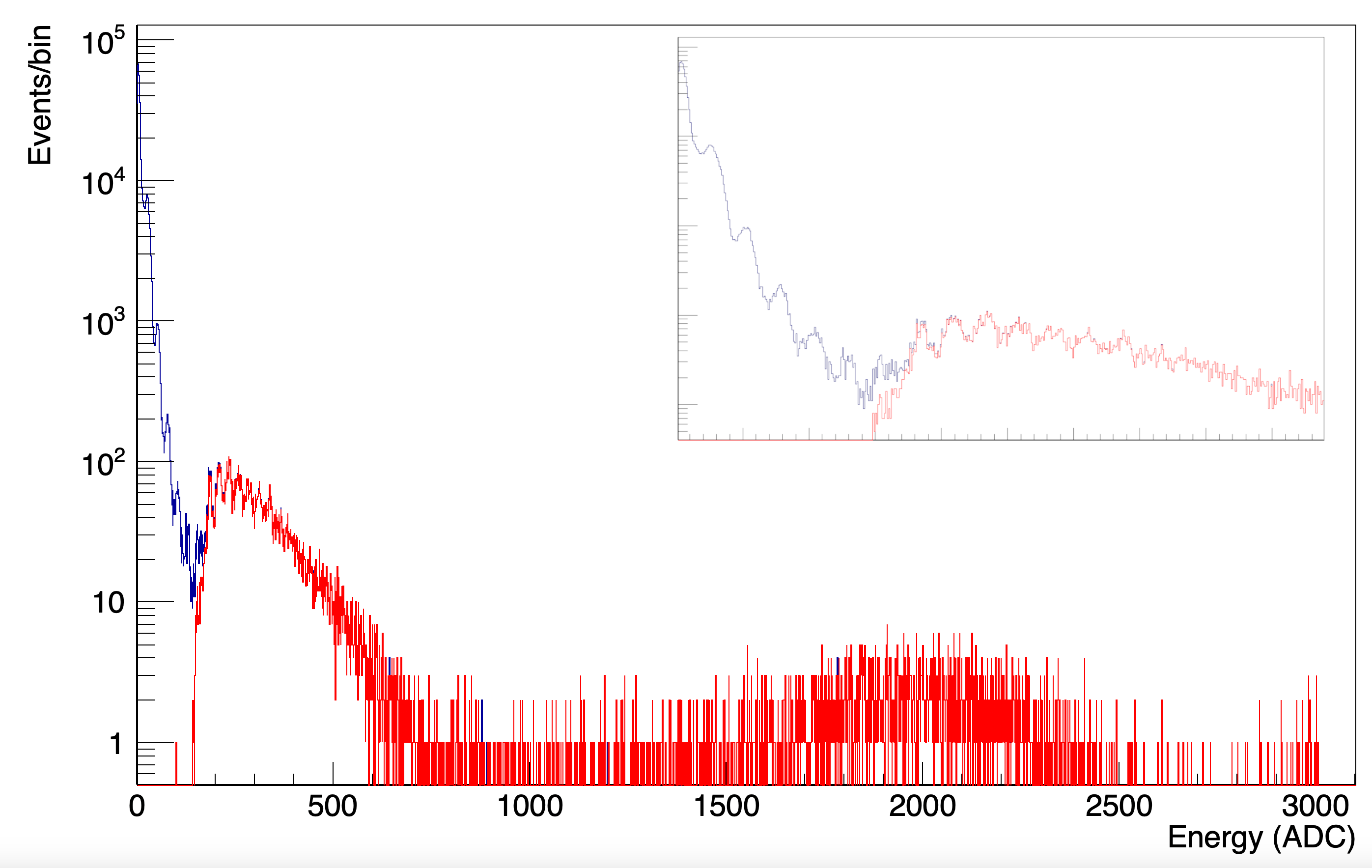}
  \caption{A spectrum from one detector channel taken with a polarimeter module using an $^{241}Am$ source after firmware updates which mitigate the issues found at ESRF. The inset shows more details of the 0 to 500 ADC range. The triggered events are shown in red while the non triggered are shown in blue. A clear separation is seen here. In addition, clear photo-electron peaks can be seen below the threshold, thereby indicating that the peak sensing is performed properly here as well.}
  \label{fig:new_spec}
\end{figure}

The issues found at high trigger rates have now been mitigated through several improvements in the front-end firmware which allow for a faster data through put. This can be seen in figure \ref{fig:new_spec} where data was taken with $\approx 2000$ triggers per second using an $^{241}Am$ source which produces primarily 59.5 keV photons. The spectrum shows a clear photo--peak, but more importantly shows no significant amount of triggers below the threshold.

The final issue in the data from ESRF was that peak sensing was not enabled in the ASIC for channels which did not issue a charge trigger. As a result, the ADC values for such channels were typically noise and were not correlated to the real energy deposition. As such, they could not be used in the polarization analysis. This caused several secondary issues. The first is that during the event selection for polarization, as discussed in section \ref{sec:60keV}, channels without a charge trigger were not taken into account. This thereby directly resulted in a lower energy threshold of $\approx$ 8 keV for both the channels selected in the polarization analysis. As a 60~keV photon has a large probability of depositing energies below 8~keV when Compton scattering all such events are lost here. 

The second result of the lack of peak sensing is related to the measurement of the trigger thresholds. The implementation of the threshold in the simulations, as discussed in section \ref{sec:simulations}, requires both the threshold mean and its width, which is a result of noise in the threshold DAC. Although determining the mean position of the threshold is possible, the width of this threshold is distorted in the final data because the ADC value assigned to all energy depositions which do not issue a trigger are wrong. As a result the width used in the simulations had to be fine tuned such that the results matched the real data, rather than using the measured threshold width.

Since the beam campaign the firmware of the polarimeter module has been updated such that peak sensing is enabled for all channels as soon as one channel exceeds the charge threshold (or when the time threshold trigger is issued). This is achieved by forcing the ASIC to start peak sensing on all channels using an external line from the FPGA. The decision from the FPGA is taken within nanoseconds after the trigger is issued by the ASIC. This does not significantly affect the spectra. The final result of this fix can be seen again in figure \ref{fig:new_spec} where clear photo-electron peaks, or fingers, can be seen also below the threshold. 

Finally, as was discussed in section \ref{sec:LY}, the scintillator material used in the polarimeter module which was used for the polarization studies in these tests is not yet the optimum. Switching this to EJ-248M, as was illustrated in figure \ref{fig:EJ248M}, will increase the light yield and thereby allow to decrease the threshold by approximately $10\%$. This material will therefore be used in the final version of the POLAR-2 polarimeter module. 

\section{Extrapolated POLAR-2 Performance}\label{sec:extrapolated_performances}

After verifying the performance of the POLAR-2 MC simulations framework using measured data, we can use this to simulate the full 100-modules POLAR-2 detector to study its expected performance. Although the simulation framework used for this is overall equal to that used to simulate the detector response for the irradiation campaign, a few modifications were made:

\begin{itemize}
    \item The number of detector modules was increased to 100 and additional detector mechanics were added to simulate the full POLAR-2 detector as it is shown in figure \ref{fig:POLAR-2}. It should be noted that the majority of the components, such as the aluminium structure below the polarimeter modules, does not affect the results. Rather, such structures are only important for future simulations of the background.
    \item The time trigger logic is added. This implies that in case two channels within a polarimeter module are above the time trigger the full module is read out. The time trigger threshold was set at 4 keV. The charge threshold was left at approximately 8 keV (equal to the values used at ESRF). It is likely, that this threshold can be lowered in the future, for example, thanks to the use of EJ-248M as the scintillator material, as well as the lower operating temperature foreseen to be used in orbit. This has to be confirmed through detailed tests. It should be noted that this is significantly lower than typical threshold of spectrometers which use SiPMs. This is possible thanks to the requirement that two detector channels trigger within a 100 ns window, thereby significantly reducing the dark noise induced triggers.
    \item Peak sensing is now enabled for all channels in a polarimeter module when either a valid time or charge trigger is issued. As a result we now see photo-electron peaks also below the trigger threshold as in figure \ref{fig:new_spec}.
\end{itemize}

\subsection{Effective Area}

The effective area of POLAR-2 is simulated by irradiating the full detector using mono-energetic photons emitted from a plane with dimensions of $800\times800\,\mathrm{mm^2}$ placed above the detector surface. The simulations were performed using mono-energetic beams with energies ranging from 10 to 800 keV. The output from the simulations is processed in the data analysis pipeline identical to that used for the single polarimeter module setup. The total effective area of the detector is then simply defined as the number of valid triggers in the detector divided by the simulated photon flux which for these simulations corresponds to $781\,\mathrm{cm^{-2}}$ for simulations below 100~keV and $312\,\mathrm{cm^{-2}}$ above this. These simulations were repeated for photons with an incoming polar angle of $30^\circ$, while keeping the $\phi$ angle at $0^\circ$, to understand the angular dependence.

\begin{figure}[!h]
  \centering
  \includegraphics[width=.95\textwidth]{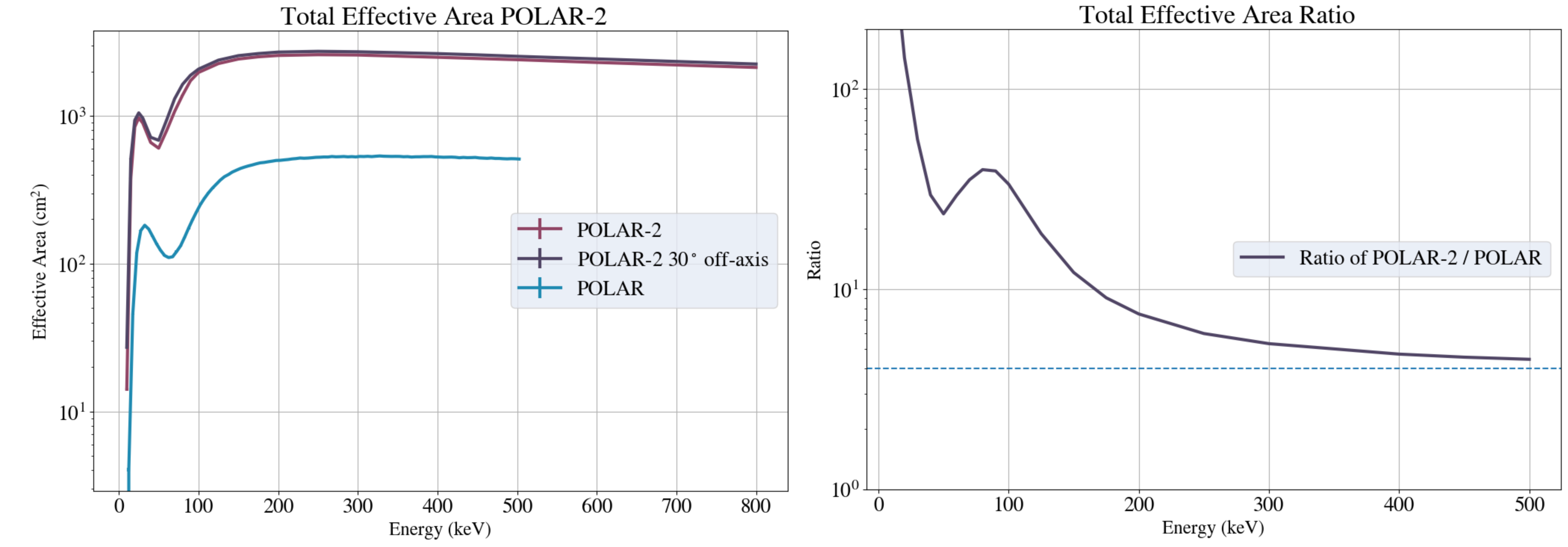}
  \caption{Left: The effective area for all events as a function of energy from simulations. The effective area for an on-axis flux for POLAR-2 is shown along with that from POLAR. In addition the effective area for an off-axis flux with an incoming angle of $\theta=30^\circ,\phi=0^\circ$. Right: the ratio of the total effective area of POLAR-2 over that of POLAR. The ration which can be directly attributed to the increase in geometrical area is indicated with the blue dotted line.}
  \label{fig:total_eff}
\end{figure}

The results are shown in figure \ref{fig:total_eff} where the effective areas are compared to that of POLAR. We can see that while above 200 keV the effective area of POLAR-2 is approximately 4 times that of POLAR, matching the increase in geometrical area, at lower energies the increase is more significant. This is thanks mainly to the higher light yield which allows for a lower detection threshold. 

The effect of the lower detection threshold ($\approx 8\,\mathrm{keV}$ for POLAR-2 compared to $\approx 15\,\mathrm{keV}$ for POLAR) also affects the dip in the effective area which can be seen for both detectors below 100~keV. This local minimum is a result of the increasing probability for photons to undergo Compton scattering with increasing energy. As photons of $\approx 20-30\,\mathrm{keV}$ still have a significant probability to undergo photo-absorption in the plastic, the energy they deposit is typically above the low energy threshold. With increasing energies the probability to deposit all the energy through photo-absorption decreases, resulting in a significant fraction of the photons to Compton scatter in one scintillator, before leaving the detector. The maximum energy deposition a 50 keV photon can deposit in the case where it backscatters out of the detector, is 8 keV. As a result, POLAR-2 will be able to measure such interactions, however, for photons with 40~keV, which can deposit only 5.5 keV, it is less likely\footnote{Due to Poisson fluctuations in the number of photo-electrons produced by a 5.5 keV deposition, as well as noise, such interactions can still trigger the detector.}. For POLAR-2 a local minimum is therefore visible around 45 keV. For POLAR, which has a higher detection threshold this minimum is at a higher energies and is also more pronounced.

\begin{figure}[!h]
  \centering
  \includegraphics[width=.95\textwidth]{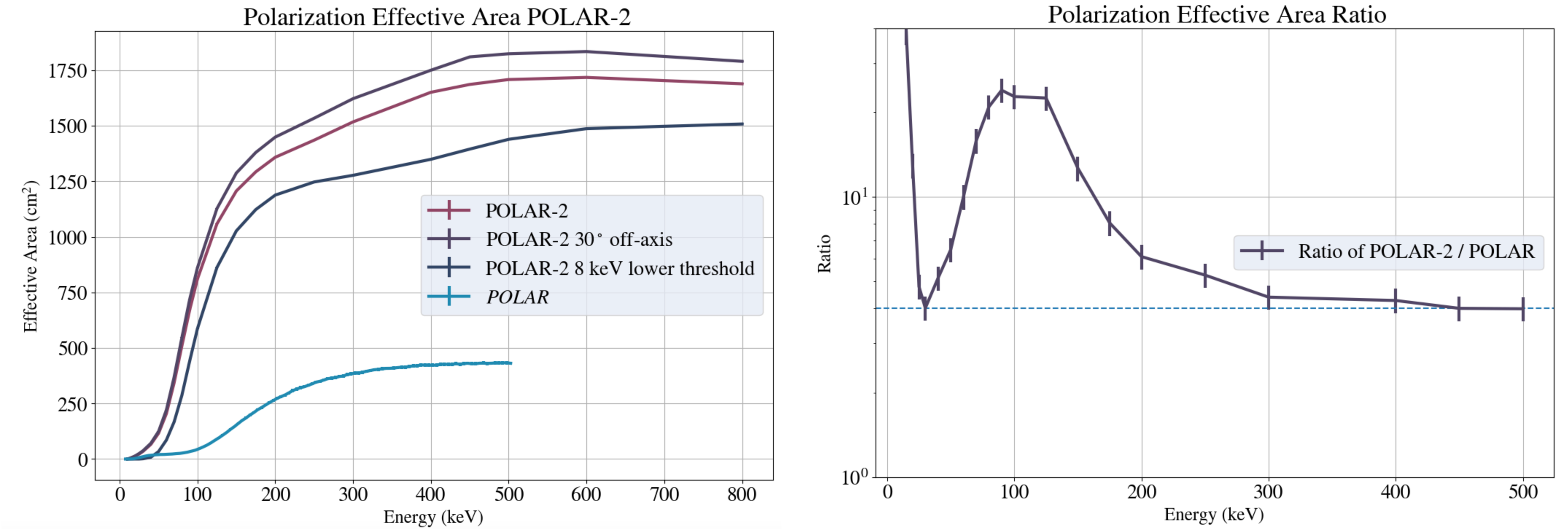}
  \caption{Left: The effective area for polarization events as a function of energy from simulations. The effective area for an on-axis flux for POLAR-2 is shown along with that from POLAR. In addition the effective area for an off-axis flux with an incoming angle of $\theta=30^\circ,\phi=0^\circ$. Finally, the effective area is shown for an on-axis flux where an additional lower energy threshold of 8 keV is set. Right: The ratio of the effective area of POLAR-2 over that of POLAR. The ratio which can be directly attributed to the increase in geometrical area is indicated with the blue dotted line.} 
  \label{fig:pol_eff}
\end{figure}

While figure \ref{fig:total_eff} shows the effective area which can be used for measuring the GRB flux, position and spectrum, figure \ref{fig:pol_eff} shows the effective area for all triggers which, using the simple event selection discussed prior, are used for polarization measurements. To produce this the total number of entries in the scattering angle distribution is used rather than the total number of triggers in the detector. As a result the effective area is lower. Again we can see that at higher energies the ratio between the two matches the increase in the geometrical area of the detector while below 200 keV the increase is much more significant.  

Also here, the effective area for a flux with an incoming angle polar of $30^\circ$ and $\phi=0$ is presented which, as expected, is slightly higher than that for the on-axis case. Finally, here we also show the effective area for polarization events when we impose an additional, offline, threshold for the two energy depositions of 8 keV. We can see that this significantly decreases the effective area for polarization events, specifically at low incoming photons energies. Such a cut, however, will increase the $\mu_{100}$ as will be discussed next.

\subsection{Modulation Factor}

\begin{figure}[!h]
  \centering
  \includegraphics[width=.7\textwidth]{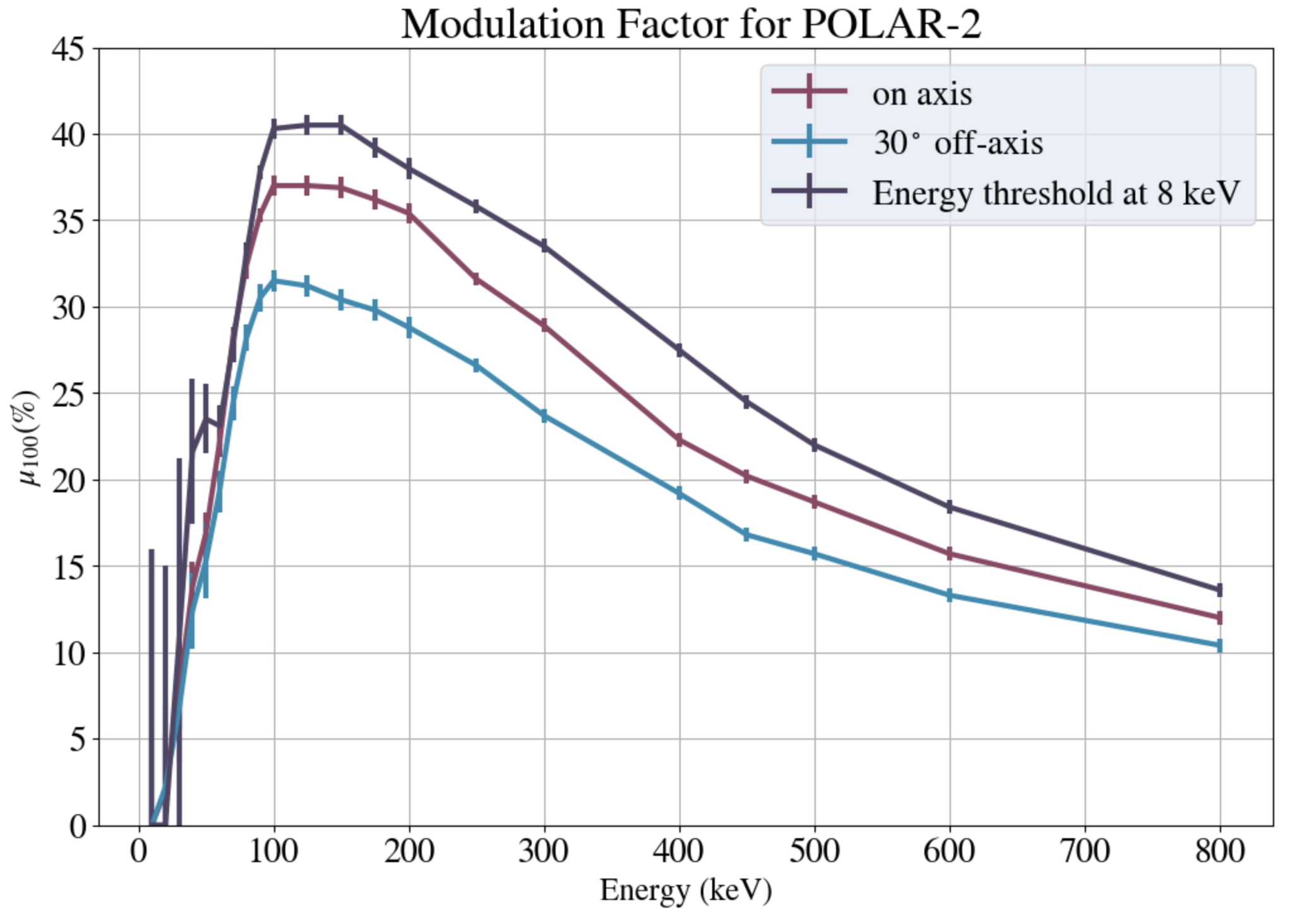}
  \caption{The modulation factor as a function of energy from simulations. The $\mu_{100}$ for an on-axis flux for POLAR-2 is shown along with that for an off-axis flux with an incoming angle of $\theta=30^\circ,\phi=0^\circ$. Finally the $\mu_{100}$ is shown for an on-axis flux where an additional lower energy threshold of 8 keV is set.}  \label{fig:mod_pol}
\end{figure}

The simulations used to produce the effective area predictions were also used to simulate the $\mu_{100}$ for POLAR-2. For this purpose the scattering angle distributions were produced for a polarized flux, corrected using the unpolarized distribution and finally fitted using a harmonic function with a period of $180^\circ$ to extract the relative amplitude of the modulation. The results of this are shown for the on-axis case and a $30^\circ$ off-axis case in figure \ref{fig:mod_pol}. As was also shown with the data from the ESRF tests in section \ref{sec:esrf_campaign}, the difference in the $\mu_{100}$ between the on-axis and off-axis case is minimal at lower energies. Only when exceeding $\approx 80\,\mathrm{keV}$ do we really start to see a significant difference. 

In addition, the $\mu_{100}$ was also calculated when applying the additional offline low energy threshold on the data of 8 keV. We can see here that this significantly increases the $\mu_{100}$ at higher energies with a peak at $41\%$ compared to $37\%$ without this cut. However, as was shown in figure \ref{fig:pol_eff}, such a cut does result in a significant decrease in the effective area. A commonly used figure of merit for polarization measurements is the Minimal Detectable Polarization (MDP) \cite{Weisskopf}:
\begin{equation}
    \mathrm{MDP} = \frac{2\sqrt{\mathrm{-ln}(1-{\rm C.L.})}}{\mu_{100}C_{\rm s}}\sqrt{C_{\rm s}+C_{\rm b}}\,.
\end{equation}

\noindent where $\mathrm{C.L.}$ is the confidence level, $C_{\rm s}$ is the number of signal events and $C_{\rm b}$ the number of background events. This figure of merit gives the minimal level of polarization of the source one can distinguish from being unpolarized using the measurement with a confidence level C.L. It should be noted that, as discussed in detail in \cite{Kole_Sun}, the MDP is not a fully accurate figure of merit to use for wide field of view polarimeters like POLAR-2. Although it does provide a good approximation on how to optimize things, it should not be over interpreted.

Although the MDP scales linearly with $\mu_{100}$ and only through $\sqrt{C_{\rm s}}$, and therefore the effective area, fully optimizing the ratio between the $\mu_{100}$ and the effective area requires assumptions on the spectrum of the incoming flux. In addition, cuts like the 8 keV lower threshold applied here, will also affect the background rate $C_{\rm b}$, however, the dependency here can only be studied accurately using real data. 

The application of the additional lower energy cut therefore only shows that an optimization using such cuts, as well as for example cuts on the distance between the two energy interactions in the detector, should be studied. However, such optimization studies require also a good understanding of the background, and therefore will be the subject of dedicated studies in the future.

Finally, it is important to address concerns regarding the increase in dark noise in the SiPMs as a result of radiation damage during operation in space. Such an incrase in dark noise would require an increase in the thresholds to limit the number of noise induced triggers. For details on this the reader is referred to the following studies performed by the POLAR-2 team \cite{Slawek, DeAngelis:2022jgd}. These studies indicate that, through the use of annealing the required increase in threshold would amount to approximately $0.75\,\mathrm{keV}$ after 2 years.

\newpage
\section{Conclusions}\label{sec:summary}

The first POLAR-2 polarimeter module was irradiated using fully polarized beams at the ID15A ESRF beam facility. The performance of the polarimeter module indicates that the detector is significantly more sensitive than the detector modules previously used in the POLAR detector. This is mainly attributable to the increase in the light yield of the detector achieved through the use of SiPMs.

During the analysis phase some shortcomings of the current design were additionally found, all of which have since been solved using upgrades of the firmware.

The performance of the detector module furthermore matches that predicted through the POLAR-2 MC simulation framework, thereby verifying its validity. As a result, this could be used to produce predictions on the performance of the full POLAR-2 detector. Through this it was found that POLAR-2 will  have an effective area for polarization which already exceeds $100\,\mathrm{cm^2}$ at 40~keV, while the data taken at ESRF additionally proved that the instrument is sensitive to polarization at these energies with a $\mu_{100}$ of $\approx 16\%$. For its predecessor POLAR the effective are only started to exceed $100\,\mathrm{cm^2}$ at 120~keV, while it had no significant sensitivity to polarization below $50\,\mathrm{keV}$.

Although the simulation framework already shows that the detector should be sensitive at energies of 30 keV, dedicated tests need to be performed to confirm this. Further improvements to the POLAR-2 design compared to that used here at ESRF, for example through changing the plastic scintillator material, will allow to further decrease the detection threshold. A future test at ESRF will therefore aim to measure its sensitivity at energies of 30~keV and potentially 25~keV.

\acknowledgments

We acknowledge the European Synchrotron Radiation Facility who provided the beam time and facilities which were crucial for this work \cite{ESRF_data}. We would in particular like to thank the team of ID15A who assisted us during the 1 week campaign.

We gratefully acknowledge the Swiss Space Office of the State Secretariat for Education, Research and Innovation (ESA PRODEX Programme) which supported the development and production of the POLAR-2 detector. M.K. and N.D.A. acknowledge the support of the Swiss National Science Foundation. National Centre for Nuclear Research acknowledges support from Polish National Science Center under the grant UMO-2018/30/M/ST9/00757. We gratefully acknowledge the support from the National Natural Science Foundation of China (Grant No. 11961141013, 11503028), the Xie Jialin Foundation of the Institute of High Energy Physics, Chinese Academy of Sciences (Grant No. 2019IHEPZZBS111), the Joint Research Fund in Astronomy under the cooperative agreement between the National Natural Science Foundation of China and the Chinese Academy of Sciences (Grant No. U1631242), the National Basic Research Program (973 Program) of China (Grant No. 2014CB845800), the Strategic Priority Research Program of the Chinese Academy of Sciences (Grant No. XDB23040400), and the Youth Innovation Promotion Association of Chinese Academy of Sciences.


 \bibliographystyle{JHEP}
 \bibliography{biblio.bib}

\providecommand{\href}[2]{#2}\begingroup\raggedright\begin{thebibliography}{10}

\bibitem{Meegan:1992xg}
C.A.~Meegan, G.J.~Fishman, R.B.~Wilson, W.S.~Paciesas, G.N.~Pendleton, J.M.~Horack et~al., \emph{{Spatial distribution of gamma-ray bursts observed by BATSE}}, \href{https://doi.org/10.1038/355143a0}{\emph{Nature} {\bfseries 355} (1992) 143}.

\bibitem{Costa:1997obd}
E.~Costa et~al., \emph{{Discovery of an X-ray afterglow associated with the gamma-ray burst of 28 February 1997}}, \href{https://doi.org/10.1038/42885}{\emph{Nature} {\bfseries 387} (1997) 783} [\href{https://arxiv.org/abs/astro-ph/9706065}{{\ttfamily astro-ph/9706065}}].

\bibitem{vanParadijs:1997wr}
J.~van Paradijs et~al., \emph{{Transient optical emission from the error box of the gamma-ray burst of 28 February 1997}}, \href{https://doi.org/10.1038/386686a0}{\emph{Nature} {\bfseries 386} (1997) 686}.

\bibitem{1997Natur.387..878M}
M.R.~{Metzger}, S.G.~{Djorgovski}, S.R.~{Kulkarni}, C.C.~{Steidel}, K.L.~{Adelberger}, D.A.~{Frail} et~al., \emph{{Spectral constraints on the redshift of the optical counterpart to the {\ensuremath{\gamma}}-ray burst of 8 May 1997}}, \href{https://doi.org/10.1038/43132}{\emph{Nature} {\bfseries 387} (1997) 878}.

\bibitem{Kouveliotou:1993yx}
C.~Kouveliotou, C.A.~Meegan, G.J.~Fishman, N.P.~Bhyat, M.S.~Briggs, T.M.~Koshut et~al., \emph{{Identification of two classes of gamma-ray bursts}}, \href{https://doi.org/10.1086/186969}{\emph{Astrophys. J. Lett.} {\bfseries 413} (1993) L101}.

\bibitem{Galama}
T.J.~Galama et~al., \emph{{Discovery of the peculiar supernova 1998bw in the error box of GRB 980425}}, \href{https://doi.org/10.1038/27150}{\emph{Nature} {\bfseries 395} (1998) 670} [\href{https://arxiv.org/abs/astro-ph/9806175}{{\ttfamily astro-ph/9806175}}].

\bibitem{Patat}
F.~Patat, E.~Cappellaro, J.~Danziger, P.~Mazzali, J.~Sollerman, T.~Augusteijn et~al., \emph{The metamorphosis of sn 1998bw}, \href{https://doi.org/10.1086/321526}{\emph{Astrophysical Journal} {\bfseries 555} (2001) 900}.

\bibitem{Eichler:1989ve}
D.~Eichler, M.~Livio, T.~Piran and D.N.~Schramm, \emph{{Nucleosynthesis, Neutrino Bursts and Gamma-Rays from Coalescing Neutron Stars}}, \href{https://doi.org/10.1038/340126a0}{\emph{Nature} {\bfseries 340} (1989) 126}.

\bibitem{LIGOScientific:2017vwq}
{\scshape LIGO Scientific, Virgo} collaboration, \emph{{GW170817: Observation of Gravitational Waves from a Binary Neutron Star Inspiral}}, \href{https://doi.org/10.1103/PhysRevLett.119.161101}{\emph{Phys. Rev. Lett.} {\bfseries 119} (2017) 161101} [\href{https://arxiv.org/abs/1710.05832}{{\ttfamily 1710.05832}}].

\bibitem{LIGOScientific:2017ync}
{\scshape LIGO Scientific, Virgo, Fermi GBM, INTEGRAL, IceCube, AstroSat Cadmium Zinc Telluride Imager Team, IPN, Insight-Hxmt, ANTARES, Swift, AGILE Team, 1M2H Team, Dark Energy Camera GW-EM, DES, DLT40, GRAWITA, Fermi-LAT, ATCA, ASKAP, Las Cumbres Observatory Group, OzGrav, DWF (Deeper Wider Faster Program), AST3, CAASTRO, VINROUGE, MASTER, J-GEM, GROWTH, JAGWAR, CaltechNRAO, TTU-NRAO, NuSTAR, Pan-STARRS, MAXI Team, TZAC Consortium, KU, Nordic Optical Telescope, ePESSTO, GROND, Texas Tech University, SALT Group, TOROS, BOOTES, MWA, CALET, IKI-GW Follow-up, H.E.S.S., LOFAR, LWA, HAWC, Pierre Auger, ALMA, Euro VLBI Team, Pi of Sky, Chandra Team at McGill University, DFN, ATLAS Telescopes, High Time Resolution Universe Survey, RIMAS, RATIR, SKA South Africa/MeerKAT} collaboration, \emph{{Multi-messenger Observations of a Binary Neutron Star Merger}}, \href{https://doi.org/10.3847/2041-8213/aa91c9}{\emph{Astrophys. J. Lett.} {\bfseries 848} (2017) L12} [\href{https://arxiv.org/abs/1710.05833}{{\ttfamily
  1710.05833}}].

\bibitem{Gill}
R.~Gill, M.~Kole and J.~Granot, \emph{{GRB Polarization: A Unique Probe of GRB Physics}}, \href{https://doi.org/10.3390/galaxies9040082}{\emph{Galaxies} {\bfseries 9} (2021) 82} [\href{https://arxiv.org/abs/2109.03286}{{\ttfamily 2109.03286}}].

\bibitem{Kole_Sun}
M.~Kole and S.~Jianchao, \emph{Gamma-ray polarimetry of transient sources with polar}, \href{https://doi.org/10.1007/978-981-16-4544-0_142-1}{\emph{Chapter in Handbook of X-ray and Gamma-ray Astrophysics} (2022) }.

\bibitem{Produit2018}
N.~Produit, T.~Bao, T.~Batsch, T.~Bernasconi, I.~Britvich, F.~Cadoux et~al., \emph{Design and construction of the polar detector}, \href{https://doi.org/10.1016/j.nima.2017.09.053}{\emph{Nuclear Instruments and Methods in Physics Research Section A: Accelerators, Spectrometers, Detectors and Associated Equipment} {\bfseries 877} (2018) 259}.

\bibitem{Zhang2019}
S.-N.~Zhang, M.~Kole, T.-W.~Bao, T.~Batsch, T.~Bernasconi, F.~Cadoux et~al., \emph{Detailed polarization measurements of the prompt emission of five gamma-ray bursts}, \href{https://doi.org/10.1038/s41550-018-0664-0}{\emph{Nature Astronomy} {\bfseries 3} (2019) 258}.

\bibitem{Kole2020}
{Kole, M.}, {De Angelis, N.}, {Berlato, F.}, {Burgess, J. M.}, {Gauvin, N.}, {Greiner, J.} et~al., \emph{The polar gamma-ray burst polarization catalog}, \href{https://doi.org/10.1051/0004-6361/202037915}{\emph{A\&A} {\bfseries 644} (2020) A124}.

\bibitem{Burgess2019}
{Burgess, J. M.}, {Kole, M.}, {Berlato, F.}, {Greiner, J.}, {Vianello, G.}, {Produit, N.} et~al., \emph{Time-resolved grb polarization with polar and gbm - simultaneous spectral and polarization analysis with synchrotron emission}, \href{https://doi.org/10.1051/0004-6361/201935056}{\emph{A\&A} {\bfseries 627} (2019) A105}.

\bibitem{NDA_2023_ICRC}
N.~De~Angelis et~al., \emph{{Energy-dependent polarization of Gamma-Ray Bursts’ prompt emission with the POLAR and POLAR-2 instruments}}, \href{https://doi.org/10.22323/1.444.0619}{\emph{PoS} {\bfseries ICRC2023} (2023) 619}.

\bibitem{Li_crab}
H.-C.~Li et~al., \emph{{Gamma-ray polarimetry of the Crab pulsar observed by POLAR}}, \href{https://doi.org/10.1093/mnras/stac522}{\emph{Mon. Not. Roy. Astron. Soc.} {\bfseries 512} (2022) 2827} [\href{https://arxiv.org/abs/2202.10877}{{\ttfamily 2202.10877}}].

\bibitem{NDA_thesis}
N.~De~Angelis, \emph{Development of the Next Generation Space-based Compton Polarimeter and Energy Resolved Polarization Analysis of Gamma-Ray Bursts Prompt Emission}, Ph.D. thesis, Département de Physique Nucléaire et Corpusculaire, Université de Genève, December 2023.
\newblock 10.13097/archive-ouverte/unige:173869.

\bibitem{EJ248}
``Eljen ej-248m datasheet.'' Last consulted in August 2023.

\bibitem{EJ200}
``Eljen ej-200 datasheet.'' Last consulted in August 2023.

\bibitem{NDA_opt}
N.D.~Angelis, F.~Cadoux, C.~Husi, M.~Kole and S.~Mianowski, ``Optical design, simulations, and characterization of a polar-2 polarimeter module.'' \textit{In preparation}. 2024.

\bibitem{3M}
``3m datasheet.'' Last accessed 26 February 2024.

\bibitem{Claryl}
``Claryl datasheet by today.'' Last accessed 26 February 2024.

\bibitem{Hamamatsu}
Hamamatsu, ``Sipm datasheet.''

\bibitem{MAPSIL_paper}
O.~Guillaumon, S.~Remaury, P.~Nabarra, P.~Guigue-Joguet and H.~Combes, ``Development of a new silicone adhesive for space use: Mapsil\textregistered qs 1123.''.

\bibitem{MAPSIL_website}
``Map space coating website - mapsil-qs1132 manufacturer.'' Last consulted in August 2023.

\bibitem{BabyMind}
``Private communication with members of the babymind collaboration.'' 2022.

\bibitem{POLAR:2018hqh}
{\scshape POLAR} collaboration, \emph{{In-Orbit Instrument Performance Study and Calibration for POLAR Polarization Measurements}}, \href{https://doi.org/10.1016/j.nima.2018.05.041}{\emph{Nucl. Instrum. Meth. A} {\bfseries 900} (2018) 8} [\href{https://arxiv.org/abs/1805.07605}{{\ttfamily 1805.07605}}].

\bibitem{Hualin}
H.~Xiao et~al., \emph{{A crosstalk and non-uniformity correction method for the space-borne Compton polarimeter POLAR}}, \href{https://doi.org/10.1016/j.astropartphys.2016.06.007}{\emph{Astropart. Phys.} {\bfseries 83} (2016) 6} [\href{https://arxiv.org/abs/1507.04474}{{\ttfamily 1507.04474}}].

\bibitem{G4}
{\scshape GEANT4} collaboration, \emph{{GEANT4--a simulation toolkit}}, \href{https://doi.org/10.1016/S0168-9002(03)01368-8}{\emph{Nucl. Instrum. Meth. A} {\bfseries 506} (2003) 250}.

\bibitem{POLAR:2017iip}
{\scshape POLAR} collaboration, \emph{{Instrument performance and simulation verification of the POLAR detector}}, \href{https://doi.org/10.1016/j.nima.2017.07.070}{\emph{Nucl. Instrum. Meth. A} {\bfseries 872} (2017) 28} [\href{https://arxiv.org/abs/1708.00664}{{\ttfamily 1708.00664}}].

\bibitem{Birks}
J.B.~Birks, \emph{{Scintillations from Organic Crystals: Specific Fluorescence and Relative Response to Different Radiations}}, \href{https://doi.org/10.1088/0370-1298/64/10/303}{\emph{Proc. Phys. Soc. A} {\bfseries 64} (1951) 874}.

\bibitem{Polar_Birks}
X.~Zhang, H.~Xiao, B.~Yu, S.~Orsi, B.~Wu, W.~Hu et~al., \emph{Study of non-linear energy response of polar plastic scintillators to electrons}, \href{https://doi.org/https://doi.org/10.1016/j.nima.2015.06.031}{\emph{Nuclear Instruments and Methods in Physics Research Section A: Accelerators, Spectrometers, Detectors and Associated Equipment} {\bfseries 797} (2015) 94}.

\bibitem{NDA_master}
N.~De~Angelis, \emph{{Studies of readout electronics and optical elements for a gamma-ray telescope}},  2019.

\bibitem{ESRF_data}
A.~Bacelj, N.~De~Angelis, A.~Elwertowska, M.~Kole, T.~Kowalski, G.~Koziol et~al., \emph{Final calibration measurements of the space based gamma-ray detector polar-2 using synchrotron radiation},  2026.
\newblock 10.15151/ESRF-ES-1092783487.

\bibitem{ID15}
``Id15a beamline details.'' Last consulted in February 2024.

\bibitem{Weisskopf}
M.C.~Weisskopf, R.F.~Elsner and S.L.~O'Dell, \emph{{On understanding the figures of merit for detection and measurement of x-ray polarization}}, \href{https://doi.org/10.1117/12.857357}{\emph{Proc. SPIE Int. Soc. Opt. Eng.} {\bfseries 7732} (2010) 77320E} [\href{https://arxiv.org/abs/1006.3711}{{\ttfamily 1006.3711}}].

\bibitem{Slawek}
{\scshape POLAR-2} collaboration, \emph{{Proton irradiation of SiPM arrays for POLAR-2}}, \href{https://doi.org/10.1007/s10686-022-09873-6}{\emph{Exper. Astron.} {\bfseries 55} (2023) 343} [\href{https://arxiv.org/abs/2210.01457}{{\ttfamily 2210.01457}}].

\bibitem{DeAngelis:2022jgd}
N.~De~Angelis et~al., \emph{{Temperature dependence of radiation damage annealing of Silicon Photomultipliers}}, \href{https://doi.org/10.1016/j.nima.2022.167934}{\emph{Nucl. Instrum. Meth. A} {\bfseries 1048} (2023) 167934} [\href{https://arxiv.org/abs/2212.08474}{{\ttfamily 2212.08474}}].

\end{thebibliography}\endgroup


\end{document}